\pdfoutput=1
\documentclass[10pt]{article}

\usepackage[utf8]{inputenc}
\usepackage{cite}
\usepackage{hyperref}
\hypersetup{colorlinks=true,linkcolor=blue2,citecolor=red2,urlcolor=green2,pdfencoding=auto,linktocpage}
\usepackage{amsmath,amssymb,amsbsy,amstext,amsthm,simplewick,amsfonts}
\usepackage{mathrsfs}
\usepackage{graphicx}
\usepackage{wrapfig}
\usepackage{upgreek}
\usepackage{bm} %
\usepackage{framed}
\usepackage{bbm}
\usepackage{textcomp}
\usepackage{adjustbox}
\usepackage{makecell}
\usepackage{tcolorbox}
\usepackage{empheq}
\usepackage[normalem]{ulem}
\usepackage{enumitem}
\usepackage{braket} 
\usepackage{array}
\usepackage{dsfont}
\usepackage{physics}
\usepackage{ulem}
\usepackage{xcolor}
\usepackage{caption}
\usepackage{subcaption}
\usepackage{tocloft}
\usepackage{tikz}
\usetikzlibrary{positioning}%
\usepackage{multirow}

\usetikzlibrary{positioning}
\usetikzlibrary{decorations.markings}
\usetikzlibrary{patterns.meta}
\tikzset{->-/.style={decoration={
  markings,
  mark=at position .5 with {\arrow{>}}},postaction={decorate}}}
\tikzset{-<-/.style={decoration={
  markings,
  mark=at position .5 with {\arrow{<}}},postaction={decorate}}}
\tikzset{-><-/.style={decoration={
  markings,
  mark=at position .33 with {\arrow{>}},,
  mark=at position .67 with {\arrow{<}}},postaction={decorate}}}
\tikzset{wb/.pic={
\filldraw[color=black, fill=white, thick] (0,0) circle (10pt);}}
\tikzset{bb/.pic={
\filldraw[color=black, fill=gray, thick] (0,0) circle (10pt);}}
\tikzset{cb/.pic={
\filldraw[color=black, fill=white, thick] (0,0) circle (10pt);
\draw[thick] (0,0)++(-45:10pt) -- ++(135:20pt);
\draw[thick] (0,0)++(-135:10pt) -- ++(45:20pt);}}

\usepackage[framemethod=default]{mdframed}

\newmdenv[backgroundcolor=gray!15,%
skipabove=5pt,%
skipbelow=5pt,%
leftmargin=2pt,%
rightmargin=2pt,%
innertopmargin=-6pt,%
innerbottommargin=5pt,%
innerleftmargin=5pt,%
innerrightmargin=5pt,%
splittopskip=0pt,%
splitbottomskip=0pt,%
linewidth=0pt,%
nobreak=true]%
{keyeqn}

\graphicspath{{Fig/}}

\definecolor{red2}{RGB}{214, 39, 40}
\definecolor{green2}{RGB}{0,170,0}
\definecolor{blue2}{RGB}{0,100,200}
\definecolor{magenta2}{RGB}{191,64,191}
\definecolor{purple2}{RGB}{112,48,160}
\definecolor{orange2}{RGB}{255,192,0}

\def\fnl{f_{\rm NL}^{\rm loc}}
\def\Tr{{\rm Tr}}

\def\d{\mathrm{d}}

\def\bfk{\mathbf{k}}
\def\bfp{\mathbf{p}}

\def\bfx{\mathbf{x}}

\def\Re{\mathrm{Re}\,}
\def\Im{\mathrm{Im}\,}
\def\S{\mathcal{S}}
\def\E{\mathcal{E}}
\def\LUV{\Lambda_{\text{UV}}}
\def\LIR{\Lambda_{\text{IR}}}
\def\Mpl{M_{\text{Pl}}}

\newcommand{\mcA}{{\mathcal{A}}}
\newcommand{\wT}{{\omega_\mathrm{T}}}

\newcommand{\kmin}{{k_\text{min}}}
\newcommand{\kmax}{{k_\text{max}}}

\numberwithin{equation}{section}

\allowdisplaybreaks[1]
\begin{document}

\begin{titlepage}
	\setcounter{page}{1} \baselineskip=15.5pt 
	\thispagestyle{empty}

     \begin{center}
		{\fontsize{18}{18}\centering {\bf{Perturbative unitarity bounds from momentum-space entanglement}}}
	\end{center}

	\vskip 18pt
	\begin{center}
		\noindent
		{\fontsize{12}{18}\selectfont Carlos Duaso Pueyo\footnote{\tt carlos.duasopueyo@sns.it}$^{,ab}$, Harry Goodhew\footnote{\tt goodhewhf@ntu.edu.tw}$^{,acd}$ Ciaran McCulloch\footnote{\tt cam235@cam.ac.uk}$^{,a}$ and Enrico Pajer\footnote{\tt enrico.pajer@gmail.com}$^{,a}$}
	\end{center}
	
	\begin{center}
		\vskip 12pt
		$^{a}$ \textit{Department of Applied Mathematics and Theoretical Physics, University of Cambridge,\\Wilberforce Road, Cambridge, CB3 0WA, UK} \\
        $^{b}$ \textit{Scuola Normale Superiore, Piazza dei cavalieri 7, 56126 Pisa, Italy} \\
        $^c$ \textit{Leung Center for Cosmology and Particle Astrophysics, Taipei 10617, Taiwan}\\
        $^d$ \textit{Center for Theoretical Physics, National Taiwan University, Taipei 10617, Taiwan}
	\end{center}

	\vskip 50pt
	\noindent\rule{\textwidth}{0.4pt}
	\noindent \textbf{Abstract} %
 Physical theories have a limited regime of validity and hence must be accompanied by a breakdown diagnostic to establish when they cease to be valid as parameters are varied. %
 For perturbative theories, estimates of the first neglected order offer valuable guidance, but one is often interested in sharp bounds beyond which perturbation theory necessarily fails. 
 In particle physics, it is common to employ the bounds on partial waves imposed by unitarity as such a diagnostic.
 Unfortunately, these bounds don't extend to curved spacetime, where scattering experiments are challenging to define. Here, we propose to use the growth of entanglement in momentum space as a breakdown diagnostic for perturbation theory in general field theories. This diagnostic can be readily used in cosmological spacetimes and does not require any flat spacetime limit. 
 More in detail, we consider the so-called \textit{purity} %
 of the reduced density operator constructed by tracing out all but one of the Fourier modes in an effective theory and we present a diagrammatic technique to compute it perturbatively. Constraints on the regime of validity of perturbation theory are then derived when the perturbative purity violates its unitarity bounds. 
 
 In flat spacetime, we compare these purity bounds to those from partial waves. We find general qualitative agreement but with remarkable differences: purity bounds can be sometimes weaker, but other times they exist when no partial wave bounds exist. We then derive purity bounds for scalar field theories in de Sitter spacetime for a variety of interactions that appear in inflationary models. Importantly, our bounds make no reference to time evolution and in de Sitter they depend exclusively on scale-invariant ratios of the physical kinematics.
	
	\noindent\rule{\textwidth}{0.4pt}

\end{titlepage}

\newpage
\setcounter{page}{2}
{
	\tableofcontents
}

\section{Introduction}\label{IntroSect}
In physics, just like in life, it is important to know when to stop. Our many different descriptions of physical phenomena are intrinsically partial and limited to specific situations and regimes. Much of the difficulty in the formulation of a scientific theory is to establish what aspects of the system under investigation are important and what are not. Because of this, it is in practice essential to be well aware of the intrinsic limitations of any given physical theory. Such limitations are conveniently separated into two camps. In the first camp live theories that contain the seed of their own downfall. This is the case of a theory that makes nonsensical predictions when extrapolated beyond its regime of validity. An example is a quantum mechanical theory in which certain interactions or particles' properties are chosen to break unitarity. The predictions of that theory cannot possibly be interpreted as probabilities and this signals that the theory has broken down. In the second camp are theories that appear to be fully self consistent, but in actuality fail to correctly describe the system because they miss important aspects of the problem. An example could be the Newtonian description of a relativistic phenomenon, such as the precession of the perihelion of Mercury or the light ring of a black hole. In this work we focus on the first camp and develop a diagnostic tool for the breakdown of effective field theories in a generic spacetime.\\

While theories can break down for many different reasons, we are interested here in the failure engendered by our attempt to solve a non-linear problem using \textit{perturbation theory}. This is a classical theme in physics and much has been written about the regime of validity of perturbation theory. At the most naive level, perturbation theory fails when we drop terms that are comparable to, or larger than, those we keep. This tautological test is never useful in practice as an exact diagnostic because if we knew precisely the next ``higher" order we would simply include it in our predictions. A more useful application of this criterion arises when we cannot or do not want to compute higher order terms but we are able to quickly estimate them. This then leads to approximate bounds on the validity of theory. A typical example of this are effective field theories where one attempts to estimate the breakdown of perturbation theory by using the power counting that organizes different operators. An archetypal, albeit quite simplistic example is when the validity of an effective field theory (EFT) is bounded by the scale that suppresses higher dimensional operators. \\

Because of the nature of the scientific endeavor and the sociological dynamics of scientific discoveries, we are surprisingly often pushed to live very close to the boundary of validity of a theory. This happens naturally when we take a modification of (some) standard model, which is well constrained by experiment X, and try to use it to explain a different phenomenon Y. To know if this is a viable option, we need to establish how ``flexible" our theory is and this more often than not pushes the theory towards the boundary of its regime of validity. Because this is such a general occurrence, it is in practice very useful to have \textit{breakdown diagnostic} tools that are as sharp as possible. A sharp bound on when a theory ceases to be self consistent provides a healthy boundary to otherwise wild speculations. \\

An important trade-off in the breakdown diagnostic business is that between precision and optimality. Certain
bounds might be very precise but sub-optimal, giving the illusion that a theory is still perfectly valid when instead it has already produced inconsistencies, but of the type that are not detected by our diagnostic. At the opposite end of the spectrum are very imprecise bounds, such as those from scaling estimates, that compensate for their lack of precision by capturing a vast gamut of ways in which mathematical self-consistency might fail. Because of this trade off, it is always useful to have a range of breakdown diagnostics that can fit different needs of precision and optimality.

\paragraph{The state of the art} In the particle physics and high energy context, many sharp bounds arise from the condition of unitarity in the quantum mechanical description. Here we use the term ``unitarity" as a catch-all expression for a variety of properties in the mathematical apparatus of quantum mechanics that ensure that the statistical predictions are meaningful, which typically means positive (classical) probabilities that add up to one. More in detail, meaningful probabilities require a variety of (some necessary and some sufficient) ingredients including a positive-definite inner product, unitary time evolution, an appropriate gauge-invariant quantization procedure, and so on. In this context, the breakdown diagnostic par excellence are \textit{partial wave unitarity bounds}. These are bounds on how strongly particles can interact in a scattering experiment. Most often these are used in a thought experiments where two particles collide to produce another 2 particles. Intuitively, there must be a bound on how strongly the incoming particles interact because unitary time evolution must preserve the norm of the incoming state vector. Because of the azimuthal symmetry of the initial conditions, this bound actually applies separately to each partial wave. These bounds have a very distinguished history and they have been successfully used to predict new physics, as for example in the exploration of the electroweak scale and the discovery of the Higgs boson.\\

Despite their usefulness in particle physics, partial wave unitarity bounds are not well suited for applications to curved spacetime, such as in cosmology and holography. The issue is that it is often difficult to precisely define scattering amplitudes in spacetimes that are not asymptotically flat (but see \cite{Mack:2009mi,Penedones:2010ue,Marolf:2012kh,Melville:2023kgd,Donath:2024utn,Melville:2024ove} for attempts in dS and AdS). One way out of this impasse is taking the flat spacetime limit of a theory in curved spacetime, hence effectively importing flat spacetime partial wave unitarity bounds to cosmology (see e.g. \cite{Baumann:2015nta,Grall:2020tqc,Melville:2019wyy,Kim_2021}). This procedure is at best an estimate of what intrinsic cosmological bounds would be, since it neglects all terms proportional to the Hubble parameter. More concerning is the fact that the class of consistent field theories in flat and curved spacetime are probably very different. Examples are supersymmetric theories, where supersymmetry is broken in a generic FLRW spacetime, or theories of a medium such as a (super)fluid or a (super)solid, which couple consistently to gravity only in curved spacetime \cite{Pajer:2020wnj}. Moreover, there is something quite puzzling about importing partial wave bounds to cosmological observables, especially in the much studied case of cosmological correlators. Amplitude bounds give us a scale at which an EFT breaks down. However, cosmological correlators often enjoy scale invariance and so only depend on the ratio of wavenumbers. \\

More generally, if we think of an EFT as a box that computes a class of observables, then we should expect that the box stops working in different regimes if we ask it to compute different observables. In fact, there is no reason to expect that the regime of validity of an EFT that computes sub-Hubble amplitudes is the same as that of an EFT that computes super-Hubble correlators. In fact, as an elucidating example, we will argue in this work that the regime of validity of EFTs in de Sitter is in general very different from that in their flat-space counterpart.

\subsection{Purity: perturbative unitarity bounds from entanglement} 

The discussion above motivated us to look for other types of perturbative unitarity bounds. In particular, we would like to work with quantities that can be defined easily in any spacetime, without specific assumptions about the space or time asymptotics. This paper will explore \textit{perturbativity bounds emerging from the growth of quantum entanglement}, as measured by the so-called purity. Before introducing these concepts, we would first like to give an intuition of why the strength of entanglement is a natural quantity to consider. The Hilbert space of a free quantum field theory is presented to us as a tensor product of harmonic oscillators corresponding to the different Fourier modes. Each Fourier mode is an energy eigenstate and so the wavefunction of the whole system is the product of single particle wavefunctions. Interactions change this picture and allow excitations to evolve into each other, relating states with different quantum numbers to one another. %
Correspondingly, the wavefunctions of different excitations become entangled. 
Hence, the growth of entanglement between these modes is a direct measurement of the strength of interactions. From this perspective, the quantum generation of primordial perturbations in the early universe can be equivalently thought of as the growth of entanglement in the wavefunction of the cosmos\footnote{Some measurements of entanglement can also decrease, see for example \cite{Colas:2022kfu,Colas:2024xjy,Burgess:2024eng} for a dedicated discussion.}. \\

\begin{figure}[h]
   \centering
   \includegraphics{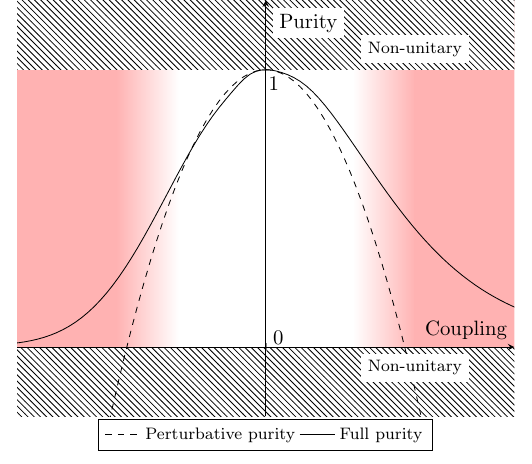}
   \caption{The purity of a single Fourier mode calculated perturbatively (dashed line) with a representative non-perturbative purity for comparison (solid line). The red shaded regions indicate the range of couplings where perturbation theory becomes unreliable, as diagnosed by when the perturbative purity (dashed line) becomes negative, hence violating the unitarity bound $0\leq \gamma \leq 1$. %
   \label{fig1}}
\end{figure}

The standard way to quantify entanglement of a pure state $\rho$ is to specify a bipartition of the Hilbert space of the system and to trace over one of its two parts. Following standard notation we will call the two parts of the bipartition \textit{system} $\mathcal{S}$ and \textit{environment} $\mathcal{E}$. The reduced density operator $\rho_R$ describing the system is
\begin{align}
    \rho_R=\Tr_\E\,\rho\,.
\end{align} 
Entanglement, then, is measured by how mixed $\rho_R $ is. A popular quantity to consider is the entanglement entropy, namely the von Neumann entropy of $\rho_R$. Here we prefer to work with a related but simpler quantity, the so-called purity $\gamma$, which is defined by
\begin{align}
    \gamma=\Tr_\S \,\rho_R^2\,.
\end{align}
From the defining properties of $\rho_R$, specifically $\rho_R\geq0$, $\Tr \rho_R=1$ and $\rho_R^\dagger=\rho_R$, one immediately sees that the purity must satisfy the unitarity bounds 
\begin{align}
       0\leq \gamma \leq 1\,.
\end{align}
While these bounds must hold non-perturbatively, it is immediate to see that they will not be satisfied in perturbation theory, simply because we are trying to approximate the function $\gamma(g)$ as a power law in some coupling constant $g$. This phenomenon is depicted in Figure~\ref{fig1}. There we show a representative\footnote{This is just a cartoon of the expected behaviour that purity decreases as we increase interactions. However, if $g$ is not directly related to the size of interaction, then the exact function $\gamma(g)$ could also be non-monotonic in $g$. Indeed a non-monotonic behaviour can be obtained for example from an interaction $f(g)\phi^n$ where $f$ is a non-monotonic function. It would be interesting to find the correct variable in terms of which $\gamma$ is necessarily monotonic in the non-perturbative theory.} non-perturbative purity $\gamma(g)$ and contrast it with its perturbative approximation. For a free theory, $g=0$, we expect $\gamma=1$, i.e. a pure state. Then, as long as $g$ is small, we expect perturbation theory to be accurate and so an analytic expansion around $g=0$ should look like 
\begin{align}
    \gamma(g)=1- \frac{g^2}{2} \left| \frac{\partial^2 \gamma}{\partial g^2} \right|+\mathcal{O}(g^3)\,,
\end{align}
where we assumed the theory remains unitary for either sign of $|g|\ll 1$, so that the linear term is absent and $\partial_g^2 \gamma\leq 0$. Clearly, when $g$ is large enough, $\gamma$ violates its unitarity bounds and this signals that this order in perturbation theory is not self consistent anymore. These purity bounds on $g$ translate into bounds on the range of validity of the EFT if we imagine keeping $g$ fixed but we extend the range of scales that are supposed to be well described by the EFT. To understand this we need to specify the bipartition we want to consider. 

\paragraph{Bounds on the range of validity of EFTs} To choose a bipartition, we take inspiration from the natural observables of the early universe, namely cosmological correlators. These correlators quite literally tell us how entangled different Fourier modes are with each other. Hence it is natural to specify a bipartition of the Hilbert space into two sets of Fourier modes and trace over one of them. To capture as much entanglement as possible and hence to get the strongest bound, we should trace over as much of the Hilbert space as possible. Therefore, for an arbitrary Fourier mode $\bfp$ of a given field\footnote{For simplicity, we often restrict ourselves to considering a single scalar field, but we consider also two scalars in Section~\ref{sec:flat_Sdphi2} and gravitons in Section~\ref{sec:GR}. Our proposal applies straightforwardly to theories with many fields of arbitrary (integer) spin.}, we define
\begin{align}
    \text{System $\mathcal{S}$}\coloneqq \phi_{\pm \bfp} \quad\quad  \text{Environment $\mathcal{E}$}\sim\{\phi_{\bfk}\}\, \text{s.t. } \bfk\neq \pm \bfp\,.
\end{align}
Then we consider the notion of the reduced density matrix $\rho_R(\bfp)=\Tr_\mathcal{E}\,\rho$ obtained by tracing over all other modes except $\bfp$. By isotropy $\rho_R(\bfp)=\rho_R(p)$. Since we want to compute $\gamma $ \textit{within} an EFT, and any EFT has a finite range of validity, we should be more careful about how we define the environment. \\

\begin{figure}
   \centering
   \includegraphics[scale=0.89]{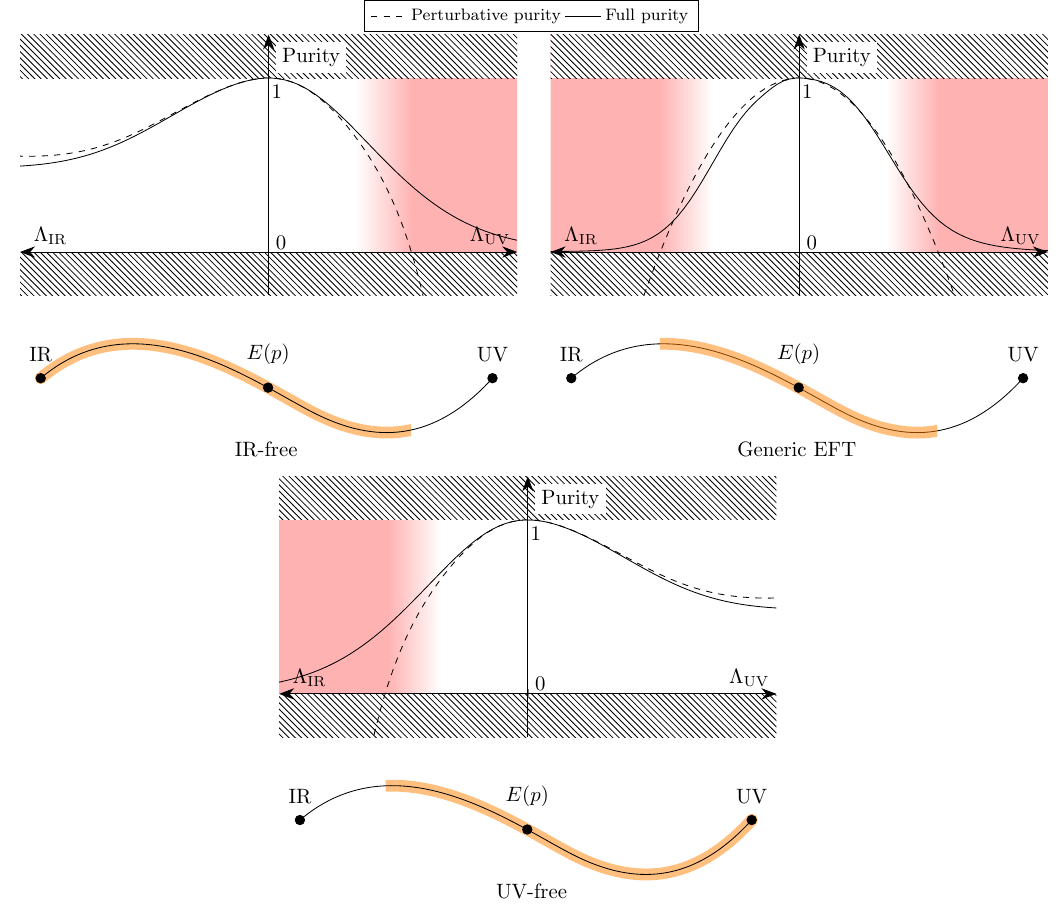}
   \caption{We plot the purity of a single Fourier mode $\bfp$ with associated energy $E(p)$ in three different EFTs as function of UV cutoff $\Lambda_\text{UV}$ and IR cutoff $\Lambda_\text{IR}$, for fixed couplings. In the left-hand side of each plot, we fix $\Lambda_\text{UV} = E(p)$ and vary $\Lambda_\text{IR} \leq E(p)$. In the right-hand side of each plot, we fix $\Lambda_\text{IR} = E(p)$ and vary $\Lambda_\text{UV} \geq E(p)$. The mid-point corresponds to $\Lambda_\text{UV} = \Lambda_\text{IR} = E(p)$, and the reduced density operator is pure as nothing has been traced out. The lower half of each plot sketches an RG trajectory in theory space. The orange highlighted region shows where the perturbative EFT is a good description, which is captured by the purity bound.\\ 
   We show three cases: \textit{(Top left)} The EFT is free in the IR. As the IR cutoff is lowered, the purity of the mode $\bfp$ approaches a nonzero limit as there is a decoupling of scales. The theory is not free in the UV; as the UV cutoff is raised, the mode $\bfp$ becomes increasingly entangled. \textit{(Top right)} The EFT is not free in the IR or the UV. It can be used only to describe an intermediate window of scales. \textit{(Bottom)} An EFT that is free in the UV (see Section~\ref{sec:flat_phicubed}). Perturbativity fails if the IR cutoff is set too low.\label{figRG}}
\end{figure}

Consider the simplest possible setup in which the validity of an EFT is controlled by a single energy scale $E$ and let $E(p)$ be the energy associated to the Fourier mode $\bfp$ by some dispersion relation. For fixed values of the couplings, the EFT is valid in some range of scales 
\begin{align}
\Lambda_{\text{IR}}\leq E\leq \Lambda_{\text{UV}}\,.    
\end{align}
In this range, the EFT is a convenient way to approximate the full theory in terms of the local, perturbative dynamics of few(er) effective degrees of freedom. This job becomes harder and harder as $ \Lambda_{\text{IR}} $ and $\Lambda_{\text{UV}}$ are moved away from $E$ and from each other and eventually the EFT breaks down when these cutoff scales cross the threshold where new degrees of freedoms mediate non-local interactions. Hence, if we want to estimate entanglement within the EFT we should restrict ourselves to
\begin{align}
   \mathcal{E}\coloneqq \{\phi_{\bfk}\}\, \text{s.t. } \bfk\neq \pm \bfp \,\, \& \,\, \Lambda_{\text{IR}}\leq E(k)\leq \Lambda_{\text{UV}}\,.
\end{align}
From the point of view of entanglement, what happens is that the EFT is describing more and more modes that are all coupled and entangled to each other, even in the vacuum of the interacting theory. Even though the couplings are kept constant, the entanglement between the mode $\bfp$ and the rest grows as $\Lambda_{\text{IR}}\to 0 $ and/or $ \Lambda_{\text{UV}} \to \infty$, simply because we are tracing over more and more modes. The reduced density matrix 
\begin{align}
    \rho_R(\Lambda_{\text{IR}},\Lambda_{\text{UV}},p)=\Tr_\E \rho=\underset{\substack{\bfk \neq\pm\bfp \\ \Lambda_{\text{IR}} \leq E(q) \leq \Lambda_{\text{UV}}}}{\Tr}   \rho
\end{align}
computed in the perturbative EFT becomes more and more mixed. As a consequence, the purity $\gamma$ of $\rho_R(p)$ decreases and eventually violates a unitarity bound (in our case $\gamma$ becomes negative). Calculating the $ \Lambda_{\text{IR,UV}}$ for which $\gamma$ computed at leading order in $g$ becomes negative gives us a sharp bound on the validity of that order in perturbation theory. But it also gives a good way to diagnose the proximity of the boundary of validity of the EFT, where the theory is inconsistent no matter how many orders in perturbation theory are included. \\

This general idea is summarized in the cartoon in Figure~\ref{figRG}. There we sketch the representative behaviour of the purity
\begin{align}
\gamma(\Lambda_{\text{IR}},\Lambda_{\text{UV}},p)=\Tr_\S \, \rho_R^2
\end{align}
in the full non-perturbative theory (continuous line) and in perturbation theory in the EFT (dashed line). The horizontal axis corresponds to varying $\Lambda_{\text{IR}}$ and $\Lambda_{\text{UV}}$. On the left of the vertical axis we vary $\Lambda_{\text{IR}}$ while keeping $\Lambda_{\text{UV}}=E(p)$ and on the right of the vertical axis we vary $\Lambda_{\text{UV}}$ while keeping\footnote{Because of these choices, the purity curve needs not be differentiable at the vertical axis in general.} $\Lambda_{\text{IR}}=E(p)$. A very schematic Renormalization Group (RG) flow underneath shows that the purity bounds are supposed to capture the interval of energies for which the full theory is well approximated by the EFT. \\

To conclude we should mention that there has been a lot of progress in understanding entanglement in quantum field theory over many years. One key difference in our analysis is that the entanglement we are discussing is across different Fourier modes, not between different regions of space, which is where most of the recent work has focused. Fourier-mode entanglement has been much less studied, although some previous literature has certainly inspired our approach \cite{Balasubramanian:2011wt}. Other works that consider entanglement to bound the strength of interactions include \cite{Aoude:2024xpx,Cheung:2023hkq,Peschanski:2016hgk,Cao:2022iqh,Kowalska:2024kbs,Brahma:2023lqm,Brahma:2023hki,Boutivas:2023mfg,Cheung_2019}. \\

\paragraph{Structure of the paper} The rest of the paper is organised as follows. In Section~\ref{sec:density matrix}, we review the description of the density matrix of a quantum field theory in the field basis, then describe a diagrammatic method to derive the purity of the reduced density matrix of a single Fourier mode. In Section~\ref{sec:perturbative unitarity}, we discuss three diagnostic tools for the breakdown of perturbation theory: perturbativity of the loop expansion, amplitudes' partial-wave unitarity, and a new purity bound based on the results of Section~\ref{sec:density matrix}. In Section~\ref{sec:flat spacetime}, we go on to apply these bounds to a range of theories on flat spacetime to compare purity bounds with more established results. We find general qualitative agreement when all diagnostic tools provide bounds, and discuss some differences. Next, Section~\ref{sec:de Sitter} applies the purity bound to quantum field theory in the Poincar\'e patch of de Sitter spacetime, where defining scattering amplitudes or estimating the size of loop corrections is more difficult. We consider interactions from the effective field theory of inflation; another interaction which we show can be perturbative only for a limited range of kinematics; and local-type non-gaussianity of a spectator scalar. These de Sitter bounds show a range of qualitative and quantitative differences from those on flat spacetime. We conclude in Section~\ref{ConclusionsSect} with a summary of our results and future directions.

\vskip 10pt
\noindent
{\it Note added:} 
As we prepared our ArXiv submission, Ref.~\cite{Ueda} appeared with a similar analysis exploiting the non-negativity of an information theoretic quantity. However, there they explored the relative entropy between the free and interacting theories whereas we focus on the purity. The extent to which these overlap is not clear to us at this time.

\paragraph{Notation and conventions}
We parameterize the field-theoretic wavefunction at some time $t_0$ (equivalently at conformal time $\eta_0$) as 
\begin{align} \label{psin}
\Psi[\phi;t_0]=\exp\left[   -\sum_{n}^{\infty}\frac{1}{n!} \int_{\bfk_{1},\dots,\bfk_{n}}\,  (2\pi)^{3} \delta_D^{(3)} \left( \sum_{a}^{n} \bfk_{a} \right) \psi_{n}(\{\bfk\};t_0) \, \phi(\bfk_{1})\dots \phi(\bfk_{n})\right]\,,
\end{align}
where to simplify the notation we have introduced
\begin{align}
\int_{\bfk} \equiv \int \frac{\d^3 k}{(2\pi)^{3}} \, .
\end{align}
We will be using conformal coordinates on a flat slicing,
\begin{equation}
    ds^2 = a(\eta)^2 (-d\eta^2+d\mathbf{x}^2 ) \, ,
\end{equation}
focusing on two spacetimes. The first is Minkowski, $a(\eta)=1$, for which we will pick the time slice $t_0=0$ without loss of generality. The second is de Sitter, $a(\eta)=-1/H\eta$, for which we will compute the wavefunction at the future boundary $\eta_0=0$. 

For 4-point scattering amplitudes with external masses $m$ we will use the Mandelstam variables
\begin{equation}
    s=-(p_1+p_2)^2 \, , \qquad t=-(p_1+p_4)^2 \, , \qquad u=-(p_1+p_3)^2=4m^2-s-t \, ,
\end{equation}
which in the center-of-mass frame ($\mathbf{p}_1=-\mathbf{p}_2$ and $p_1^{(0)}=p_2^{(0)}\equiv E$) become
\begin{equation} \label{eq:com}
    \left\{ \begin{aligned}
        s&\doteq 4E^2 \, , \\
        t&\doteq -2\left(E^2-m^2\right) (1-\cos\theta) \, , \\
        u&\doteq -2\left(E^2-m^2\right) (1+\cos\theta) \, ,
    \end{aligned} \right.
\end{equation}
with $\theta$ the angle between $\mathbf{p}_1$ and $\mathbf{p}_3$. 

Finally, following the conventions in \cite{BBBB}, we will sometimes write the symmetric three-point wavefunction coefficients in terms of the elementary symmetric polynomials,
\begin{align}\label{esp}
    k_T\equiv k_1+k_2+k_3 \, , \quad e_2\equiv k_1 k_2+k_1 k_3+k_2 k_3 \, , \quad e_3\equiv k_1k_2k_3 \, .
\end{align}

\subsection{Summary of the results} For the convenience of the reader, in the following we summarize our main results:
\begin{itemize}
    \item We derived a diagrammatic representation of perturbative contributions to the purity $\gamma$ of the reduced density matrix $\rho(p)$ of a single Fourier mode $\bfp$. The result is remarkably simple. Moreover, our analysis applies also to perturbative calculations of $\Tr \rho^N$ for any $N$. We get
    \begin{equation} 
        \frac{{\rm Tr}\, \rho_R^N}{\left( {\rm Tr}\, \rho_R \right)^N} = \exp \left( - N \sum D^{(c)} \right) = \left[ \frac{{\rm Tr}\, \rho_R^M}{\left( {\rm Tr}\, \rho_R \right)^M} \right]^{N/M} \qquad \text{for } N,M\geq 2 \, ,
    \end{equation}
    where $D^{(c)}$ is the set of diagrams defined in~\eqref{eq:Ddiagdef}.
    \item We compared our perturbative unitarity bounds from purity, or \textit{purity bounds} for short, to partial wave bounds in flat space for simple cubic interactions of a massive scalar with itself and with another massive scalar. For general interactions we found that the bounds are qualitatively similar and can be related to the scaling expected by estimating higher order contributions. One should keep in mind that perturbation theory may break down in different regimes for different observables. For a more in depth discussion of this issue see the text around \eqref{below320}. 
    \item Purity bounds can be weaker but also much stronger than partial wave bounds. An extreme example is the field redefinition of a free theory. Since amplitudes are field-redefinition invariant, partial waves do not provide any bounds. Conversely, purity detects the physical effect that different Fourier modes of the interacting ``redefined" field are coupled to each other and correctly diagnoses the breakdown of perturbation theory
    \item We used our purity bounds to detect the range of validity of perturbative scalar field theories in de Sitter spacetime, focusing on interactions that are relevant for inflation, such as those arising in the effective field theory of inflation \cite{Cheung:2007st}. We compare these bounds to the previous literature. These examples showcase the importance of \textit{``cosmology native" bounds}, which do not rely on a flat spacetime limit. 
    \item In contrast to flat-space amplitudes, where bounds are given on (dimensionless) ratios of coupling constants and the center of mass energy, in cosmology bounds involve coupling constants, the Hubble parameter and the ratio of wavenumbers. Perturbation theory fails when, for a fixed coupling constant and fixed Hubble, we push to extreme ratios of kinematics, such as for example in the squeezed limit of the three point function (see Figure~\ref{fig:squeezed cutoff}) or for folded configurations in the case of a Bogoliubov initial state (see Figure~\ref{fig:bogo kinematics}).
    \item We present a preliminary investigation of our purity bound in the case of dynamical gravity. Intriguingly, our final result \eqref{finalGR} matches what would be expected from requiring that the maximum entropy for a given region of space is that of a black hole of equal size.
\end{itemize}

\section{Purity in quantum field theory}
\label{sec:density matrix}

In this section, after briefly introducing the concepts of a reduced density matrix and purity, we derive diagrammatic rules for the calculation of purity in quantum field theory, having in mind a bipartition of the Hilbert space into two sets of Fourier modes. To this end, we start by working in a finite volume. Taking the infinite volume limit results in a dramatic simplification and our final results is \eqref{eq:resPurity}, where the \textit{purity diagrams} $D^{(c)}$ are defined in \eqref{eq:Ddiagdef} and below. 

The reader who is mostly interested in purity bounds on the validity of perturbation theory and their pragmatic implementation may skip this section and move directly to Section~\ref{sec:perturbative unitarity}.

\subsection{Density matrix and the wavefunction}

Our staring point is a quantum effective field theory. Throughout this work we will assume that we are in a vacuum state of this \textit{interacting} EFT, which we denote by $|\Omega\rangle$. The associated density matrix of this pure state is simply the projector
\begin{equation}
    \rho = |\Omega\rangle \langle \Omega | \, .
\end{equation}
This operator is Hermitian, $\rho^\dagger=\rho$, and positive semi-definite, $\bra{\psi}\rho\ket{\psi} \geq 0$. However we find it convenient \textit{not} to normalize $\rho$, i.e. in general we have $\Tr \rho \neq 1$. The technical reason is that the trace is divergent in the infinite volume limit and it is useful to keep track of this divergence explicitly. The connection between the density matrix and the wavefunction~\eqref{psin} is made explicit in the basis of field eigenstates, which satisfy $\hat{\phi}(\bfx)\ket{\phi}=\phi(\bfx)\ket{\phi}$, where, just for this equation, we have explicitly indicated the operator $\hat\phi$ with a hat, but we will drop this in the following. We are free to insert two resolutions of the identity in terms of such eigenstates,
\begin{equation}
    \rho = \int \mathcal{D}\phi \mathcal{D}\bar{\phi} \, |\phi\rangle \langle\phi |\rho|\bar{\phi}\rangle \langle\bar{\phi}| = \int \mathcal{D}\phi \mathcal{D}\bar{\phi} \, \rho_{\phi\bar{\phi}} \, |\phi\rangle \langle\bar{\phi}|\,.
\end{equation}
From this, we see immediately that the components $\rho_{\phi\bar \phi}$ of the operator $\rho$ in the field basis are
\begin{equation} \label{eq:dmatrixField}
    \rho_{\phi\bar{\phi}} = \langle\phi |\rho|\bar{\phi}\rangle = \langle\phi |\Omega\rangle \langle \Omega|\bar{\phi}\rangle = \Psi [\phi] \Psi[\bar{\phi}]^* \, ,
\end{equation}
where $\Psi[\phi] \equiv \langle\phi |\Omega\rangle$ is the field-theoretic wavefunction corresponding to the state $\ket{\Omega}$. \\

We will work with a theory of a single real scalar field $\phi$. Reality implies that in momentum space we have $\phi_{\bfk}=\phi^*_{-\bfk}$. To better handle the contributions from the different momenta and properly handle infinities, we will put the theory on a torus of volume $L^3$. In practice, while $L<\infty$ all of our expressions are finite and the trace operation over the Hilbert space is well defined. Then, we notice that the single-Fourier mode purity we compute remains finite as we take the volume to infinity. This appears to provide a self-consistent procedure.

The wavefunction~\eqref{psin} at some fixed time $t$ is then
\begin{equation} \label{eq:wfcoefs}
	\Psi [\phi] = \frac{1}{\mathcal{N}} \, \exp \left[ -\sum_{n=2}^{\infty} { \frac{1}{n!} \frac{1}{L^{3n}} \sum_{\mathbf{k}_1,\ldots,\mathbf{k}_n} L^3 \, \delta_{\mathbf{k}_1+\ldots+\mathbf{k}_n,\mathbf{0}} \, \psi_n (\mathbf{k}_1,\ldots,\mathbf{k}_n) \, \phi_{\mathbf{k}_1} \cdots \phi_{\mathbf{k}_n} } \right] \, .
\end{equation}
Unless otherwise specified, the sums and products over $\bfk$ run over all momenta (appropriately quantized in the finite volume case). As we mentioned, we will not assume that $\rho$ or $\Psi$ are normalized to unity. Instead, we rescale $\Psi$ by the norm $\mathcal{N}$ of the \textit{free} theory vacuum $|0\rangle$,
\begin{equation}
    \mathcal{N}^{\,2} \equiv \langle 0 | 0 \rangle = \left( \prod_{\bfk} \int \mathcal{D} \phi_{\bfk} \right) \exp \left( -\sum_{\bfk} { \frac{\text{Re}\, \psi_2(\bfk)}{L^{3}} \, |\phi_{\mathbf{k}}|^2 } \right) \, .
\end{equation}
As a consequence of this, the trace of $\rho$ is not fixed to unity, but rather
\begin{equation} \label{eq:Trrhodef}
    {\rm Tr}\, \rho = \int \mathcal{D} \phi \, \rho_{\phi\phi} = \int \mathcal{D} \phi \left| \Psi[\phi] \right|^2 \, .
\end{equation}
In analogy to the parameterization of $\Psi$, we introduce the following parameterization of the \textit{diagonal}\footnote{In principle we could also have introduced a parameterization that includes the off-diagonal elements, $ \rho_{\phi\bar{\phi}} \sim \exp(-\sum \rho_{nm}\phi^n \bar{\phi}^m) $, but we will not need this notation in the following.} elements of $\rho$,
\begin{equation} \label{eq:rhocoefs}
	\rho_{\phi\phi} = \frac{1}{\mathcal{N}^2} \, \exp \left[ -\sum_{n=2}^{\infty} { \frac{1}{n!} \frac{1}{L^{3n}} \sum_{\mathbf{k}_1,\ldots,\mathbf{k}_n} L^3 \, \delta_{\mathbf{k}_1+\ldots+\mathbf{k}_n,\mathbf{0}} \, \rho_n (\mathbf{k}_1,\ldots,\mathbf{k}_n) \, \phi_{\mathbf{k}_1} \cdots \phi_{\mathbf{k}_n} } \right] \, ,
\end{equation}
where
\begin{equation} \label{eq:rhodef}
    \rho_n (\mathbf{k}_1, \ldots, \mathbf{k}_n) \equiv \psi_n (\mathbf{k}_1, \ldots, \mathbf{k}_n) + \psi_n^* (-\mathbf{k}_1, \ldots, -\mathbf{k}_n) \, .
\end{equation}
We should stress that time evolution plays no role in this story. We assume that we have been given the wavefunction of some $\ket{\Omega}$ state of an EFT at some time and we quantify its entanglement at that time. In practice, if one is given a theory via a Lagrangian, one should first compute $\Psi$ via the familiar path integral and then make contact with our analysis. \\

Now we want to specify a bipartition of the Hilbert space. We choose one non-vanishing mode $\bfp\neq 0$ as our system $\mathcal{S}$ and all other modes $\bfk \neq \bfp$ as the environment $\mathcal{E}$. As mentioned earlier, if the EFT has a finite range of validity we have to restrict the environment to modes that are within the remit of the EFT
\begin{align} \label{eq:defEnv}
    \mathcal{E}\coloneqq \{\phi_{\bfk}\}\, \text{s.t. } \bfk\neq \pm \bfp \,\, \& \,\, \Lambda_{\text{IR}}\leq E(q)\leq \Lambda_{\text{UV}}\,.
\end{align}
It will be convenient to separately denote the dependence of the wavefunction on the environment modes and the system modes $\phi_{\pm\bfp}$ in the following way,
\begin{equation}
    \Psi[\phi_{\bfk},\phi_{\pm\bfp}] \quad\text{with } \bfk \in \mathcal{E} \text{ and } \pm\bfp \in \mathcal{S} \, .
\end{equation}
The reduced density matrix $\rho_R$ of $\phi_{\pm\bfp}$ is obtained from $\rho$ by tracing out the modes in the environment, $\rho_R = {\rm Tr}_{\mathcal{E}} \, \rho$, which in the field basis means
\begin{equation} \label{eq:rhoR}
    (\rho_R(p))_{\phi\bar{\phi}} = \left( \prod_{\bfk \in \mathcal{E}} \int \mathcal{D} \phi_{\bfk} \right) \rho_{\phi\bar{\phi}} \Big|_{\substack{\phi_{\bfk}=\bar{\phi}_{\bfk} \\ \text{ for } \bfk\in\mathcal{E}}} = \left( \prod_{\bfk \in \mathcal{E}} \int \mathcal{D} \phi_{\bfk} \right) \Psi[\phi_{\bfk},\phi_{\pm\bfp}] \Psi[\phi_{\bfk},\bar{\phi}_{\pm\bfp}]^* \, .
\end{equation}
Notice that this ``single-mode" reduced density matrix $\rho_R$ is not anymore a functional, but rather just a function of the two variables $\phi_\bfp$ and $\phi_{-\bfp}$. From now on, we will work with the density matrix components in the field basis. We will sometimes drop the field subscripts to avoid clutter. To conclude, we recall that purity is a particular measure of how mixed a given state described by $\rho_R$ is. It is defined by
\begin{align} \label{eq:puritydef}
\gamma(\Lambda_{\text{IR}},\Lambda_{\text{UV}},p)=\Tr_\S \, \rho_R^2\,.
\end{align}
We will often omit its arguments when this does not engender confusion. 

As pointed out in~\cite{Balasubramanian:2011wt}, the interpretation of the entanglement entropy of a single Fourier mode is subtle. It cannot be understood as just the entropy of a region of Fourier space in the limit in which it is narrowed down to a single momentum $\bfp$, since the former is divergent and the latter is not. The same applies to the purity that we compute in this work. The correct interpretation, according to~\cite{Balasubramanian:2011wt}, is that the entropy $S_{EE}$ of a single mode equals the entropy density (in position space) of the modes in an infinitesimal range $\d^3\bfp$ around $\bfp$. This should similarly apply to the purity that we compute.

\subsection{A diagrammatic approach} \label{sec:diagrams}

Here we present a step-by-step diagrammatic procedure that streamlines the computation of the reduced density matrix $\rho_R$ and any ${\rm Tr} \, \rho_R^N$ for the bipartition defined in the previous subsection. Although straightforward, this method can still be time-consuming due to the proliferation of diagrams, but we will show that it can be greatly simplified if the theory lives in an infinite-volume space, which is the case of interest for us. In that limit, the diagrams contributing to ${\rm Tr} \, \rho_R^N$ can be readily drawn by following a simple set of rules.

We want to compute traces (or, in the field basis language, path integrals) of the density matrix~\eqref{eq:dmatrixField} in perturbation theory. This means that we will expand in the wavefunction coefficients $\psi_n$ with $n\geq 3$, and the integral will be given by all possible Wick contractions among the field profiles $\phi$ and $\bar{\phi}$ in the expansion. The combinatorics of such a computation can be conveniently represented by diagrams formed by lines and blobs at the vertices where lines meet. A blob with $n$ lines attached corresponds to a wavefunction coefficient $\psi_n$. There will be three types of blob corresponding to the different ways in which wavefunction coefficients appear, namely
\begin{gather}
\vcenter{\hbox{\begin{tikzpicture}
    \coordinate (w) at (-100pt,0);
    \coordinate (g) at (100pt,0);
    \draw[thick] (w) -- ++(60:40pt);
    \draw[thick] (w) -- ++(-60:40pt); 
    \draw (w) pic {wb};
    \filldraw[color=black] (w)++(-25:25pt) circle (0.5pt); 
    \filldraw[color=black] (w)++(0:25pt) circle (0.5pt); 
    \filldraw[color=black] (w)++(25:25pt) circle (0.5pt); 
    \draw[thick,<-,color=gray] (w)++(6pt,22pt) -- ++(60:13pt);
    \node[color=gray] at ([shift={(29pt,32pt)}]w) {$\mathbf{k}_1$};
    \draw[thick,<-,color=gray] (w)++(6pt,-22pt) -- ++(-60:13pt);
    \node[color=gray] at ([shift={(29pt,-32pt)}]w) {$\mathbf{k}_n$};
    \node[right] at ([shift={(45pt,0)}]w) {$= - \dfrac{\psi_n (\mathbf{k}_1, \ldots, \mathbf{k}_n)}{L^{3n}} \, ,$};
    \draw[thick] (g) -- ++(60:40pt);
    \draw[thick] (g) -- ++(-60:40pt); 
    \draw (g) pic {bb};
    \filldraw[color=black] (g)++(-25:25pt) circle (0.5pt); 
    \filldraw[color=black] (g)++(0:25pt) circle (0.5pt); 
    \filldraw[color=black] (g)++(25:25pt) circle (0.5pt); 
    \draw[thick,<-,color=gray] (g)++(6pt,22pt) -- ++(60:13pt);
    \node[color=gray] at ([shift={(29pt,32pt)}]g) {$\mathbf{k}_1$};
    \draw[thick,<-,color=gray] (g)++(6pt,-22pt) -- ++(-60:13pt);
    \node[color=gray] at ([shift={(29pt,-32pt)}]g) {$\mathbf{k}_n$};
    \node[right] at ([shift={(45pt,0)}]g) {$= - \dfrac{\psi_n^* (-\mathbf{k}_1, \ldots, -\mathbf{k}_n)}{L^{3n}} \, ,$};
\end{tikzpicture}}} \label{eq:wbblobs} \\
\vcenter{\hbox{\begin{tikzpicture}
    \coordinate (c) at (0,0);
    \draw[thick] (c) -- ++(60:40pt);
    \draw[thick] (c) -- ++(-60:40pt); 
    \draw (c) pic {cb};
    \filldraw[color=black] (c)++(-25:25pt) circle (0.5pt); 
    \filldraw[color=black] (c)++(0:25pt) circle (0.5pt); 
    \filldraw[color=black] (c)++(25:25pt) circle (0.5pt); 
    \draw[thick,<-,color=gray] (c)++(6pt,22pt) -- ++(60:13pt);
    \node[color=gray] at ([shift={(29pt,32pt)}]c) {$\mathbf{k}_1$};
    \draw[thick,<-,color=gray] (c)++(6pt,-22pt) -- ++(-60:13pt);
    \node[color=gray] at ([shift={(29pt,-32pt)}]c) {$\mathbf{k}_n$};
    \node[right] at ([shift={(45pt,0)}]c) {$= - \dfrac{\rho_n (\mathbf{k}_1, \ldots, \mathbf{k}_n)}{L^{3n}} %
    \, ,$};
\end{tikzpicture}}} \label{eq:cblob}
\end{gather}
where $\rho_n$ and $\psi_n$ were defined in~\eqref{eq:rhocoefs} and~\eqref{eq:wfcoefs}.
A line that connects two blobs or starts and ends in the same blob represents the Wick contraction of a Fourier mode. There will be three different types of such lines depending on whether the mode is generic and can have any momentum (solid line), belongs to the environment and hence can only have momentum $\bfk\in\mathcal{E}$ (dashed), or belongs to the system and has the selected momentum $\pm\bfp$ (arrowed):
\begin{align} 
& \vcenter{\hbox{\begin{tikzpicture}
    \coordinate (a) at (45pt,0pt);
    \draw[thick] (a)++(-70pt,0) -- ++(60pt,0);
    \draw[thick,->,color=gray] (a)++(-50pt,-8pt) -- ++(20pt,0);
    \node[color=gray] at ([shift={(-22pt,-10pt)}]a) {$\bfk$};
    \node[right] at (a) {$= \,\, \dfrac{L^3}{\rho_2(\bfk)} = \dfrac{L^3}{2\text{Re}\, \psi_2(\bfk)} \qquad {\rm for} \quad \bfk\in \mathcal{S} \cup \mathcal{E} \,, $}; 
\end{tikzpicture}}} \label{eq:solidline} \\
& \vcenter{\hbox{\begin{tikzpicture}
    \coordinate (a) at (45pt,0pt);
    \draw[thick,dashed] (a)++(-70pt,0) -- ++(60pt,0);
    \draw[thick,->,color=gray] (a)++(-50pt,-8pt) -- ++(20pt,0);
    \node[color=gray] at ([shift={(-22pt,-10pt)}]a) {$\bfk$};
    \node[right] at (a) {$= \,\, \dfrac{L^3}{\rho_2(\bfk)} = \dfrac{L^3}{2\text{Re}\, \psi_2(\bfk)} \qquad {\rm for} \quad \bfk\in\mathcal{E} \, ,$};
\end{tikzpicture}}} \label{eq:dashedline} \\
& \vcenter{\hbox{\begin{tikzpicture}
    \coordinate (a) at (45pt,0pt);
    \draw[thick,->-] (a)++(-70pt,0) -- ++(60pt,0);
    \draw[thick,->,color=gray] (a)++(-50pt,-8pt) -- ++(20pt,0);
    \node[color=gray] at ([shift={(-22pt,-10pt)}]a) {$\bfp$};
    \node[right] at (a) {$= \,\, \dfrac{L^3}{\rho_2(\bfp)} = \dfrac{L^3}{2\text{Re}\, \psi_2(\bfp)}\qquad {\rm for} \quad \bfp\in\mathcal{S}  \, .$};
\end{tikzpicture}}} \label{eq:arrowedline}
\end{align}
The diagrams that contribute to the reduced density matrix $\rho_R$ will also display external lines connected to black or white blobs. They represent the selected field modes $\phi_{\pm\bfp}$ and $\bar{\phi}_{\pm\bfp}$, which are not yet traced out in $\rho_R$:
\begin{equation} \label{eq:extlines}
\vcenter{\hbox{\begin{tikzpicture}
    \coordinate (w) at (-75pt,0);
    \coordinate (g) at (75pt,0);
    \coordinate (wl) at (-75pt,-35pt);
    \coordinate (gl) at (75pt,-35pt);
    \draw[thick,-<-] (w)++(0:10pt) -- ++(0:40pt);
    \draw (w) pic {wb};
    \node[right] at ([shift={(60pt,0)}]w) {$= \, \phi_{\mathbf{p}} \,\, ,$};
    \draw[thick,-<-] (g)++(0:10pt) -- ++(0:40pt);
    \draw (g) pic {bb};
    \node[right] at ([shift={(60pt,0)}]g) {$= \, \bar{\phi}^{\, *}_{-\mathbf{p}} \,\, ,$};
    \draw[thick,->-] (wl)++(0:10pt) -- ++(0:40pt);
    \draw (wl) pic {wb};
    \node[right] at ([shift={(60pt,0)}]wl) {$= \, \phi_{-\mathbf{p}} \,\, ,$};
    \draw[thick,->-] (gl)++(0:10pt) -- ++(0:40pt);
    \draw (gl) pic {bb};
    \node[right] at ([shift={(60pt,0)}]gl) {$= \, \bar{\phi}^{\, *}_{\mathbf{p}} \,\, .$};
\end{tikzpicture}}}
\end{equation}
Finally, all diagrams will obey these simple rules:
\begin{equation} \label{eq:diagrules}
    \parbox{0.90\textwidth}{\centering 
    \begin{itemize}
        \item Every blob comes with a Kronecker delta that imposes momentum conservation, cf.~\eqref{eq:wfcoefs} and~\eqref{eq:rhocoefs}.
        \item Momenta running in loops of solid lines~\eqref{eq:solidline} must be summed over all $\bfk\in \mathcal{S} \cup \E$.
        \item Momenta running in loops of dashed lines~\eqref{eq:dashedline} must be summed over all $\bfk\in\mathcal{E}$.
        \item Each diagram is divided by its symmetry factor.
    \end{itemize}
    }
\end{equation}
Having set the ground, we are ready to calculate the quantities of interest.

\paragraph{Tr $\boldsymbol{\rho_R}$} It is convenient to start by computing the norm of the reduced density matrix, which is given by its trace. It can be done even before having an explicit expression for $\rho_R$, since this trace is actually equal to the trace of the full density matrix $\rho$. Then, using~\eqref{eq:Trrhodef} and~\eqref{eq:rhocoefs},
\begin{equation}
    \begin{aligned}
            {\rm Tr}\, \rho_R & = {\rm Tr}_{\mathcal{S}}\, ({\rm Tr}_{\mathcal{E}} \, \rho) = \int \mathcal{D} \phi \, \rho_{\phi\phi} %
            \\
            & = \frac{1}{\mathcal{N}^{\,2}} \left( \prod_{\bfk} \int \mathcal{D} \phi_{\bfk} \right) \exp \left[ -\sum_{n=2}^{\infty} { \frac{1}{n!} \frac{1}{L^{3n}} \sum_{\mathbf{k}_1,\ldots,\mathbf{k}_n} L^3 \, \delta_{\mathbf{k}_1+\ldots+\mathbf{k}_n,\mathbf{0}} \, \rho_n (\mathbf{k}_1,\ldots,\mathbf{k}_n) \, \phi_{\mathbf{k}_1} \cdots \phi_{\mathbf{k}_n} } \right] \, .
    \end{aligned}
\end{equation}
We are interested in computing the trace in perturbation theory, so we expand the integrand in the coefficients $\rho_n$ with $n\geq 3$ getting
\begin{equation} \label{eq:traceexpansion}
    \begin{aligned}
            {\rm Tr}\, \rho_R &=  \, \frac{1}{\mathcal{N}^{\,2}} \left( \prod_{\bfk} \int \mathcal{D} \phi_{\bfk} \right) \exp \left( -\sum_{\bfk} { \frac{\rho_2 (\mathbf{k})}{2L^{3}} \, |\phi_{\mathbf{k}}|^2 } \right) \\[2pt]
            & \times \left[ 1 - \sum_{n=3}^{\infty} { \frac{1}{n!} \frac{1}{L^{3n}} \sum_{\mathbf{k}_1,\ldots,\mathbf{k}_n} L^3 \, \delta_{\mathbf{k}_1+\ldots+\mathbf{k}_n,\mathbf{0}} \, \rho_n (\mathbf{k}_1,\ldots,\mathbf{k}_n) \, \phi_{\mathbf{k}_1} \cdots \phi_{\mathbf{k}_n} } \right. \\
            & \qquad\qquad \left. + \frac{1}{2} \left( \sum_{n=3}^{\infty} { \frac{1}{n!} \frac{1}{L^{3n}} \sum_{\mathbf{k}_1,\ldots,\mathbf{k}_n} L^3 \, \delta_{\mathbf{k}_1+\ldots+\mathbf{k}_n,\mathbf{0}} \, \rho_n (\mathbf{k}_1,\ldots,\mathbf{k}_n) \, \phi_{\mathbf{k}_1} \cdots \phi_{\mathbf{k}_n} } \right)^2 + \ldots \right] \, .
    \end{aligned}
\end{equation}

As announced, we will represent the result of this integral using diagrams. Notice that since we are looking at the diagonal of $\rho$, only the combinations $\rho_n$ appear in~\eqref{eq:traceexpansion}, so the diagrams will only contain crossed blobs~\eqref{eq:cblob}. Furthermore, since we are integrating over all modes $\bfk\in\S\cup \E$ without distinction, the diagrams will only contain solid lines~\eqref{eq:solidline}. We conclude that the integral will be given by all possible diagrams consisting of solid lines joining crossed blobs (up to the desired order in the couplings) and subject to the rules~\eqref{eq:diagrules}. Let us call such diagrams \textit{trace diagrams}, denoting them collectively by $D_T$, and let $D_T^{(c)}\subset D_T$ be the subset of connected ones. Then we have the general result
\begin{equation} \label{eq:Trrho}
    {\rm Tr}\, \rho_{\rm R} = 1 + \sum D_T = \exp \left( \sum D_T^{(c)} \right) \, .
\end{equation}

Let us illustrate all of this with an example. For a theory which has only a cubic wavefunction coefficient we have, at $\mathcal{O}(\psi_3^2)$,
\begin{equation} \label{eq:TrrhoPsi3}
\begin{aligned}
    {\rm Tr}\, \rho_{\rm R} = 1 & + 
    \frac{1}{12} \, \vcenter{\hbox{\begin{tikzpicture}
    \coordinate (o) at (0,0);
    \coordinate (c1) at ([shift={(-30pt,0)}]o);
    \coordinate (c2) at ([shift={(30pt,0)}]o);
    \draw[thick] (c1) -- (c2);
    \draw[thick] (c1) to[out=-90,in=-90] (c2);
    \draw[thick] (c1) to[out=90,in=90] (c2);
    \draw (c1) pic {cb};
    \draw (c2) pic {cb};
    \end{tikzpicture}}}
    + \frac{1}{8} \hspace{-25pt} \vcenter{\hbox{\begin{tikzpicture}
    \coordinate (o) at (0,0);
    \coordinate (c1) at ([shift={(-30pt,0)}]o);
    \coordinate (c2) at ([shift={(30pt,0)}]o);
    \draw[thick] (c1) -- (c2);
    \draw[thick] (c1) to[out=45,in=135,distance=60pt] (c1);
    \draw[thick] (c2) to[out=45,in=135,distance=60pt] (c2);
    \path (c1) to[out=-90,in=-90] (c2);
    \draw (c1) pic {cb};
    \draw (c2) pic {cb};
    \end{tikzpicture}}} \hspace{-25pt} \\[-5pt]
    = 1 & + \frac{1}{12 \, L^3} \sum_{\bfk_1,\bfk_2} \frac{\rho_3 (\mathbf{k}_1, \mathbf{k}_2, -\mathbf{k}_1-\mathbf{k}_2) \, \rho_3 (-\mathbf{k}_1, -\mathbf{k}_2, \mathbf{k}_1+\mathbf{k}_2)}{\rho_2(\mathbf{k}_1) \, \rho_2(\mathbf{k}_2) \, \rho_2(\mathbf{k}_1+\mathbf{k}_2)} \\
    & + \frac{1}{8 \, L^3} \sum_{\bfk_1,\bfk_2} \frac{\rho_3 (\bfk_1, -\bfk_1, \mathbf{0}) \, \rho_3 (\bfk_2, -\bfk_2, \mathbf{0})}{\rho_2(\bfk_1) \, \rho_2(\bfk_2) \, \rho_2(\mathbf{0})} \, .
\end{aligned}
\end{equation}
We show the trace diagrams for a theory with a quartic wavefunction coefficient in~\eqref{eq:TrrhoPsi4}.
\paragraph{Reduced density matrix $\boldsymbol{\rho}_R$} The next step is to obtain the reduced density matrix $\rho_R$, which is given by the path integral~\eqref{eq:rhoR},
\begin{equation} \label{eq:rhoRpathint}
    \rho_R = \left( \prod_{\bfk \in \mathcal{E}} \int \mathcal{D} \phi_{\bfk} \right) \Psi[\phi_{\bfk},\phi_{\pm\bfp}] \Psi[\phi_{\bfk},\bar{\phi}_{\pm\bfp}]^* \, .
\end{equation}
Similarly to what we did in~\eqref{eq:traceexpansion}, the wavefunctions in the integrand can be expanded perturbatively in the coefficients $\psi_n$ with $n\geq 3$. Then the result of the integral can also be expressed diagrammatically. Notice that we are taking the diagonal of $\rho$ in the subspace of modes with momentum $\bfk\in\mathcal{E}$, so the wavefunction coefficients that only come with such modes will appear in the combination~\eqref{eq:rhodef}, and hence will be represented by a crossed blob~\eqref{eq:cblob}. Also, we integrate precisely over all these modes, so the lines indicating Wick contractions (which have both ends attached to blobs) can only be dashed lines~\eqref{eq:dashedline}. Conversely, there is no integration over the modes with momentum $\pm\bfp$ so they are represented by the arrowed lines~\eqref{eq:extlines}, which have one free end. In~\eqref{eq:extlines}, the field modes $\phi_{\pm\bfp}$ (ket) and $\bar{\phi}_{\pm\bfp}$ (bra) are associated to white and black blobs respectively, accounting for the fact that the former enter through $\Psi[\phi_{\bfk},\phi_{\pm\bfp}]$ and the latter trough $\Psi[\phi_{\bfk},\bar{\phi}_{\pm\bfp}]^*$ in the path integral~\eqref{eq:rhoRpathint}.

We conclude that the contributions to the reduced density matrix $\rho_R$ can be conveniently represented by all possible diagrams consisting of blobs and lines (up to the desired order in the couplings) that satisfy rules~\eqref{eq:diagrules} and for which:
\begin{equation} \label{eq:diagrulesrhoR}
    \parbox{0.90\textwidth}{\centering 
    \begin{itemize}
        \item Internal lines (both ends connected to blobs) must be dashed.
        \item External lines (only one end is connected to a blob) must be arrowed. One must sum over all possible arrow directions consistent with overall momentum conservation of the diagram.
        \item Blobs only connected to dashed lines must be crossed.
        \item Blobs connected to external lines must be black or white. One must sum over all possible colorings of such blobs.
    \end{itemize}
    }
\end{equation}
We will call the set of all such diagrams \textit{reduced diagrams} $D_R$, with the connected ones forming the subset $D_R^{(c)} \subset D_R$. Finally, the reduced density matrix will also contain the Gaussian factor of the $\pm\bfp$ sector. Taking all of this into account, the result for $\rho_R$ is
\begin{equation} \label{eq:rhoR2}
    \rho_R = \frac{1}{\tilde{\mathcal{N}}^{\,2}} \cdot \exp \left( - \frac{\psi_2 (\mathbf{p})}{L^3} |\phi_{\mathbf{p}}|^2 - \frac{\left(\psi_2 (\mathbf{p}) \right)^*}{L^3} |\bar{\phi}_{\mathbf{p}}|^2 \right) \cdot \exp \left( \sum D_R^{(c)} \right) \, ,
\end{equation}
where
\begin{equation}
    \tilde{\mathcal{N}}^{\,2} \equiv \int \mathcal{D} \phi_{\bfp} \, \exp \left( - \frac{2\text{Re}\, \psi_2(\bfp)}{L^{3}} \, |\phi_{\bfp}|^2 \right) \, .
\end{equation}
We show explicit graphical representations of $\rho_R$ for two different theories in~\eqref{eq:rhoRPsi3} and~\eqref{eq:rhoRPsi4}. Here we just illustrate the above discussion with a couple of individual diagrams that can contribute to $\rho_R$:
\begin{align}
    \frac{1}{2} \,\, \vcenter{\hbox{\begin{tikzpicture}
    \coordinate (o) at (0,0);
    \coordinate (c1) at ([shift={(-30pt,0)}]o);
    \coordinate (c2) at ([shift={(30pt,0)}]o);
    \draw[thick,dashed] (c1) -- (c2);
    \draw[thick,->-] (c1)++(110:10pt) -- ++(110:30pt);
    \draw[thick,-<-] (c1)++(70:10pt) -- ++(70:30pt);
    \draw[thick,dashed] (c2) to[out=45,in=135,distance=60pt] (c2);
    \draw (c1) pic {wb};
    \draw (c2) pic {cb};
    \end{tikzpicture}}} \hspace{-21pt}
    & = \, \frac{\phi_{\bfp} \phi_{-\bfp}}{2 \, L^6} \cdot \sum_{\bfk\in\mathcal{E}} \frac{\psi_3 (\bfp,-\bfp, \mathbf{0}) \, \rho_3 (\bfk,-\bfk,\mathbf{0})}{\rho_2(\mathbf{0}) \, \rho_2(\bfk)} \, , \label{eq:rhoR1std} \\[7pt]
    \frac{1}{36} \,\, \vcenter{\hbox{\begin{tikzpicture}
    \coordinate (o) at (0,0);
    \coordinate (c1) at ([shift={(-30pt,0)}]o);
    \coordinate (c2) at ([shift={(30pt,0)}]o);
    \draw[thick,dashed] (c1) -- (c2);
    \draw[thick,-<-] (c1)++(90:10pt) -- ++(90:30pt);
    \draw[thick,-<-] (c1)++(115:10pt) -- ++(115:30pt);
    \draw[thick,-<-] (c1)++(65:10pt) -- ++(65:30pt);
    \draw[thick,->-] (c2)++(90:10pt) -- ++(90:30pt);
    \draw[thick,->-] (c2)++(115:10pt) -- ++(115:30pt);
    \draw[thick,->-] (c2)++(65:10pt) -- ++(65:30pt);
    \draw (c1) pic {wb};
    \draw (c2) pic {bb};
    \end{tikzpicture}}} \hspace{5pt}
    & = \, \frac{(\phi_{\bfp} \bar{\phi}^*_{\bfp})^3}{36 \, L^{15}} \cdot \frac{\left| \psi_4 (\bfp,\bfp,\bfp,-3\bfp) \right|^2}{\rho_2(3\bfp)} \, . \label{eq:rhoR2ndd}
\end{align}

\paragraph{Tr $\boldsymbol{\rho_R^2}$} We are now ready to compute the trace of the squared density matrix. In the field basis it is given by
\begin{equation} \label{eq:trrhosq}
    {\rm Tr}_\S\, \rho_{\rm R}^2 = \int \mathcal{D}\phi_{\pm\mathbf{p}} \mathcal{D}\bar{\phi}_{\pm\mathbf{p}} \, (\rho_{\rm R})_{\phi\bar{\phi}} \, (\rho_{\rm R})_{\bar{\phi}\phi} \, .
\end{equation}
We have to integrate over the field modes $\phi_{\pm\bfp}$ and $\bar{\phi}_{\pm\bfp}$, which graphically corresponds to connecting external arrowed lines in the $D_R$ diagrams with each other to represent Wick contractions of these modes. The resulting diagrams will have no external lines, only dashed or arrowed internal lines~\eqref{eq:dashedline} and~\eqref{eq:arrowedline}.

The integral~\eqref{eq:trrhosq} is then conveniently computed by substituting in the integrand the diagrammatic expressions corresponding to the two $\rho_R$'s (up to the desired order in the couplings), taking into account that the second $\rho_R$ has the labels $\phi$ and $\bar{\phi}$ switched. Graphically this means that its white blobs will be connected to fields $\bar{\phi}_{\pm\bfp}$ and its black blobs to fields $\phi_{\pm\bfp}$. Then, one performs all possible combinations of contractions between lines with the same label (meaning contractions of the type $\phi$-$\phi$ or $\bar{\phi}$-$\bar{\phi}$, but not $\phi$-$\bar{\phi}$). Graphically this implies that:
\begin{equation} \label{eq:diagrulesTrrhosq}
    \parbox{0.90\textwidth}{\centering 
    \begin{itemize}
        \item Contractions within the same diagram are only allowed if the lines depart from blobs of the same colour.
        \item Contractions between two disconnected diagrams that both belong to the same $\rho_R$ are only allowed if the lines depart from blobs of the same colour.
        \item Contractions between two disconnected diagrams that belong to different $\rho_R$'s are only allowed if the lines depart from blobs of different colour.
    \end{itemize}
    }
\end{equation}
Let us see an example of the diagrams that result from this tracing. The term in the integrand that only contains the diagram in~\eqref{eq:rhoR1std} (coming from either of the density matrices) will result in
\begin{equation} 
\begin{aligned}
    {\rm Tr}_\S\, \rho_R^2 \, & \supset \, \int \mathcal{D}\phi_{\pm\mathbf{p}} \mathcal{D}\bar{\phi}_{\pm\mathbf{p}} \left(
    \frac{1}{2} \,\, \vcenter{\hbox{\begin{tikzpicture}
    \coordinate (o) at (0,0);
    \coordinate (c1) at ([shift={(-30pt,0)}]o);
    \coordinate (c2) at ([shift={(30pt,0)}]o);
    \draw[thick,dashed] (c1) -- (c2);
    \draw[thick,->-] (c1)++(110:10pt) -- ++(110:30pt);
    \draw[thick,-<-] (c1)++(70:10pt) -- ++(70:30pt);
    \draw[thick,dashed] (c2) to[out=45,in=135,distance=60pt] (c2);
    \draw (c1) pic {wb};
    \draw (c2) pic {cb};
    \end{tikzpicture}}} \hspace{-25pt} \right) = \,
    \frac{1}{2} \hspace{-25pt} \vcenter{\hbox{\begin{tikzpicture}
    \coordinate (o) at (0,0);
    \coordinate (c1) at ([shift={(-30pt,0)}]o);
    \coordinate (c2) at ([shift={(30pt,0)}]o);
    \draw[thick,dashed] (c1) -- (c2);
    \draw[thick,->-] (c1) to[out=45,in=135,distance=60pt] (c1);
    \draw[thick,dashed] (c2) to[out=45,in=135,distance=60pt] (c2);
    \draw (c1) pic {wb};
    \draw (c2) pic {cb};
    \end{tikzpicture}}} \hspace{-25pt} \\[8pt]
    & = \frac{1}{2 \, L^3} \sum_{\bfk\in\mathcal{E}} \frac{\psi_3 (\bfp,-\bfp, \mathbf{0}) \, \rho_3 (\bfk,-\bfk,\mathbf{0})}{\rho_2(\mathbf{0}) \, \rho_2(\bfk) \, \rho_2(\bfp)} \, ,
\end{aligned}
\end{equation}
whereas the term which only contains the diagram~\eqref{eq:rhoR2ndd} will give zero, as it has an odd number of lines coming out of each type of blob,
\begin{equation} 
    {\rm Tr}\, \rho_R^2 \, \supset \, \int \mathcal{D}\phi_{\pm\mathbf{p}} \mathcal{D}\bar{\phi}_{\pm\mathbf{p}} \left(
    \frac{1}{36} \,\, \vcenter{\hbox{\begin{tikzpicture}
    \coordinate (o) at (0,0);
    \coordinate (c1) at ([shift={(-30pt,0)}]o);
    \coordinate (c2) at ([shift={(30pt,0)}]o);
    \draw[thick,dashed] (c1) -- (c2);
    \draw[thick,-<-] (c1)++(90:10pt) -- ++(90:30pt);
    \draw[thick,-<-] (c1)++(115:10pt) -- ++(115:30pt);
    \draw[thick,-<-] (c1)++(65:10pt) -- ++(65:30pt);
    \draw[thick,->-] (c2)++(90:10pt) -- ++(90:30pt);
    \draw[thick,->-] (c2)++(115:10pt) -- ++(115:30pt);
    \draw[thick,->-] (c2)++(65:10pt) -- ++(65:30pt);
    \draw (c1) pic {wb};
    \draw (c2) pic {bb};
    \end{tikzpicture}}} \hspace{5pt}
    \right) = 0 \, .
\end{equation}

In \eqref{eq:TrRho2Psi3_1} and \eqref{eq:TrRho2Psi4_1} we show the explicit graphical representations of ${\rm Tr}\, \rho_R^2$ for two different theories. There we notice that the resulting diagrams can be combined into a form (see \eqref{eq:TrRho2Psi3_2} and~\eqref{eq:TrRho2Psi4_2}) that makes the following explicit: they are \textit{almost} equal to two times the trace diagrams $D_T$ that appeared in ${\rm Tr}\, \rho_R$. This is general and can be easily understood: if the contraction rules~\eqref{eq:diagrulesTrrhosq} did not apply and all contractions were allowed, we would find the result to be (two times, since we integrate two copies of $\rho_R$) all possible diagrams $D_T$. However, the rules~\eqref{eq:diagrulesTrrhosq} forbid some specific contractions and therefore some diagrams cannot appear. Let us denote these \textit{forbidden diagrams} as $D_F$. The result is then
\begin{equation} \label{eq:Trrhosq}
    {\rm Tr}_S \, \rho_R^2 = 1 + 2 \left( \sum D_T - \sum D_F \right) = \exp \left[ 2 \left( \sum D_T^{(c)} - \sum D_F^{(c)} \right) \right] \, ,
\end{equation}
with $D_F^{(c)}\subset D_F$ the subset of connected forbidden diagrams.\footnote{Notice that the set $D_F$ includes all connected forbidden diagrams, $D_F^{(c)}$, and all disconnected diagrams that contain at least one connected forbidden piece, i.e. at least one piece belonging to $D_F^{(c)}$.}. We now have all the necessary ingredients to compute the purity~\eqref{eq:puritydef}. Taking into account the correct normalization of $\rho_R$, and simply combining~\eqref{eq:Trrho} and~\eqref{eq:Trrhosq} we have
\begin{equation} \label{eq:Puritydiags}
    \gamma = \frac{{\rm Tr}\, \rho_{\rm R}^2}{\left( {\rm Tr}\, \rho_{\rm R} \right)^2} = \frac{\exp \Big[ 2 \Big( \sum D_T^{(c)} - \sum D_F^{(c)} \Big) \Big]}{\exp \Big( 2 \sum D_T^{(c)} \Big)} = \exp \left( - 2 \sum D_F^{(c)} \right) \, .
\end{equation}
In the following we discuss how this results simplifies when we take the infinite volume limit.

\subsection{Purity and the infinite volume limit} 

We can now determine the scaling of a general diagram (without external legs) with the spatial volume $L^3$, to see what their behaviour is for infinite volume.

Each connected diagram will contain a factor $L^{-3N_V}$, i.e. $N_V$ negative powers of the spacetime volume. This number can be counted easily by noticing, from~\eqref{eq:wfcoefs} that every blob $\psi_n$ comes with $n-1$ powers of inverse volume and, from~\eqref{eq:solidline}-\eqref{eq:arrowedline}, that every line comes with a power of volume. Making use of the usual identities that relate the number of loops, vertices, and lines of a connected graph, this implies
\begin{equation}
    N_V = l-1 \, ,
\end{equation}
where $l$ is the number of loops of the diagram. We can also count the number $N_S$ of sums over momenta in a diagram. It will simply be given by the number $l$ of loops minus the number $I_a$ of internal arrowed lines, which have fixed momentum $\pm\mathbf{p}$. Hence\footnote{\label{ft:p0} There is an important caveat here if $\mathbf{p}=0$: an internal line with zero momentum can join two otherwise disconnected subdiagrams without fixing any loop momentum. In that case $I_a$ in~\eqref{eq:nsums} counts the number of internal lines with $\mathbf{p}=0$ except those which divide the diagram into two disconnected diagrams when cut.

In any case, the rest of results of this section do not apply when $\mathbf{p}=0$: the contributions in that case are not correctly captured by the diagrams that we define and the result for the purity is in fact divergent.}
\begin{equation} \label{eq:nsums}
    N_S = l-I_a \, .
\end{equation}
Finally, notice that when taking the infinite volume limit we do the usual substitution
\begin{equation}
    \frac{1}{L^3} \sum_{\mathbf{k}} \qquad\xrightarrow[L\to\infty]{}\qquad \int_{\mathbf{k}} \, ,
\end{equation}
so every sum ``absorbs'' one inverse volume factor. The number of remaining powers of (infinite) volume of a diagram will then be $N_S - N_V = 1-I_a$. Hence, in the infinite volume limit the scaling behaviour of a diagram with $I_a$ lines of fixed momentum $\pm\mathbf{p}$, $D_{I_a}$, is determined purely by this number:
\begin{equation} \label{eq:volumef}
    \lim_{L\to\infty} D_{I_a} \,\, \sim \,\, L^{3(1-I_a)} \, ,
\end{equation}
where the caveat of footnote~\ref{ft:p0} applies in the $\mathbf{p}=0$ case. This can be understood intuitively: diagrams with $I_a=0$ contain integrals over all modes, so that one is actually integrating over all space. The result should then diverge. But when choosing a line to be $\pm\mathbf{p}$ (i.e. $I_a=1$) one is effectively choosing a specific scale at which to look at the problem, eliminating one integration over all space. The result then becomes finite.

\paragraph{Purity} We found a general expression for purity in~\eqref{eq:Puritydiags}. Notice that the counting~\eqref{eq:volumef} implies that the trace diagrams $D_T^{(c)}$, which have $I_a=0$, are divergent in the infinite volume limit. From that point of view, their cancellation in~\eqref{eq:Puritydiags} is essential to get a finite result for the purity. In fact, the forbidden diagrams $D_F^{(c)}$ that contribute to the purity have $I_a>0$ and are therefore finite (or vanish) in the infinite volume limit! More precisely, in that limit we can forget about diagrams with $I_a>1$ and keep only those with $I_a=1$. Additionally, for these $I_a=1$ diagrams we can convert dashed lines into solid lines (since the difference is a zero-measure couple of points in the integration over momenta). Therefore:
\begin{equation} \label{eq:resPurity}
    \text{Infinite volume:} \qquad \gamma = \exp \left( - 2 \sum D^{(c)} \right) \, ,
\end{equation}
where the set $D^{(c)}$, that we will call \textit{purity diagrams}, is constructed by picking the subset of diagrams in $D_F^{(c)}$ that have $I_a=1$ and substituting dashed lines by solid lines in them. But admittedly this definition is not very useful yet, because it requires us to first compute $\rho_R$ and the diagrams $D_F^{(c)}$ explicitly, which can be tedious.

Fortunately there is a much more direct way of building the set $D^{(c)}$. It consists of all diagrams of the schematic form 
\begin{equation} \label{eq:Ddiagdef}
    D^{(c)} \,\, \ni \quad \vcenter{\hbox{\begin{tikzpicture}
        \coordinate (o) at (0,0);
        \path (o) +(135:30pt) coordinate (w);
        \path (o) +(45:30pt) coordinate (b);
        \filldraw[color=white, fill=white, postaction={pattern={Lines[angle=45, distance=4mm, line width=2mm]},pattern color=lightgray}](o) circle[radius=30pt];
        \draw[thick] (o) circle[radius=30pt];
        \draw[thick,-><-] (w) .. controls(-30pt,60pt) and (30pt,60pt) .. (b);
        \draw (w) pic {wb};
        \draw (b) pic {bb};
    \end{tikzpicture}}} \quad ,
\end{equation}
where the double-arrowed line indicates that $\bfp$ can flow in either direction, and the big striped blob represents any diagram that satisfies the following properties:
\begin{itemize}
    \item It can only contain solid lines and crossed blobs.
    \item It cannot have any line that is forced to carry momentum $\pm\bfp$ after momentum conservation is imposed.
    \item There must be a continuous path between the white and the black blobs.\footnote{In fact there must be at least two such paths, otherwise the previous condition is violated.}
    \item The whole diagram~\eqref{eq:Ddiagdef} must be connected.
\end{itemize}
This characterization provides a fast and efficient way of writing down the contributions to the purity of a mode with momentum $\pm\bfp$, at the desired order in perturbation theory.

We show now some examples. For a theory with only a three-point wavefunction coefficient $\psi_3$ and at $\mathcal{O}(\psi_3^2)$, the purity diagrams are just
\begin{equation}
    D^{(c)} = \left\{
    \frac{1}{2} \, \vcenter{\hbox{\begin{tikzpicture}
    \coordinate (o) at (0,0);
    \coordinate (c1) at ([shift={(-30pt,0)}]o);
    \coordinate (c2) at ([shift={(30pt,0)}]o);
    \draw[thick] (c1) -- (c2);
    \draw[thick] (c1) to[out=-90,in=-90] (c2);
    \draw[thick,->-] (c1) to[out=90,in=90] (c2);
    \draw (c1) pic {wb};
    \draw (c2) pic {bb};
    \end{tikzpicture}}} \, ,
    \frac{1}{2} \, \vcenter{\hbox{\begin{tikzpicture}
    \coordinate (o) at (0,0);
    \coordinate (c1) at ([shift={(-30pt,0)}]o);
    \coordinate (c2) at ([shift={(30pt,0)}]o);
    \draw[thick] (c1) -- (c2);
    \draw[thick] (c1) to[out=-90,in=-90] (c2);
    \draw[thick,-<-] (c1) to[out=90,in=90] (c2);
    \draw (c1) pic {wb};
    \draw (c2) pic {bb};
    \end{tikzpicture}}} \, \right\} \, ,
\end{equation}
and therefore the (infinite volume) purity at that order is 
\begin{equation} \label{eq:purityPsi3}
\begin{aligned}
    \gamma = \exp \left( - 2 \sum D^{(c)} \right) & \simeq 1 - 2 \sum D^{(c)} = 1 
    - \vcenter{\hbox{\begin{tikzpicture}
    \coordinate (o) at (0,0);
    \coordinate (c1) at ([shift={(-30pt,0)}]o);
    \coordinate (c2) at ([shift={(30pt,0)}]o);
    \draw[thick] (c1) -- (c2);
    \draw[thick] (c1) to[out=-90,in=-90] (c2);
    \draw[thick,->-] (c1) to[out=90,in=90] (c2);
    \draw (c1) pic {wb};
    \draw (c2) pic {bb};
    \end{tikzpicture}}}
    - \vcenter{\hbox{\begin{tikzpicture}
    \coordinate (o) at (0,0);
    \coordinate (c1) at ([shift={(-30pt,0)}]o);
    \coordinate (c2) at ([shift={(30pt,0)}]o);
    \draw[thick] (c1) -- (c2);
    \draw[thick] (c1) to[out=-90,in=-90] (c2);
    \draw[thick,-<-] (c1) to[out=90,in=90] (c2);
    \draw (c1) pic {wb};
    \draw (c2) pic {bb};
    \end{tikzpicture}}} \\
    & = 1 - 2 \int_{\mathbf{k}} {\frac{\left| \psi_3 (\mathbf{p},\mathbf{k},-\mathbf{p}-\mathbf{k}) \right|^2}{2\text{Re}\, \psi_2(\mathbf{p}) \, 2\text{Re}\, \psi_2(\mathbf{k}) \, 2\text{Re}\, \psi_2(\mathbf{p}+\mathbf{k})}} \, .
\end{aligned}
\end{equation}
Henceforth every mention of the purity refers to the infinite volume limit. We see that only the black and white blobs of~\eqref{eq:Ddiagdef} appear in these diagrams, crossed blobs only start to show up at higher orders. At $\mathcal{O}(\psi_3^3)$ there are no new diagrams but at $\mathcal{O}(\psi_3^4)$ we get
\begin{equation}
    D^{(c)} \supset \left\{
    \frac{1}{2} \, \vcenter{\hbox{\begin{tikzpicture}
    \coordinate (o) at (0,0);
    \coordinate (c1) at ([shift={(-30pt,0)}]o);
    \coordinate (c2) at ([shift={(30pt,0)}]o);
    \coordinate (c3) at ([shift={(-30pt,-30pt)}]o);
    \coordinate (c4) at ([shift={(30pt,-30pt)}]o);
    \draw[thick] (c1) to[out=-25,in=-155] (c2);
    \draw[thick,-><-] (c1) to[out=25,in=155] (c2);
    \draw[thick] (c1) -- (c3);
    \draw[thick] (c2) -- (c4);
    \draw[thick] (c3) to[out=-25,in=-155] (c4);
    \draw[thick] (c3) to[out=25,in=155] (c4);
    \draw (c1) pic {wb};
    \draw (c2) pic {bb};
    \draw (c3) pic {cb};
    \draw (c4) pic {cb};
    \end{tikzpicture}}} \, ,
    \frac{1}{2} \, \vcenter{\hbox{\begin{tikzpicture}
    \coordinate (o) at (0,0);
    \coordinate (c1) at ([shift={(-30pt,0)}]o);
    \coordinate (c2) at ([shift={(30pt,0)}]o);
    \coordinate (c3) at ([shift={(0,-50pt)}]o);
    \coordinate (c4) at ([shift={(0,-20pt)}]o);
    \draw[thick,-><-] (c1) -- (c2);
    \draw[thick] (c1) -- (c3);
    \draw[thick] (c2) -- (c3);
    \draw[thick] (c1) -- (c4);
    \draw[thick] (c2) -- (c4);
    \draw[thick] (c3) -- (c4);
    \draw (c1) pic {wb};
    \draw (c2) pic {bb};
    \draw (c3) pic {cb};
    \draw (c4) pic {cb};
    \end{tikzpicture}}} \, ,
    \frac{1}{2} \, \vcenter{\hbox{\begin{tikzpicture}
    \coordinate (o) at (0,0);
    \coordinate (c1) at ([shift={(-30pt,0)}]o);
    \coordinate (c2) at ([shift={(30pt,0)}]o);
    \coordinate (c3) at ([shift={(0,-30pt)}]o);
    \coordinate (c4) at ([shift={(55pt,-30pt)}]o);
    \draw[thick] (c1) to[out=-25,in=-155] (c2);
    \draw[thick,-><-] (c1) to[out=25,in=155] (c2);
    \draw[thick] (c1) -- (c3);
    \draw[thick] (c2) -- (c3);
    \draw[thick] (c3) -- (c4);
    \draw[thick] (c4) to[out=45,in=135,distance=60pt] (c4);
    \draw (c1) pic {wb};
    \draw (c2) pic {bb};
    \draw (c3) pic {cb};
    \draw (c4) pic {cb};
    \end{tikzpicture}}} \hspace{-25pt} \right\} \, .
\end{equation}
We show the purity diagrams $D^{(c)}$ for other theories in~\eqref{eq:purityPsi4} and~\eqref{eq:purityPsi6}.

\paragraph{Tr $\boldsymbol{\rho_R^N}$} The path integral that computes this quantity is
\begin{equation} \label{eq:trrhoN}
    {\rm Tr}_\S\, \rho_{\rm R}^N = \int \mathcal{D}\phi_{\pm\mathbf{p}}^{(1)} \cdots \mathcal{D}\phi_{\pm\mathbf{p}}^{(N)} \, (\rho_{\rm R})_{\phi^{(1)}\phi^{(2)}} \, (\rho_{\rm R})_{\phi^{(2)}\phi^{(3)}} \cdots (\rho_{\rm R})_{\phi^{(N)}\phi^{(1)}} \, ,
\end{equation}
so we now have an integrand with $N>2$ copies of the reduced density matrix with their adjacent indices contracted. Just like in the ${\rm Tr} \, \rho_R^2$ case, this leads to a series of rules for the possible contractions between arrowed lines in the diagrams, so the result can also be written as the trace diagrams $D_T$ minus a set of diagrams forbidden by these rules. It is hard to find a general description of the result in the case of finite spatial volume $L^3$, so we will focus on the $L\to\infty$ limit.

As we already mentioned, \eqref{eq:volumef} implies that for infinite volume the diagrams $D_T$ diverge. However, we are ultimately interested in the normalised quantity ${\rm Tr} \, \rho_R^N/({\rm Tr} \, \rho_R)^N$, where these divergences are cancelled out by the denominator. Only the aforementioned forbidden diagrams remain, and again from~\eqref{eq:volumef} we know that just those with one internal arrowed line, $I_a=1$, give a nonzero result when $L\to\infty$. This means that in this limit only the single contractions between two adjacent matrices $\rho_R$ in~\eqref{eq:trrhoN} will contribute, and hence the type of diagrams that result will be the same as for the purity (just with a different overall factor coming from the fact that the trace now consists of $N$ index contractions instead of $2$). We find then the remarkable result
\begin{equation} \label{eq:resTrrhoN}
    \text{Infinite volume:} \qquad \frac{{\rm Tr}_\S\, \rho_R^N}{\left( {\rm Tr}_\S\, \rho_R \right)^N} = \exp \left( - N \sum D^{(c)} \right) \qquad \text{for } N\geq 2 \, .
\end{equation}
This is, all $N$-traces of the reduced density matrix of a momentum mode $\pm\bfp$ are given by a single quantity, which is exponentiated to the $N$-th power. Notice that this implies the following relation between traces of $\rho_R$,
\begin{equation}\label{surp}
    \frac{{\rm Tr}_\S\, \rho_R^N}{\left( {\rm Tr}_\S\, \rho_R \right)^N} = \left[ \frac{{\rm Tr}_\S\, \rho_R^M}{\left( {\rm Tr}_\S\, \rho_R \right)^M} \right]^{N/M} \qquad \text{for } N,M\geq 2 \, .
\end{equation}
A few comments on this result are in order:
\begin{itemize}
    \item This result~\eqref{surp} holds to all orders in perturbation theory. 
    \item The infinite volume limit is essential to obtain it. For a theory in finite spatial volume, \eqref{eq:resTrrhoN} is, in general, not satisfied, and more than one quantity is needed to characterize the $N$-traces of $\rho_R$.
    \item The relation in~\eqref{surp} constitutes a very strong constraint on what $\rho_R$ can be. For a finite dimensional Hilbert space it would imply that $\rho_R$ must be a pure state, so that $\rho_R^N=\rho_R$. When the Hilbert space is infinite dimensional this is not the case anymore, but still this relation puts severe constraints. We leave further investigation of this aspect for future work. 
\end{itemize}

\subsection{Relation to previous literature} 

We would like to comment on the relation between the diagrammatic formalism introduced in~\cite{Colas:2024ysu} to compute the perturbative purity in field theory and the technique presented here. The main technical difference is that here we start from a situation in which the dynamical evolution has already been solved for, and it is encoded in some known wavefunction $\Psi[\phi]$. Our diagrams  express ${\rm Tr}_\S \, \rho_R^N$ in terms of momentum integrals over the wavefunction coefficients. In contrast, \cite{Colas:2024ysu} starts from an interaction Hamiltonian and defines a set of Feynman rules to express the purity in terms of time and momentum integrals of unequal-time correlators of system and environment. The setup of that paper does not assume a specific bipartition, and can in principle be used for our choice of system and environment. The result for $\gamma$ would in that case be a sum of integrals of unequal-time correlators of the field, which should map to our expression in terms of wavefunction coefficients. \\

On a different note, we would like to make a connection between our results and those of~\cite{Balasubramanian:2011wt}. The authors of that paper computed the entanglement entropy and mutual information between different sets of momentum modes of a QFT in Minkowski. At leading order in $\lambda$, their result for the entanglement entropy of a single mode with momentum $\bfp$ in a $\lambda\phi^n$ theory in a finite volume $V$ can be expressed in terms of our own diagrams $D_F$ as\footnote{At leading order the diagrams $D_F$ have only two blobs and are all connected, so there is no difference between writing $D_F$ or $D_F^{(c)}$ in~\eqref{eq:EntEntrBalas}.}$^{,}$\footnote{Strictly speaking the derivation in~\cite{Balasubramanian:2011wt} focused on the $\phi^3$ and $\phi^4$ cases, but from their comments on the $n\geq 5$ case we expect that~\eqref{eq:EntEntrBalas} should still hold. Moreover, their system is the set of all low-energy modes, $q<\mu$ for some Wilsonian cutoff $\mu$. Conversely our system is a single mode $\bfp$. A consequence of this difference is that their entanglement entropy is extensive and only their entropy \textit{density} is finite in the infinite-volume limit. Conversely our Renyi and entanglement entropies for a single mode are all finite as $V\to \infty$.}
\begin{equation} \label{eq:EntEntrBalas}
    S_{EE} = - \lambda^2 \log\left(\lambda^2\right) \, \sum_{\rm L.O.} D_F^{(c)} + \mathcal{O}(\lambda^2) \, ,
\end{equation}
where one should restrict the sum to only the leading order diagrams, which go as $\lambda^2$, and we have stripped that $\lambda$ dependence off the diagrams adding it to the prefactor. It was already noticed there that~\eqref{eq:EntEntrBalas} does not diverge when going to the continuum. Using our diagram notation the result that they found is
\begin{equation} \label{namely}
    \lim_{V\to \infty} S_{EE} = - \lambda^2 \log\left(\lambda^2\right) \, \sum_{\rm L.O.} D^{(c)} + \mathcal{O}(\lambda^2) \, ,
\end{equation}
as one would expect from~\eqref{eq:volumef}. This means that, at leading order in perturbation theory, the diagrams (or else the momentum integrals) that contribute to ${\rm Tr}_\S\, \rho_R^N$ are the same that contribute to the entanglement entropy ${\rm Tr}_\S \left( \rho_R \log\rho_R \right)$. This is to be expected because of the general relation between Renyi entropies $H_n$ and the entanglement entropy $S$,
\begin{align}
    -\Tr(\rho_R \log \rho_R)\equiv S_{EE} =\lim_{n\to1}H_n \equiv \lim_{n\to 1}\frac{1}{1-n}\log(\Tr \rho_R^n)\,.
\end{align}
If we use~\eqref{eq:resTrrhoN} to compute the Renyi entropies in perturbation theory and naively take the $n\to 1$ limit we find the nonsensical result $S_{EE} \sim 1/(1-1) \sim \infty$. As explained in \cite{Costa:2022bvs}, the issue is that the perturbative calculation neglects terms of order $\lambda^{2n}$, which instead become important in the calculation of the entanglement entropy which involves the limit $n\to 1$. There is a nifty trick to bypass this issue, which we review here following \cite{Costa:2022bvs}. To begin with, imagine diagonilizing $\rho_R$. The eigenvalues $p_j$ for $j=0,1,\dots$ must be $(1-\lambda^2 \sum_i a_i)$ for $j=0$ and $(\lambda^2 a_i)$ for $i=1,2,\dots$, because now we impose $\Tr \rho_R=1$ and $\rho_R$ must be pure for $\lambda=0$. Then, we can compute the entropies
\begin{align}
    S_{EE} &= - \sum_{i=0}^{\infty} p_i \log p_i \nonumber \\
    &=- \lambda^2 \sum_{i=1}^{\infty} a_i(\lambda) \log (\lambda^2 a_i(\lambda))
    - \left( 1 - \lambda^2 \sum_{i=1}^{\infty} a_i(\lambda) \right) \log \left( 1 - \lambda^2 \sum_{i=1}^{\infty} a_i(\lambda) \right)\\
    &\simeq  - \lambda^2 \log \lambda^2 \sum_i a_i(0) + \mathcal{O}(\lambda^2)\,, \nonumber \\
    H_n(\rho_R) &= \frac{1}{1-n} \log \left( \sum_{j=0}^{\infty} p_j^n \right) \simeq \frac{n \lambda^2}{n-1} \sum_{i}^{\infty} a_i(0)\,.
\end{align}
As we can see, from $H_n$ we can compute $S_{EE}$ to leading order in $\lambda$ with the substitution $n/(n-1)\to -\log \lambda^2$. This trick confirms that our result for $\Tr \rho_R^n$ agrees with the entanglement entropy computed in~\cite{Balasubramanian:2011wt,Costa:2022bvs}, namely~\eqref{namely}.\\

Finally, let us recover an interesting property of Gaussian states. It has been known for a long time that the entanglement entropy of single-mode Gaussian states only depends on one quantity: the determinant of the covariance matrix~\cite{Agarwal:1971zza}.\footnote{The case of two-mode Gaussian states was studied in~\cite{Serafini:2003ke}, where it was shown that the purity still depends only on a single quantity, but the entanglement entropy depends on two.} The state $\rho_R$ of a single mode $\pm\bfp$ that we study in this paper is not Gaussian for an interacting theory, neither at finite nor at infinite volume, as one can easily see from~\eqref{eq:rhoR2} and the example diagrams $D_R^{(c)}$. But, as explained in the previous subsection, in the infinite volume limit only the subset of diagrams $D^{(c)}\subset D_F^{(c)}$ contributes to $\Tr\,\rho_R^N$, and in fact these diagrams come exclusively from the Gaussian part of $\rho_R$. From this point of view, $\rho_R$ is effectively Gaussian when it comes to computing its $N$-traces in the infinite volume case, and hence it makes sense that all of them are determined by a single quantity, cf.~\eqref{eq:resTrrhoN}. In addition, the effective Gaussianity of $\rho_R$ offers a new way to compute the von Neumann entropy $S_{EE}$. We can use the fact that single-mode Gaussian states satisfy the following relation between $S_{EE}$ and the purity $\gamma$~\cite{Agarwal:1971zza},
\begin{equation}
    S_{EE} = \frac{1-\gamma}{2\gamma} \log \left( \frac{1+\gamma}{1-\gamma} \right) - \log \left( \frac{2\gamma}{1+\gamma} \right) \, .
\end{equation}
Plugging in our expression for the purity, Eq.~\eqref{eq:resPurity}, we obtain again the result of~\cite{Balasubramanian:2011wt} at leading order in the coupling, namely~\eqref{namely}.

\paragraph{Coupled harmonic oscillators} To conclude it is interesting to compare our result above to the exact result for the entanglement of two harmonic oscillators coupled as in 
\begin{align}
    \mathcal{L}=-\frac12 \left( \dot x_A^2+\dot x_B^2 + M^2 x_A^2+M^2 x_B^2\right)-l x_A x_B
\end{align}
Tracing over $x_B$ one finds that the reduced density matrix for $x_A$ obeys \cite{Bombelli:1986rw,Nishioka:2018khk,Costa:2022bvs}
\begin{align}
    \frac{\Tr_A \rho_R^n}{(\Tr_A \rho_R)^n}=\frac{(1-\xi)^n}{1-\xi^n} \,, \quad \text{ for } \quad \xi = \left( \frac{\left( M^2 + l \right)^{\frac{1}{4}} - \left( M^2 - l \right)^{\frac{1}{4}}}{\left( M^2 + l \right)^{\frac{1}{4}} + \left( M^2 - l \right)^{\frac{1}{4}}} \right)^2\,.
\end{align}
Note that this result in a Gaussian theory does not obey the infinite-volume relation in~\eqref{surp}.

\section{Perturbative unitarity bounds} \label{sec:perturbative unitarity}

In this section, we will be discussing three different diagnostics for the breakdown of perturbation theory in a quantum field theory. In Section~\ref{sec3p1} we consider the most direct diagnostic, namely the estimate of higher order corrections. We provide explicit expressions for the perturbative contributions to amplitudes, correlators and the field theoretic wavefunction. Then, in Section~\ref{sec3p2}, we present a lightening review of perturbative unitarity bounds from the partial wave decomposition of scattering amplitudes in flat spacetime. This will be our main benchmark when exploring the bounds from purity. Finally, in Section~\ref{sec3p3}, we come to the main topic of this work and discuss how to use the perturbative calculation of purity as a breakdown diagnostic for perturbation theory.

\subsection{Perturbativity of the loop expansion}\label{sec3p1}

In cosmology, we are typically interested in the breakdown of perturbativity in in-in correlators, and we will estimate one-loop corrections to low-point correlators such as the power spectrum. Occasionally, in the de Sitter examples, we find it more convenient to formulate this constraint using the $\psi_2$ wavefunction coefficient. Since wavefunction coefficients are precursors to correlators, we expect that the constraints arising from perturbativity of the wavefunction coefficients' loop expansion should capture also the breakdown of perturbativity in correlators.

\paragraph{Superficial degree of divergence}

Calculating loop corrections explicitly is difficult in general; one way of estimating the size of loop corrections \textit{in Minkowski spacetime} is through their superficial degree of divergence. We will estimate this degree of divergence in a power counting scheme where each operator of dimension $N$ is accompanied by a Wilson coefficient and a factor $\Lambda^{4-N}$. A general Feynman diagram computing an amplitude, a correlator, or a wavefunction coefficient will have $E$ external lines, $V$ vertices, and $L$ loops. At each vertex $v$, $I_v$ internal lines and $E_v$ external lines meet, with $D_v$ derivatives each producing some momentum $q$. There are a total of $I$ internal lines, $E$ external lines, and $D$ derivatives. The superficial degree of divergence is obtained by taking all momentum scales in a diagram to the cutoff $\LUV$ simultaneously. It represents the approximate scaling of the diagram with the cutoff. \\
For completeness, we describe the calculation of the superficial degree of divergence for amplitudes, correlators, and wavefunction coefficients, beginning with the most familiar of the three.
\begin{itemize}
\item \textit{Amplitudes:}
   A textbook result \cite{Weinberg:1995,burgess2020introduction} is that a scattering amplitude $\mcA$ has superficial degree of divergence
\begin{equation}
   \mcA \sim \Lambda^{4-E}  \left( 4\pi \right)^{-2L}  \left( \frac{\LUV}{\Lambda} \right)^{2 + 2L + D - 2 V} \, .
\end{equation}
This comes from writing
\begin{equation}
   \mcA = \int \frac{\dd^4 \bfp_1}{(2\pi)^4} \dots \frac{\dd^4 \bfp_L}{(2\pi)^4} \, \prod_v \Lambda^{4 - I_v - E_v -D_v} q^{D_v} \prod_I G_I(p_I) \, ,
\end{equation}
then treating each $\dd^4 \bfp$, $q$, and $G_I$ as $\LUV^4$, $\LUV$, and $\LUV^{-2}$, respectively, and applying the Euler identity
\begin{equation}
   V - I + L = 1 \, .
\end{equation}
\item \textit{Correlators:} The in-in$=$in-out result \cite{Donath:2024utn} relates an $E$-point Fourier-space in-in correlator to an $E$-point in-out amplitude $\mcA$ in the same theory:
\begin{equation}
   \ev{\varphi^E} = \int \frac{\dd \omega_1}{2\pi} \dots \frac{\dd \omega_E}{2\pi} \, 2\pi \delta_D(\wT) \prod_{E}\left( \frac{i}{\omega^2 - k^2 - m^2} \right) \mcA \, .
\end{equation}
As the integrals over the $\omega$s correspond to an inverse Fourier transform which fixes the time of the correlator, they are not truncated at $\LUV$.
The residue theorem suggests that each $\omega$ should count as $\sqrt{k^2 + m^2}$, except for one of the integral measures, which is removed by the energy-conserving Dirac delta.
All of the factors of $2\pi$ appearing due to the residue theorem cancel with the factors in the integral measures.
In that case,
\begin{equation}
   \ev{\varphi^E} \sim  (4\pi)^{-2L} \frac{\Lambda^{4-E}}{\left( \sqrt{k^2 + m^2} \right)^{E + 1}} \left( \frac{\LUV}{\Lambda} \right)^{2+  2 L + D - 2V} \, .
\end{equation}
\item \textit{Wavefunction:} A wavefunction coefficient $\psi_E$ with $E$ external lines at $L$ loops takes the form of an integral
\begin{equation}
   \psi_E \sim \int \dd t_1 \dots \dd t_V \prod_E K_k(t) \int_{ \bfp_1 \dots  \bfp_L} \prod_v \Lambda^{4 - I_v - E_v -D_v} q^{D_v} \prod_I G_I(t, t') \, ,
\end{equation}
Simply taking all loop and $q$ momenta to the cutoff $\LUV$, and counting each time integral as $\LUV^{-1}$, the wavefunction coefficient scales as
\begin{equation}
   \psi_E \sim (2\pi^2)^{-L} \Lambda^{3-E} \left( \frac{\LUV}{\Lambda} \right)^{D + 2L - 2V + 1} \, .
\end{equation}
The factor of $(2\pi^2)^{-L}$ comes from the $(2\pi)^{-3}$ associated with each momentum integral measure and a factor of $4\pi$ for each angular integral.
The result is clearly too crude, since it does not account for the derivatives that act on external lines and cannot be taken to the cutoff.
One way to account for this is to exchange an appropriate number of factors of $\LUV$ for factors of $k$ or $E_k$, an external momentum or energy.
\end{itemize}

\subsection{A brief review of partial-wave perturbative unitarity bounds}\label{sec3p2}

Unitarity---that is, the conservation of probability---demands that the outgoing probability flux in a scattering process is not larger than the incoming flux. It is therefore not surprising then that unitarity implies bounds on the size of scattering amplitudes. There are many ways to make these bounds precise; in this work we will use one of them, the partial wave unitarity bound, that is particularly useful for $2\to 2$ scattering. We review it briefly in what follows.

To implement this bound, four-point scattering amplitudes are expanded in partial waves,
\begin{equation}
    \mathcal{A}_4 = 16\pi \sum_{l=0}^{\infty} \, (2l+1) \, a_l \, P_l(\cos{\theta}) \, ,
\end{equation}
with $\theta$ the angle between two ingoing particles in the center of mass frame and $P_l$ the Legendre polynomial of order $l$, whose orthogonality allows us to extract the partial wave coefficients as follows:
\begin{equation}
    a_l = \frac{1}{32\pi} \int_{-1}^1 { \d(\cos{\theta}) \, P_l (\cos{\theta}) \, \mathcal{A}_4} \, .
\end{equation}
Unitarity then requires for all coefficients $a_l$:
\begin{equation} \label{eq:pwbC}
    |a_l| \leq 1 \, , \qquad \left| \Re a_l \right| \leq \frac{1}{2} \, , \qquad 0\leq \Im a_l \leq 1 \, .
\end{equation}
A scattering amplitude can, and often does, violate these bounds when computed at some finite order in perturbation theory. This usually happens when the coupling or the energy become too large, and signals a breaking of perturbation theory: higher-order contributions have become important and need to be added to the S-matrix in order to restore unitarity. Alternatively, the violation starting at a given energy might signal the need for new physics appearing at that scale.

In this work we will apply the partial wave bounds~\eqref{eq:pwbC} to tree-level amplitudes, which are purely real. Therefore, we will simply use
\begin{equation} \label{eq:pwb}
    \left| \Re a_l \right| < \frac{1}{2} \qquad \forall \,\, l \, .
\end{equation}

It has been so far difficult to define scattering amplitudes on cosmological spacetimes, but see \cite{Marolf:2012kh,Melville:2023kgd,Donath:2024utn,Melville:2024ove} for some proposals. Hence, partial wave bounds have been used to constrain cosmological theories only assuming a flat space limit \cite{Baumann:2011su,Baumann:2015nta,Kim_2021}. On a related note, progress has been made in studying scattering amplitudes with explicitly and spontaneously broken Lorentz invariance on flat spacetime and applying the resulting unitarity constraints to inflationary theories \cite{Pajer:2020wnj,Grall:2020tqc}.
We will occasionally refer to these bounds when comparing the purity bound with existing results.

\subsection{Perturbative unitarity bounds from purity}\label{sec3p3}

Much like the values of the partial wave coefficients $a_l$ are constrained by unitarity in the sense that probabilities are non-negative and sum to unity, the reduced density matrix $\rho_R$ enjoys bounds due to unitarity. In particular, the reduced density matrix $\rho_R$ is self-adjoint, and when diagonalised its elements have the interpretation of probabilities for the system to be found in the states which are its eigenvectors. If we properly normalize, $\Tr \rho_R = 1$, the Cauchy-Schwarz inequality then implies that for $N>1$,
\begin{equation}
   0 \leq \Tr_\S(\rho_R^N) \leq 1 \, ,
\end{equation}
with $\Tr_\S(\rho_R^N) = 0$ for a highly-mixed state in an infinite-dimensional Hilbert space, and $\Tr_\S(\rho_R^N) = 1$ when the reduced system is in a product state with its environment.
In particular, the purity $\gamma = \Tr_\S (\rho_R^2)$ satisfies
\begin{equation}
   \gamma   \geq 0 \, .
   \label{eq:gamma is bounded}
\end{equation}
As discussed in Section~\ref{sec:density matrix}, the purity  can be calculated perturbatively; in a theory with a cubic interaction, the leading contribution is given in terms of the  $\psi_3$ wavefunction coefficient as
\begin{equation} \label{eq:puritycubic}
    \gamma (\bfp) = 1 - 2 \int_{\mathbf{k}} {\frac{\left| \psi_3 (\mathbf{p},\mathbf{k},-\mathbf{p}-\mathbf{k}) \right|^2}{2\text{Re}\, \psi_2(\mathbf{p}) \, 2\text{Re}\, \psi_2(\mathbf{k}) \, 2\text{Re}\, \psi_2(\mathbf{p}+\mathbf{k})}} \, .
\end{equation}
Since $\psi_3$ is proportional to the cubic coupling $g$, the purity takes the form
\begin{equation} \label{eq:purityschem}
   \gamma = 1 - 2 g^2 I
\end{equation}
for some positive $I$. At this order in perturbation theory, $\gamma \leq 1$ is satisfied automatically. Demanding that $\gamma \geq 0$ then places bounds on the strength of the cubic coupling and on the domain of the integral, and so on the range of kinematics for which the theory is valid:
\begin{equation}
   \abs{g} \leq \frac{1}{\sqrt{2I}} \, .
\end{equation}

Constraints on scattering amplitudes from quantum entanglement have been studied in the past, for example in the interesting work \cite{Aoude:2024xpx}. In that case, the authors computed the purity of one particle's reduced density matrix after two-to-two elastic forward scattering of flavoured scalars. They calculated $1 - \gamma$ in perturbation theory via the S-matrix and imposed the constraint that $1 - \gamma \geq 0$. This excludes non-unitary interactions and considers a situation where the final reduced state is almost pure. The approach we take in the present work is somewhat different: we require that $1 - \gamma \leq 1$ for a single Fourier mode, and so our constraints arise when the reduced state is highly mixed. Our constraints are not intended to rule out truly non-unitary theories; rather, we diagnose the breakdown of perturbative unitarity. It is interesting that constraints can arise at both extremes of the allowed range of purity. \\

In practical calculations, we will use the following result repeatedly. The $\psi_3(\bfp, \bfk, -\bfp-\bfk)$ and $\psi_2$ wavefunction coefficients appearing in the purity~\eqref{eq:puritycubic} will only depend on the magnitudes of the three momenta $p$, $k$, and $\lvert-\bfp-\bfk\rvert$. Therefore, it is convenient to perform a simple change of variables and express the integral for the purity as
\begin{align}
    \gamma (\bfp) = 1 - 2 \int_{k_1=0}^\infty \int_{k_2=\lvert k_1-p\rvert}^{k_1+p} \frac{\d k_1 \d k_2 }{(2\pi)^2} \cdot \frac{k_1k_2}{p} \cdot \frac{\left| \psi_3 (p,k_1,k_2) \right|^2}{2\text{Re}\, \psi_2(p) \, 2\text{Re}\, \psi_2(k_1) \, 2\text{Re}\, \psi_2(k_2)} \, .
\end{align} \\
We note that for $N > 1$, in the infinite-volume limit,
\begin{align}
   \Tr_\S(\rho_R^N) &= 1 - N \int_{\mathbf{k}} {\frac{\left| \psi_3 (\mathbf{p},\mathbf{k},-\mathbf{p}-\mathbf{k}) \right|^2}{2\text{Re}\, \psi_2(\mathbf{p}) \, 2\text{Re}\, \psi_2(\mathbf{k}) \, 2\text{Re}\, \psi_2(\mathbf{p}+\mathbf{k})}} \\
   &= 1 - N g^2 I \, ,
\end{align}
so that the analogous bound on the cubic coupling is
\begin{equation}\label{below320}
   \abs{g} \leq \frac{1}{\sqrt{NI}} \, .
\end{equation}
This appears concerning: it seems that by taking $N$ to be arbitrarily large, we can derive arbitrarily strong constraints on the size of $g$. We believe that this is not cause for alarm because $\Tr(\rho_R^N)$, $N \gg 2$, contains a very large number of operator insertions which can be contracted in very many ways, even at leading order in $g$. This may signal a rapid breakdown of perturbation theory in the calculation of very high-point correlations, which are less relevant in cosmology, where we usually aim to compute two-, three- and four-point correlations. This effect may be analogous to (but not the same as) the breakdown of perturbative unitarity at tree level in very high-multiplicity scattering processes in theories of self-interacting scalars, for example $1\rightarrow N$ scattering, or the production of $N$ Higgs bosons from quarks in the Standard Model \cite{Goldberg:1990qk, Argyres:1992np, Argyres:1992nq}. This is not usually considered to be a problem for perturbative calculations of few-particle scattering.
\\
There is a close relation between violation of the purity bound~\eqref{eq:gamma is bounded} and the breakdown of the loop expansion described in Section~\ref{sec3p1}. The power spectrum at one loop can be written in terms of wavefunction coefficients as\footnote{The final term arises from the coupling of each mode of the scalar field to its zero mode. For massless fields in de Sitter spacetime, if this coupling does not vanish, the correct treatment of the zero mode can include nonperturbative effects --- see \cite{Rajaraman:2010xd} for a nice explanation in Euclidean signature. In the presence of an IR cutoff, or when all interactions have sufficiently many derivatives, the final term is not expected to contribute.} (see e.g.~\cite{Melville:2021lst,Cespedes:2023aal})
\begin{align}\label{P1loop}
P(p)&=\frac{1}{2\Re \psi_2(p)}\left[1+ \frac{\Re \psi_2^{(1L)}(p)}{\Re \psi_2(p)}      -\int_{\bfk}\frac{\Re\psi_4(\bfp,-\bfp,\bfk,-\bfk)}{4\Re \psi_2(p)\left(\Re\psi_2(k)\right)^2}  \right. \nonumber \\
& \quad \left. +\int_{\bfk} \frac{(\Re \psi_3(p,k,|\bfp+\bfk|))^2}{4\Re\psi_2(k)\Re \psi_2(p)\Re\psi_2(|\bfp+\bfk|)}  + \int_{\bfk} \frac{\Re \psi_3(p, p, 0) \, \Re \psi_3(0, k, k)}{\Re \psi_2(p) \Re \psi_2(0) \Re \psi_2(k)}\right]\,.
\end{align}
where $\psi_2$ and $\psi_2^{(1L)}$ are the tree-level and one-loop contributions to the quadratic wavefunction respectively. When $\psi_3$ is real, the first classical loop term coming from $\psi_3$ is equal to the integral for the purity~\eqref{eq:puritycubic}. In the absence of a cancellation between $\psi_2^{(1L)}$ and the classical loops of $\psi_3$ and $\psi_4$, the purity bound diagnoses when the one-classical-loop correction to the power spectrum becomes \textit{of order one}. Violation of the purity bound is therefore a diagnostic of the breakdown of the loop expansion. One advantage of this approach is that, of the contributions to the one-loop power spectrum, the term in common with the purity is typically the easiest to calculate.

\section{Flat spacetime}
\label{sec:flat spacetime}

\begin{table}
\begin{center}
   \renewcommand{\arraystretch}{1.5}
\begin{tabular}[t]{c c c c c}
   \multicolumn{2}{c}{Theory} & Partial waves & Purity \\ \hline 
   \multirow{2}*{\hyperref[sec:flat_phicubed]{$\displaystyle g \phi^3/3!$}} & Massive &\multirow{2}*{$\displaystyle g^2 < \frac{12\pi}{5} m^2$ } & $\displaystyle g^2 < 24^2 m^2$ \\
   & Massless & & $\displaystyle g^2 \lesssim 23^2 \LIR^2$ \\
   \multirow{2}*{\hyperref[sec:flat_phidphi2]{$\displaystyle g \, \phi \, \partial_\mu \phi \partial^\mu \phi / 2 $}} & Massive &$\displaystyle g^2 m^2 < 32\pi/19$  & $\displaystyle g^2 m^2 < 64 $  \\
   & Massless & No bound & $\displaystyle g^2 \lesssim 24 \pi^2 \LIR / \LUV^3$ \\
   {\hyperref[sec:GR]{General Relativity}} & Massless & --- & $\displaystyle \frac{\LIR}{\LUV} > \left( \frac{\LUV}{45\pi^2\Mpl}\right)^2$  \\
   \multirow{2}*{\hyperref[sec:flat_Sdphi2]{$\displaystyle g \, S \, \partial_\mu \phi \partial^\mu \phi^* / 2 $}} & Massive & \multirow{2}*{$\displaystyle g^2 E^2 < 16 \pi$}  &  \multirow{2}*{$\displaystyle g^2 \LUV^2 < \frac{768\pi^2}{7}$ }   \\
   & Massless 
\end{tabular}
\renewcommand{\arraystretch}{1}
\label{tab:flat-space summary}
\caption{Summary of the perturbativity bounds enjoyed by theories on flat space. In the $g \phi \partial_\mu \phi \partial^\mu \phi/2$ theory, partial wave unitarity provides no bound because the theory is free after a field redefinition, and amplitudes are invariant under field redefinitions. The wavefunction is not invariant, and a purity bound is therefore found. Throughout, $\LUV$ and $\LIR$ are the UV and IR cutoffs, respectively, of each theory; $m$ is the mass of the $\phi$ field; and $E$ is the energy in the centre-of-mass frame of scattering processes. The purity is evaluated for a mode $p$; we then minimise the purity with respect to $p$ to find the strongest possible bound on the coupling $g$ and the cutoffs, which we report in the table. When $\LIR$ can be taken to zero or $\LUV$ can be taken to infinity without requiring $g=0$, we do so in order to simplify the reported bound. In that case, the theory is weakly interacting in the IR or UV, respectively: see Figure~\ref{figRG}.  }
\end{center}
\end{table}

In this section, we compute the purity of a single mode $\mathbf{p}$ in various cubic scalar theories in flat space to leading order in the couplings, cf.~\eqref{eq:puritycubic}. Then, we study the bound that positivity of the purity implies for perturbation theory in each case. Since we are in Minkowski, we are able to compare these bounds with the perturbative unitarity bounds derived from amplitude partial waves.

To compute the purity we need the two-point wavefunction coefficient of a scalar field of mass $m$ in flat space, which is
\begin{equation}
    \psi_2 (k) = E \equiv \sqrt{m^2+k^2} \, ,
\end{equation}
and the cubic one $\psi_3$, which is theory dependent. The purity is hard to compute for general mass $m$ of the field and momentum $\mathbf{p}$ of the chosen mode, so in the following we focus on two specific regimes---one is the massless limit, $m=0$, and in the other we keep the mass finite but choose $p\equiv |\bfp| \ll m$.

\subsection{Super-renormalizable theories: $\phi^3$} \label{sec:flat_phicubed}

\begin{figure}
    \centering
    \includegraphics[width=0.6\textwidth]{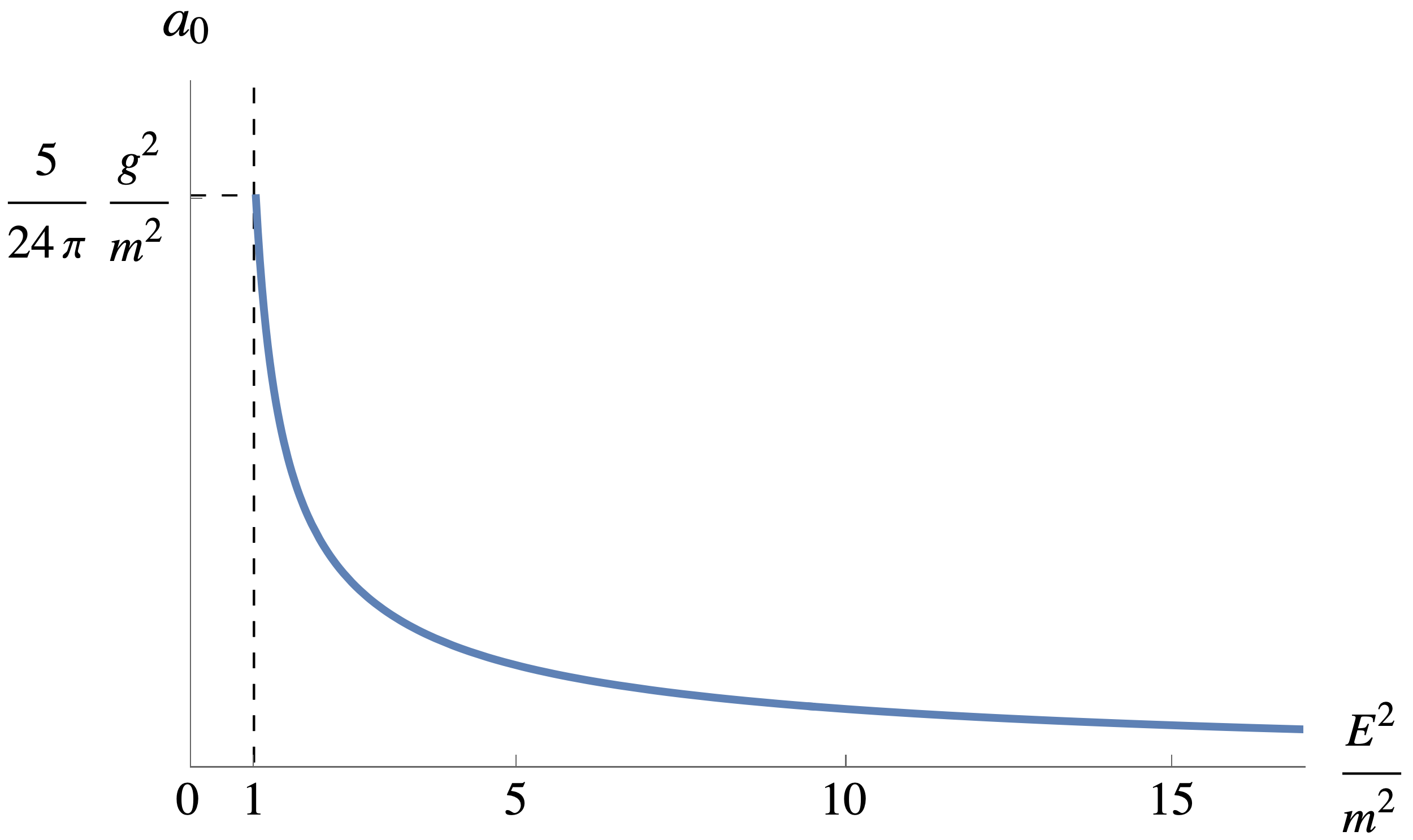}
    \caption{Partial wave coefficient $a_0$ of the tree-level $2\to 2$ scattering amplitude of scalars $\phi$ in the cubic theory~\eqref{eq:phicubed}. The kinematic variable $E$ is the energy in the center-of-mass frame.}
    \label{fig:a0_phicubed}
\end{figure}

We start by looking at a scalar with a cubic coupling in Minkowski,
\begin{equation} \label{eq:phicubed}
    \mathcal{L} = -\frac{1}{2} (\partial_{\mu}\phi )^2 - \frac{m^2}{2} \phi^2 - \frac{g}{3!}\phi^3 \, .
\end{equation}
The tree-level $2\to 2$ amplitude in this theory is simply
\begin{equation} 
    \mathcal{A}_{2\rightarrow 2}= - g^2 \left(\frac{1}{s-m^2}+\frac{1}{t-m^2}+\frac{1}{u-m^2}\right) \, ,
\end{equation}
and the corresponding partial wave coefficient $a_0$ in the center-of-mass frame (cf.~\eqref{eq:com}) is
\begin{equation} \label{eq:pwc_phicubed}
        a_0 = - \frac{1}{16\pi} \frac{g^2}{m^2} \left[ \frac{2}{4\frac{E^2}{m^2}-1}-\frac{1}{\frac{E^2}{m^2}-1}\log \left(4\frac{E^2}{m^2}-3\right)\right]\, ,
\end{equation}
which is plotted in Figure~\ref{fig:a0_phicubed} as a function of energy. 
When the coupling is smaller than the mass, in particular for
\begin{equation} \label{eq:gmlim}
    \frac{|g|}{m} < \sqrt{\frac{12\pi}{5}} \sim 2.7 \, ,
\end{equation}
the partial wave coefficient stays below $1/2$ and the theory is weakly coupled at all energies (see Figure~\ref{fig:a0_phicubed}). This is precisely what one would expect by naively comparing the free theory $\partial \phi^2+ m^2 \phi^2$ with the interaction $g\phi^3$ and estimating $\phi\sim E$. As we increase the coupling and go beyond the limit~\eqref{eq:gmlim}, we find a violation of perturbative unitarity at low energies.

We now compare this partial wave bound with the one coming from positivity of the purity~\eqref{eq:puritycubic}. To compute it, we need the three-point wavefunction coefficient for this theory:
\begin{equation}
    \psi_3 (k_1,k_2,k_3) = \frac{g}{E_1+E_2+E_3} \, .
\end{equation}

\begin{figure}
    \centering
    \includegraphics[width=0.6\textwidth]{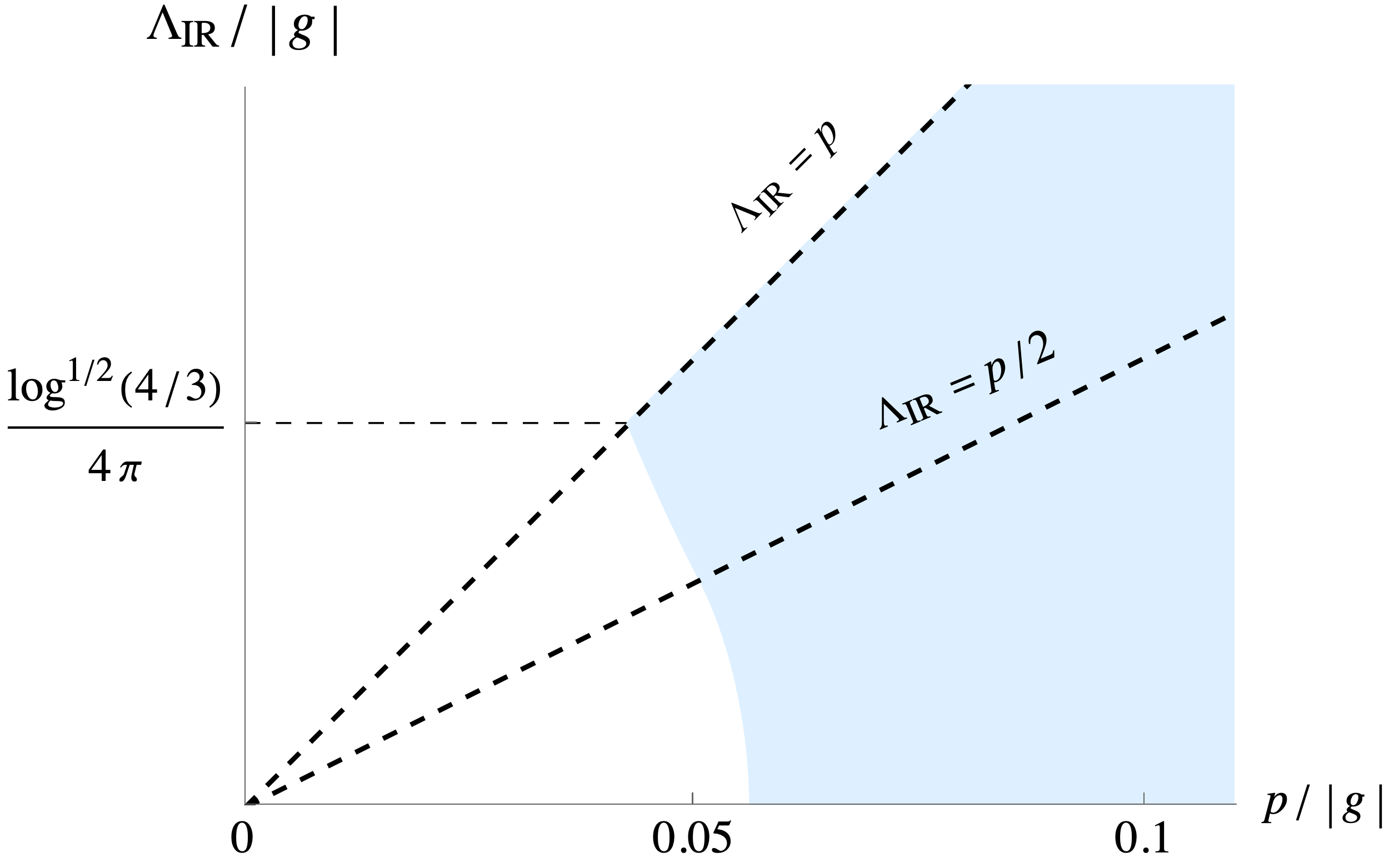}
    \caption{We have shaded (in blue) the region in the plane of energies $p$ and $\LIR$ for which the purity~\eqref{eq:phi3m0purity} satisfies the unitarity bound $\gamma\geq 0$. For $\LIR< |g| \log^{1/2} (4/3)/4\pi$ the line $\LIR=p$ is no longer included in the shaded region. Therefore, the momentum $p$ of the system cannot be taken all the way down to $\LIR$, which signals the breakdown of the EFT.}
    \label{fig:reg_phicubed}
\end{figure}

\paragraph{Purity: the massless case} We start by considering the case $m = 0$. We know that this case will be subtle because the theory becomes gap-less and can probe arbitrary low energies where the relevant coupling $g\phi^3$ becomes very big. Using $\LUV$ and $\LIR$ cutoffs in the purity integral, we get\footnote{Conservation of momentum entering $\psi_3$ and the triangle inequality imply that the integration region has a different shape depending on the hierarchy between $p$ and $2\LIR$, hence the different results for each case. The upper limit of $\LUV/2$ is required to ensure a fixed hierarchy between $\LUV-p$ and $\LIR$ for the same reason. This limit could be extended to $\LUV$ by considering different integration limits. However, we are ultimately interested in the limit $p\rightarrow \LIR$ and so it is unnecessary to extend the allowed range of $p$ in that way. \label{fn:intregions}}
\begin{equation} \label{eq:phi3m0purity0}
    \gamma (\bfp) = \begin{cases} 
    1 - \left( \dfrac{g}{4\pi p} \right)^2 \left[ \dfrac{1}{2} - \dfrac{\LIR}{p} + \log \left( \dfrac{2\, \LUV \, (
    p+\LIR)}{p \, (
    p+2\LUV)} \right) \right],& 2\LIR < p \leq \dfrac{\LUV}{2} \, , \\[15pt]
    1 - \left( \dfrac{g}{4\pi p} \right)^2 \log \left( \dfrac{4\, \LUV \, (
    p+\LIR)}{(p+2\LIR) (
    p+2\LUV)} \right),& \LIR \leq p \leq 2 \LIR \, .
    \end{cases}
\end{equation}
The coupling $g\phi^3$ is relevant in perturbation theory; in the UV the dimension-3 interaction gets arbitrarily smaller than the dimension-4 kinetic term, and the theory becomes free. This is manifest in~\eqref{eq:phi3m0purity0} because the limit $\LUV\to\infty$ yields a finite result,
\begin{equation} \label{eq:phi3m0purity}
    \gamma (\bfp) = \begin{cases} 
    1 - \left( \dfrac{g}{4\pi p} \right)^2 \left[ \dfrac{1}{2} - \dfrac{\LIR}{p} + \log \left( 1 + \dfrac{\LIR}{p} \right) \right],& 2\LIR < p \, , \\[15pt]
    1 - \left( \dfrac{g}{4\pi p} \right)^2 \log \left( 2\dfrac{p+\LIR}{p+2\LIR} \right),&\LIR \leq p \leq 2 \LIR \, .
    \end{cases}
\end{equation}
On the other end of the spectrum, notice that when $\LIR\to 0$ the bound $\gamma\geq 0$ requires $g$ to vanish for an arbitrarily small $p$. This indicates that the theory becomes strongly coupled in the IR, so we should keep $g$ fixed and try to find a lower bound for $\LIR$. Figure~\ref{fig:reg_phicubed} shows the energy scales $\{p,\LIR\}$ that yield $\gamma\geq 0$ for fixed $g$. We see that for $\LIR< |g| \log^{1/2} (4/3)/4\pi$ there is a lower bound on $p$ which is actually above the IR cutoff, which is inconsistent. We therefore require
\begin{equation} \label{eq:phi3m0bound}
    \LIR \geq \dfrac{|g|}{4\pi} \log^{1/2} (4/3) \simeq \frac{|g|}{23} \, .
\end{equation}
We conclude that, for some fixed and finite $g$, this EFT has a limited range of validity in the IR, but unlimited in the UV. This is an example of ``UV-free" theories described in the bottom panel of Figure~\ref{figRG}.\\

\paragraph{Purity: the massive case} The integral becomes hard when we add the extra variable $m\neq 0$, so we only solve it in the limit of small momentum $p$ (which in turn requires making the IR cutoff arbitrarily low, $\LIR\to 0$). Taking $p \ll m$ and working to leading order in $g$, the purity and the corresponding bound $\gamma\geq 0$ are
\begin{equation} \label{eq:pbms_phicubed}
    \gamma (\bfp) = 1 - \frac{1}{2} \left( \frac{1}{2}+\frac{4\sqrt{3}-9}{18}\pi \right) \left( \frac{g}{2\pi m} \right)^2 + \mathcal{O} \left( \frac{p}{m} \right) \qquad\Rightarrow\qquad \frac{|g|}{m} \lesssim 24 \, .
\end{equation}
Here we also see that the integral converges in the UV and so we simply take $\LUV \to \infty$. We get a very similar result to the massless case~\eqref{eq:phi3m0bound}, with the mass now playing the role of the IR cutoff. Since this result is only valid for soft kinematics, we must compare it with the partial wave bound evaluated at $E\simeq m$. From the above discussion we see that this is actually where we get the strongest bound---the theory is weakly coupled at that energy only if~\eqref{eq:gmlim} is satisfied. We conclude that the partial wave and purity bounds have the same scaling but the former is more constraining by a factor of about 9.

\subsection{Free theories and field redefinitions: $\phi \partial_\mu \phi \partial^\mu \phi$} \label{sec:flat_phidphi2}

We now consider a cubic coupling with two derivatives,
\begin{equation} \label{eq:phidphisq}
    \mathcal{L} = -\frac{1}{2} (\partial_{\mu}\phi )^2 - \frac{m^2}{2} \phi^2 - \frac{g}{2} \phi (\partial_{\mu}\phi)^2 \, .
\end{equation}
This is a particularly interesting case because in the massless case this theory is actually free, as one can see by performing the field redefinition
\begin{equation} \label{eq:fieldredef}
    \phi \quad\longrightarrow\quad \phi(\varphi) = \frac{1}{g} \left( 1+\frac{3g}{2}\varphi \right)^{2/3} - \frac{1}{g} \, .
\end{equation}
Amplitudes are field-redefinition invariant and therefore the S-matrix is trivial at all energies and for all values of $g$ for $m=0$. Therefore there is no partial wave bound whatsoever! As we will discuss shortly, this will be different for the purity bound, since the wavefunction is not field redefinition invariant. This is a feature, not a bug. In the calculation of cosmological correlators, the choice of field is typically dictated by symmetries (see e.g. \cite{Green:2020ebl}) and one does not wish to perform field redefinitions. 

When $m\neq 0$ the theory is not free. The field redefinition~\eqref{eq:fieldredef} trades the cubic derivative interaction for an infinite tower of polynomial couplings. More in detail
\begin{equation} \label{eq:massivefieldredef}
    \begin{aligned}
        \mathcal{L} (\phi) \quad\longrightarrow\quad \mathcal{L} \Big[ \phi(\varphi) \Big] & = -\frac{1}{2} (\partial_{\mu}\varphi )^2 - \frac{m^2}{2} \phi^2 \\
        & = -\frac{1}{2} (\partial_{\mu}\varphi )^2 - \frac{m^2}{2} \varphi^2 + \frac{gm^2}{2} \varphi^3 - \frac{19g^2m^2}{96} \varphi^4 + \ldots
    \end{aligned}
\end{equation}
The tree-level $2\to 2$ amplitude is
\begin{equation}
    \mathcal{A}_{2\rightarrow 2}=-\frac{g^2}{4}\left[\frac{(s+2m^2)^2}{s-m^2}+\frac{(t+2m^2)^2}{t-m^2}+\frac{(u+2m^2)^2}{u-m^2}\right] \, ,
\end{equation}
the c.o.m.~partial wave at $l=0$ for this amplitude is 
\begin{align}
    a_0=\frac{g^2m^2}{32\pi}\left(\frac{38E^2-5m^2}{4E^2-m^2}+\frac{9m^2}{4(E^2-m^2)}\log\left(\frac{m^2}{4E^2-3m^2}\right)\right),
\end{align}
which asymptotes to its maximum value at high energies. This implies the following partial wave bound on the coupling:
\begin{equation} \label{eq:pub_phidphisq}
    \max(\lvert a_0\rvert )=\lim_{E\rightarrow\infty} \lvert a_0\rvert =\frac{19g^2m^2}{64\pi}\qquad \Rightarrow\qquad g^2m^2 < \frac{32\pi}{19} \sim 5 \, .
\end{equation}

\paragraph{Purity: the massless case} Let us now compare this amplitude's bound with the unitarity bound we get from purity. The three-point wavefunction coefficient for this theory for any mass is
\begin{equation} \label{eq:wfcphidphisq}
    \psi_3 (k_1,k_2,k_3) = \dfrac{g}{E_1+E_2+E_3} \left( \sum_{a<b} E_aE_b + \dfrac{1}{2} \sum_a k_a^2 \right) \,.
\end{equation}
Let us start with the $m=0$ case. As we have seen above, the theory is free and all scattering amplitudes vanish. However, the wavefunction is not necessarily Gaussian, as we can see by taking the massless limit of~\eqref{eq:wfcphidphisq}:
\begin{equation}
    \psi_3 (k_1,k_2,k_3) = \frac{g}{2} k_T \neq 0\, .
\end{equation}
Notice that this function has no total energy pole, as expected from the fact that the corresponding  amplitude vanishes\footnote{This is the case because in Minkowski the total-energy pole is of order one. Conversely, in de Sitter the order of the pole grows with the dimension of the interaction \cite{Pajer:2020wnj} and even when the residue of the leading pole vanishes subleading poles could be non-zero \cite{Grall:2020ibl,Bonifacio:2021azc}}. The reason why $\psi_3$ is non-zero is that the wavefunction, as well as the density matrix and therefore the purity, are not field-redefinition invariant; they depend on the field basis because this is what specifies the bipartition between system and environment. In the $\phi$ basis in which the Lagrangian takes the form~\eqref{eq:phidphisq} (with $m=0$) the purity is formally divergent, so we introduce UV and IR cutoffs in the integration. The result is %
\begin{align}\nonumber
    &\gamma (\mathbf{p}) = \\&\begin{cases} 
    1 - \left( \dfrac{gp}{8\pi} \right)^2 \left( \dfrac{8\LUV^3}{3p^3}+\dfrac{2\LUV^2}{p^2}+\dfrac{2\LUV}{3p} -\dfrac{31}{12} - \dfrac{8\LIR^2}{p^2} - \dfrac{16\LIR^3}{3p^3} - \dfrac{4\LIR^4}{3p^4} \right), & 2\LIR < p \leq \dfrac{\LUV}{2} \, , \\[8pt]
    1 - \left( \dfrac{gp}{8\pi} \right)^2 \dfrac{2(\LUV-\LIR)-p}{3p} \left( 4 \dfrac{\LUV^2+\LUV\LIR+\LIR^2}{p^2}+\dfrac{5\LUV+7\LIR}{p}+\dfrac{7}{2} \right), & \LIR \leq p \leq 2 \LIR \, ,
    \end{cases}
\end{align}
for which the discussion in Footnote~\ref{fn:intregions} also applies. Following a similar analysis to the previous subsection, we find that $\gamma\geq 0$ implies
\begin{equation} \label{eq:pbphidphisqm0}
        \LUV^3 \lesssim 24\pi^2 \, \frac{\LIR}{g^2} \, .
\end{equation}
We see that this perturbative EFT, in the $\phi$ basis, can only describe a window of scales between $\LIR$ and $\LUV$. It is then an example of the ``generic EFT'' displayed in the upper-right panel of Figure~\ref{figRG}. Interestingly, we see that the allowed perturbative window $\LUV/\LIR$ shrinks as either the coupling or the UV cutoff are raised.

\paragraph{Purity: the massive case} We look at the $p\ll m$ limit, where the purity integral can be performed (as long as we take $\LIR\to 0$). In the $\phi$ basis, the purity and the corresponding $\gamma \geq 0$ bound are
\begin{equation} \label{eq:pb_phidphisq}
    \gamma (\bfp) = %
    1 - \frac{g^2\LUV^3}{6(2\pi)^2m} + \ldots %
    \qquad\Rightarrow\qquad g^2m^2 \lesssim 24\pi^2 \left(\frac{m}{\LUV}\right)^3 \, ,
\end{equation}
where $\LUV \gg m$ is again the UV cutoff of the theory and the ellipsis denotes subleading terms in $p/m$ and  $m/\LUV$. Notice that this is the same result as~\eqref{eq:pbphidphisqm0}, with the mass $m$ playing the role of the IR cutoff $\LIR$.

The UV divergence of the purity is a consequence of the interaction having two derivatives. But recall that a field redefinition removes the derivatives and makes the theory polynomially coupled, cf.~\eqref{eq:massivefieldredef}. In the $\varphi$ field basis, the purity at order $g^2$ only gets a contribution from the $\varphi^3$ interaction, and we can recover the result from Section~\ref{sec:flat_phicubed} to get
\begin{equation}
    \gamma (\bfp) = 1 - \frac{9}{2(2\pi)^2} \left( \frac{1}{2}+\frac{4\sqrt{3}-9}{18}\pi \right) g^2m^2 + \mathcal{O} \left( \frac{p}{m} \right) \qquad\Rightarrow\qquad g^2m^2 \lesssim 64 \, .
\end{equation}
This result is not UV divergent, and the purity bound can be more directly compared to the partial wave bound~\eqref{eq:pub_phidphisq}---we see that the latter is an order of magnitude more constraining. We nevertheless expect the dependence on the UV cutoff $\LUV$ to reappear at higher orders in $g$, as the non-renormalizable couplings $\varphi^5$, $\varphi^6$, etc. start to contribute to the purity.

\subsection{General relativity} \label{sec:GR}

In this section, we present a preliminary investigation of our purity bounds applied to general relativity. While we ignore a few conceptual issues, we find this exploratory analysis intriguing because our purity bound ends up recovering the same scaling as one would obtain from demanding that a black hole has the maximum entropy for a region of a given size\footnote{E.P. would like to thank Xi Tong for pointing this out.} (as formalised by a series of increasingly more general bounds \cite{Bousso:1999xy,Bekenstein:1980jp,Fischler:1998st}). 

Our approach consists of computing the purity from the cubic interaction of gravitons in GR. This is quite naive for a variety of reasons\footnote{We are thankful to D. Skinner and R. Soni for pointing out some of these issues to us.}. First, it is known that the computation of entanglement in gauge theories is considerably more complicated because of the existence of constraint equations (see e.g. \cite{Casini:2013rba,Ghosh:2015iwa,Soni:2015yga}). In the interacting theory these constraints couple different Fourier modes. Second, the graviton field is not gauge invariant and so our single-Fourier mode purity depends on the gauge choice. Here we have in mind that some physical apparatus selects a preferred set of coordinates and the graviton field we consider is defined in that gauge. It would be interesting to find a more covariant formulation of our analysis.  \\ 

Graviton interactions in general relativity have a very similar form to the previous scalar example because they also involve two derivatives and three fields to lowest order in $\Mpl^{-1}$. For spinning fields, we choose our system $\S$ to be a single helicity mode, which we will take to have helicity $+2$ and momentum $\bfp$. The purity picks up contributions from all possible helicities over which we sum,
\begin{align}
    \gamma(\bfp,+2)=1-\frac{1}{4}\sum_{hh'}\int\frac{d^3k}{(2\pi)^3}\frac{\left\lvert\psi_3^{+hh'}(p,k,\lvert\textbf{p}-\textbf{k}\rvert)\right\rvert^2}{\Re\psi_2^+(p)\Re\psi_2^h(k)\Re\psi_2^{h'}(\lvert\bfp-\bfk\rvert)}.
\end{align}
If we canonically normalize the graviton, the quadratic wavefunction coefficient is simply
\begin{align}
    \psi_2^h=k.
\end{align}
With this normalization, the interaction strength of the dimension-five operator $g\gamma \partial\gamma^2$ is tied to the Planck scale by $g=1/\Mpl$. This makes the connection to the scalar model in the previous section explicit.

Expanding the purity using the fact that the tensor structure in the wavefunction coefficient is identical to \cite{Maldacena:2011nz} we find
\begin{align}
    \gamma^+(\bfp)&=1-\frac{1}{\Mpl^2}\int\frac{d^3k}{(2\pi)^3}\frac{I_1^2I_2^2I_3^2(k_T^8+I_1^8+I_2^8+I_3^8)}{16384e_3^5}\\&=1-\left(\frac{\LUV}{\pi\Mpl}\right)^2\left(\frac{1}{90}\frac{\LUV}{p}-\frac{4301}{196608}+\mathcal{O}\left(\frac{p}{\LUV},\frac{\LIR}{\LUV}\right)\right),
\end{align}
where $I_i=k_T-2k_i$ and the integral has split into different regions according to the size of $p$ just like in the scalar case. However, in the limit that we take $\LUV\gg \LIR,p$ both branches give the same result.

The strictest purity bound for this expression then comes from taking the momentum to be as low as possible, i.e. $p=\LIR$. Thus, the corresponding purity bound (keeping just the leading order term in $\LUV/\LIR$) is
\begin{align}\label{finalGR}
    \left(\frac{\LIR}{\LUV}\right)>\frac{1}{90\pi^2}\left(\frac{\LUV}{\Mpl}\right)^2,
\end{align}
At face value, this bound can be interpreted in various ways. First and unsurprisingly, since $\LIR<\LUV$, this bound tells us that $\LUV\ll \Mpl$, as expected. Second, and also unsurprisingly, we can choose $\LUV$ such that $\LIR$ is arbitrarily small, capturing the fact that GR becomes free in the infrared. Third and more interestingly, the purity bound also tells us that the width of the window of validity of perturbation theory in GR must be finite, and it gets narrower as we push $\LUV$ towards $\Mpl$. In other words, if we want our theory to probe deep in the IR we need a correspondingly low UV cutoff. 

A few comments about the physical interpretation of our result are in order. First, in the limit in which the mode $\bfp$ is very different in size compared to the two UV modes with which it is entangled, one expects that the effect of the former on the latter is just a change of coordinates to leading order in derivatives. This is indeed very similar to the arguments underlying the well-studied cosmological soft theorems first discovered in \cite{Maldacena:2002vr}. Therefore, even though perturbation theory might be breaking down, the physical coupling among hierarchically separated modes should be easy to describe non-perturbatively. Indeed, it amounts to a change of coordinates and can be made explicit e.g. using Fermi normal coordinates in flat space or conformal Fermi coordinates in FLRW spacetimes \cite{Pajer:2013ana,Dai:2015rda}. 

Second, let us mention that the discussion of partial wave bounds for general relativity is complicated by the spin of the particles involved and by the IR divergences associated to massless gravitons and we hence omit it here. 

Third, we notice that the scaling of our final purity bound for GR, \eqref{finalGR}, is intriguingly equivalent to an entropy bound for black holes. More concretely, recall that entropy in QFT grows with the volume, so dimensionally
\begin{align}
S_\text{QFT} = (L \LUV)^3  \,,   
\end{align}
where $L$ is the size of the system under consideration. It is natural to identify this size as the IR cutoff of our EFT, so $L = 1/\LIR$. But in a theory with dynamical gravity, a black hole has the maximum possible entropy. For a black hole of area $L^2$ that is
\begin{align}
    S_\text{BH}= \frac{A}{4G_N} \sim L^2 \Mpl^2 = \left(\frac{\Mpl}{\LIR}\right)^2\,.
\end{align}
For consistency we should demand $S_\text{QFT} < S_\text{BH}$. This gives
\begin{align}
    \frac{\LIR}{\LUV} > \left(\frac{\LUV}{\Mpl}\right)^2\,.
\end{align}
This should be familiar because it happens to be the same scaling we found from our purity bound in \eqref{finalGR}, up to the very large factor of $90 \pi^2$. Notice that both bounds have to do with entanglement, as measured by purity or entropy. Note also that the black hole discussion is non-perturbative in $1/\Mpl$, while the purity bound is perturbative. It is intriguing that nevertheless the scaling appears to be the same. It would be interesting to investigate this further.

\subsection{Effective field theories: $S\partial_\mu\phi\partial^\mu\phi^*$} \label{sec:flat_Sdphi2}

Now we study a theory with a complex scalar field $\phi$ of mass $m$ coupled to a real scalar field $S$ of mass $M>m$:
\begin{equation} \label{eq:Sdphisq}
    \mathcal{L} = -\frac{1}{2} |\partial_{\mu}\phi |^2 - \frac{m^2}{2} |\phi|^2 -\frac{1}{2} (\partial_{\mu}S )^2 - \frac{M^2}{2} S^2 - g S |\partial_{\mu}\phi|^2 \, .
\end{equation}
The amplitude for the $\phi\phi^*\rightarrow\phi\phi^*$ process is, at tree level,
\begin{equation}
    \mathcal{A}_{\phi\phi^*\rightarrow\phi\phi^*} = -\frac{g^2}{4}\left[\frac{(s+2m^2)^2}{s-M^2}+\frac{(t+2m^2)^2}{t-M^2}\right] \, .
\end{equation}
This amplitude grows unbounded with the center-of-mass energy $E$, as do its partial waves. The first partial wave is
\begin{align}\nonumber
    a_0&=-\frac{g^2}{64\pi}\left(\frac{4(2E^2+m^2)^2}{4E^2-M^2}+M^2+6m^2-2E^2+\frac{(2m^2+M^2)^2\log\left(\frac{M^2}{4(E^2-m^2)+M^2}\right)}{4(E^2-m^2)}\right)\\&=-\frac{g^2E^2}{32\pi}-\frac{g^2(5m^2+M^2)}{32\pi}+\mathcal{O}\left(\frac{1}{E}^2\right).
\end{align}
So, respecting the unitarity condition $|\Re \, a_0|<1/2$ implies the following bound:
\begin{equation} \label{eq:pwb_Sdphisq}
    E^2 \lesssim \frac{16\pi}{g^2} \qquad\text{for $E\gg M,m$} \, .
\end{equation}

Besides the amplitude, in this theory we have also computed the 1-loop correction to the two-point wavefunction coefficient. The integral diverges so we need to introduce a UV cutoff $\LUV$, which must satisfy $\LUV\gg M$ such that the heavy scalar remains within the partial UV completion. We get
\begin{align}
    \psi_2^{\text{(1-loop)}} (k) = \frac{g^2 \LUV^2 (4k^2+3m^2)}{24\pi^2\sqrt{k^2+m^2}} \, .
\end{align}
Another perturbativity bound on the energy is obtained by requiring this correction to be smaller than the tree-level wavefunction coefficient:
\begin{align} \label{eq:wlb_Sdphisq}
    \psi_2^{\text{(1-loop)}} (k) < \psi_2^{\text{(tree)}} (k) = \sqrt{k^2+m^2} \qquad\Rightarrow\qquad \LUV^2 < \frac{24\pi^2}{g^2}\frac{(k^2+m^2)}{(4k^2+3m^2) }\sim \frac{7\pi^2}{g^2}  \, .
\end{align}
Hence the perturbativity of the loop expansion gives a similar result to partial wave unitarity, from which it differs by a factor of about 2.\\

We will now compute the purity bound for this theory in the two usual regimes, and will compare it with the bounds above. Assigning the momentum $k_1$ to the field $S$, the three-point wavefunction coefficient of $S\phi\phi^*$ is
\begin{equation}
    \psi_{S\phi\phi^*} (k_1,k_2,k_3) = \frac{g}{\Lambda} \cdot \frac{2\sqrt{k_2^2+m^2}\sqrt{k_3^2+m^2}-k_1^2+k_2^2+k_3^2}{2(\sqrt{k_1^2+M^2}+\sqrt{k_2^2+m^2}+\sqrt{k_3^2+m^2})} \, .
\end{equation}
In this case we are still computing the purity of a mode of momentum $\bfp$ of the field $\phi$, but the environment now consists of all the remaining modes of $\phi$ plus all the modes of $S$, which is fully integrated out. The relevant formula will still be~\eqref{eq:puritycubic}, bearing in mind that the mode $\bfp$ must be assigned to a field $\phi$, while $\bfk$ and $-\bfp-\bfk$ correspond indistinctly to the other $\phi$ and to $S$.

\paragraph{Purity: the massless case} We start with the case in which the complex scalar $\phi$ is massless, $m=0$. To regulate the integral we again cut it off at an energy $\LUV\gg M$. The selected mode $\bfp$ can therefore have any magnitude in the range $0<p<\LUV$. The result for general $p$ is complicated and barely informative, but in the limit $p\ll\LUV$, the purity and the bound $\gamma\geq 0$ are
\begin{equation}
    \gamma (\bfp) = 1 - \frac{g^2p\LUV}{24\pi^2} \left[ 1 + \mathcal{O} \left( \frac{p}{\LUV} \right) \right] \qquad\Rightarrow\qquad \LUV\lesssim \frac{24\pi^2}{pg^2} \, .
\end{equation}
The structure of the bound suggests that we will find a strongest constraint by pushing $p \to \LUV$. In that case the result is still too complicated, so we show only the leading term in the $\LUV\gg M$ expansion:
\begin{equation}
    \gamma(|\bfp|=\LUV) = 1-\frac{7g^2\LUV^2}{768\pi^2} \left[ 1 + \mathcal{O} \left( \frac{M}{\LUV} \right) \right] \qquad\Rightarrow\qquad \LUV^2 \lesssim \frac{768\pi^2}{7g^2} \, .
    \label{eq:pub_Sdphisq_massless}
\end{equation}
The purity bound tells us that perturbation theory only makes sense up to an energy $\LUV \sim 33/g$. This goes in the same direction as the partial wave bound~\eqref{eq:pwb_Sdphisq}, which is nevertheless a few times more constraining since it finds the strong coupling scale at $E\sim 7/g$. %

\paragraph{Purity: the massive case} Now we take the complex scalar $\phi$ to be massive. We can then consider two interesting limits. We can take $p\gg m$, in which case the mass is irrelevant and we return to massless discussion around~\eqref{eq:pub_Sdphisq_massless}. Alternatively, we can consider the limit of small momentum, $p \ll m$. The purity and the corresponding $\gamma\geq 0$ bound are
\begin{equation}
    \gamma (\bfp) = 1 - \frac{g^2m\LUV}{32\pi^2} + \ldots \qquad\Rightarrow\qquad \LUV\lesssim \frac{32\pi^2}{mg^2} \, ,
\end{equation}
where the ellipsis denotes terms suppressed by $p/m$, $m/\LUV$, or $M/\LUV$. We expect the coupling and the mass to satisfy $mg\ll 1$ on general EFT grounds. Therefore, we can argue that the above purity bound will be looser than the we one from $p\gg m$. Since the EFT description should hold for all modes below the cutoff $\LUV$, one should use the stronger purity bound in the ``massless" limit $p\gg m$, which as we discussed is a factor of a few weaker than the partial wave and loop perturbativity bounds.

\section{de Sitter spacetime}\label{sec:de Sitter}

In this section, we explore how the purity constraint that $\gamma\geq0$ restricts the allowed interactions in a de Sitter spacetime without any reference to flat space. This is important because previous attempts to derive such constraints have relied upon a flat spacetime limit. In other words the bounds on cosmological models have been derived from asking that they consistently describe flat space physics for modes that are so short to hardly feel the expansion of the universe. While this is certainly an informative analysis, it is possible for a theory that is completely valid in de Sitter to have no equivalent theory in flat spacetime and so we must be careful when extrapolating such results. Indeed we will present evidence here that perturbative unitarity bounds on EFTs that describe super-Hubble cosmological correlators can be parametrically different from their flat spacetime counterpart.

For concreteness, we consider theories of a massless scalar field in de Sitter, for which the real part of the two-point wavefunction coefficient at tree level takes the form
\begin{equation}
    \Re \psi_2 (k) =\frac{k^3}{H^2} \, ,
\end{equation}
with $H\equiv \dot{a}/a$ the Hubble constant.

\subsection{The effective field theory of inflation} \label{sec:EFToI}

We start by studying the fluctuations of a minimally coupled scalar around a fixed de Sitter background through the effective field theory of inflation (EFToI), whose action is given by~\cite{Cheung:2007st}
\begin{align} \label{eq:EFToI}
    S = \int \d t \, \d^3x \, a^3 \left[ \frac{1}{2}\dot{\phi}^2-\frac{1}{2a^2}(\partial_i\phi)^2 +\frac{g_1}{3!} \dot{\phi}^3+\frac{g_2}{2 a^2} \dot{\phi} (\partial_i\phi)^2 +\dots \right] \, ,
\end{align}
where we neglected higher derivatives and higher multiplicity interactions. Here $a$ is the time-dependent scale factor, which in de Sitter is $a=-(H\eta)^{-1}=e^{Ht}$. The three-point wavefunction coefficient is
\begin{equation}
    \psi_3 (k_1,k_2,k_3) = - \frac{g_1}{H} \frac{2e_3^2}{k_T^3} - \frac{g_2}{2H} \left( 12 \frac{e_3^2}{k_T^3} - 4\frac{e_2e_3}{k_T^2} - 4\frac{e_2^2}{k_T} + 11e_3 - 3e_2k_T +k_T^3 \right) \, ,
    \label{eq:EFTI psi3}
\end{equation}
where the elementary symmetric polynomials are symmetric combinations of the norms of the wavevectors $\bfk_a$ for $a=1,2,3$ defined in~\eqref{esp}. The squeezed limit for 
\begin{align}
    \bfk_1&=\bfk_s-\bfk_l/2\,,& \bfk_2&=-\bfk_s-\bfk_l/2 \,, & \bfk_3&=\bfk_l\,,
    \label{eq:symmetric squeezed limit}
\end{align}
is found to be 
\begin{equation}\label{EFTsqueezed}
   \lim_{k_l\to 0}\psi_3(k_1,k_2,k_3) =- \frac{g_1 -8g_2 + 5 g_2 \cos^2\theta}{4H} k_s^3 \left[ \left( \frac{k_l}{k_s} \right)^2 +  \mathcal{O}\left( \frac{k_l}{k_s} \right)^3  \right]\,,
\end{equation}
where $\theta$ is the angle between $\bfk_s$ and $\bfk_l$. Because this vanishes as $k_l^2$ in the squeezed limit, the purity for a single mode $\bfp$ of the scalar is found to be finite and equal to
\begin{equation}
    \gamma = 1-\frac{H^4}{6400\pi^2} \left( \frac{331}{18} g_1^2 + \frac{22959}{2} g_2^2 - 879 g_1g_2 \right) \, .
    \label{eq:EFTI purity}
\end{equation}
A remarkable difference emerges from our flat spacetime analysis: the integral appearing in the purity calculation converges both in the UV and in the IR. Hence, for ``sub-Hubble" couplings, $g_1,g_2\ll \mathcal{O}(50) H^{-2}$, this does not impose any bounds on $\LUV$ and $\LIR$. This is in stark contrast with our flat space results in Section~\ref{sec:flat_phidphi2} for the not so different interaction $\phi \partial\phi^2$.

\begin{figure}
\begin{subfigure}{.5\textwidth}
    \centering
    \includegraphics[width=.95\linewidth]{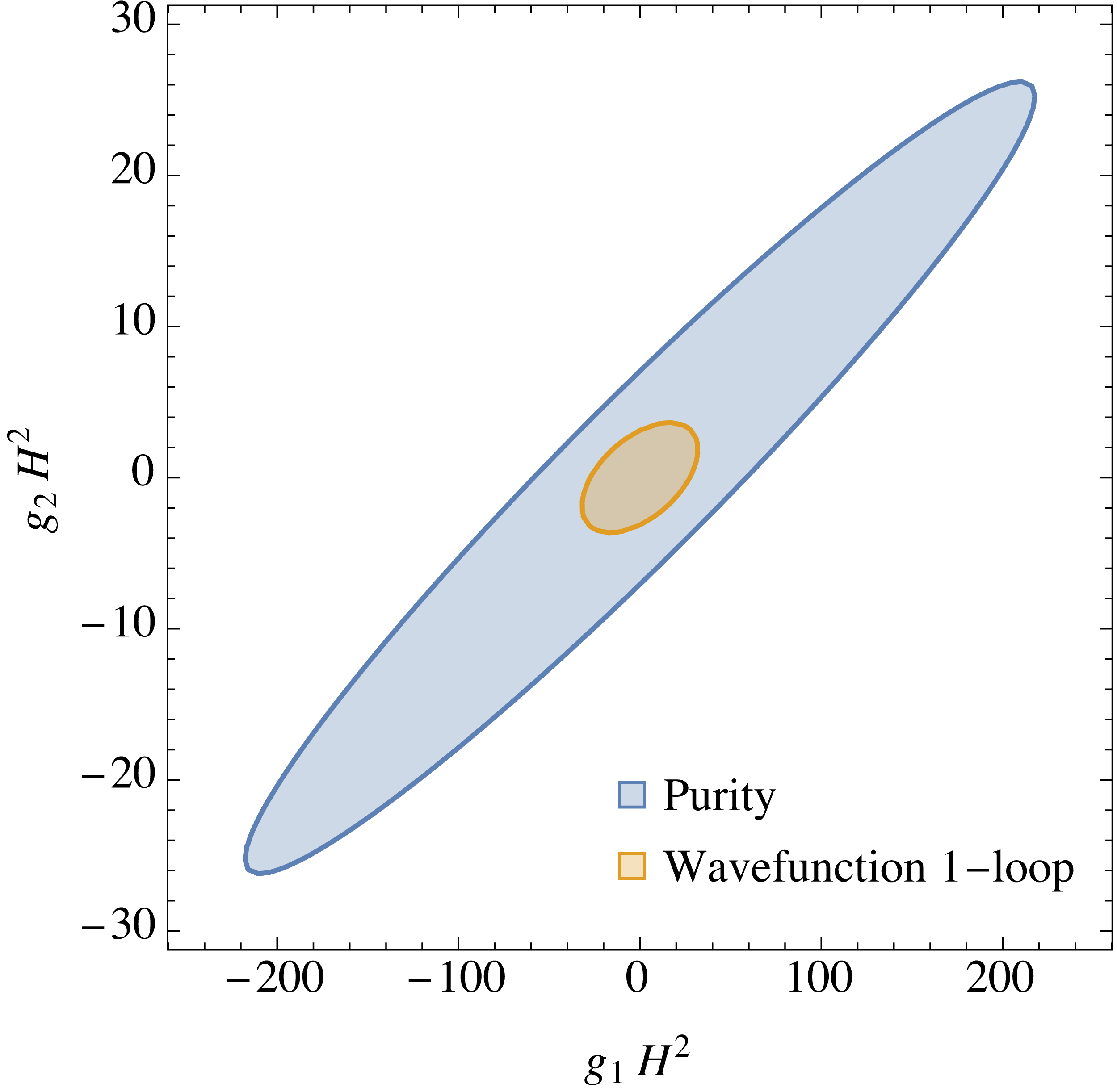}
\end{subfigure}
\begin{subfigure}{.5\textwidth}
    \centering
    \includegraphics[width=.95\linewidth]{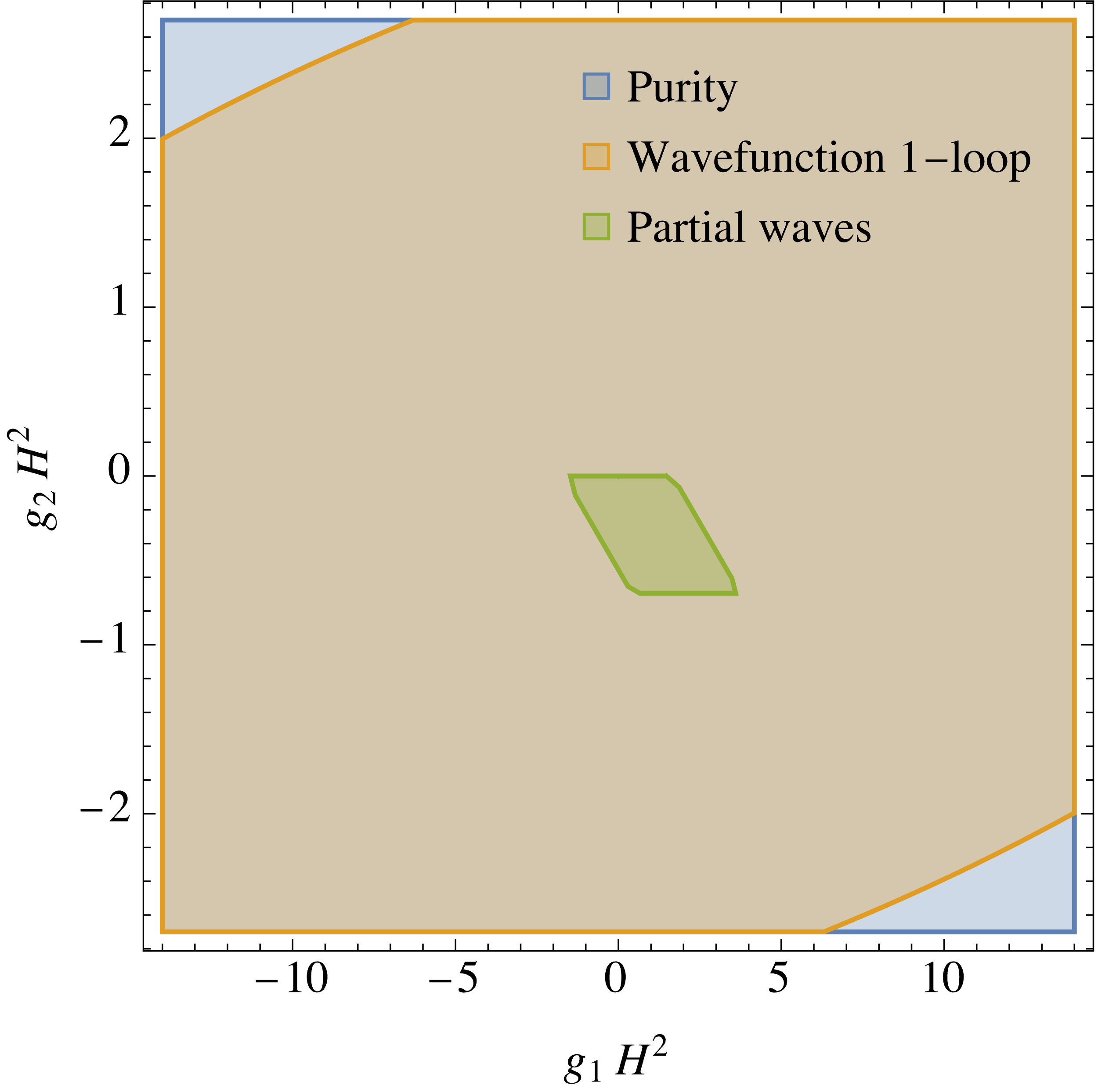}
\end{subfigure}
\caption{Constraints on parameter space for the EFToI~\eqref{eq:EFToI}. The left plot compares the purity bound~\eqref{eq:pb_EFToI} (blue) with the 1-loop wavefunction bound found in~\cite{Melville:2021lst}, i.e.~\eqref{eq:1lwb_EFToI} (brown). The right plot zooms in the region around the origin and adds on the partial wave bounds of~\cite{Grall:2020tqc}, i.e.~\eqref{eq:pwb_EFToI} (green).}
\label{fig:EFToI}
\end{figure}

The purity bound $\gamma\geq 0$ then implies, at this order in perturbation theory, that only an elliptic region of the parameter space $\{g_1,g_2\}$ is allowed, namely
\begin{equation} \label{eq:pb_EFToI}
    \gamma \geq 0 \qquad\Rightarrow\qquad g_1^2 + 69g_2^2 - 16g_1g_2 \lesssim \frac{3435}{H^4} \, .
\end{equation}
We can now compare the purity bound to other bounds on the size of the EFToI operators found in previous literature. On the one hand, the parameters $\{g_1,g_2\}$ were also constrained to an elliptic region in~\cite{Melville:2021lst} by demanding the 1-loop contribution to the two-point wavefunction coefficient to be smaller than the tree-level one. The result was
\begin{equation} \label{eq:1lwb_EFToI}
    \psi_2^{\text{(1-loop)}} <\psi_2^{\text{(tree)}} \qquad\Rightarrow\qquad g_1^2 + 77g_2^2 - 9g_1g_2 \lesssim \frac{754}{H^4} \, .
\end{equation}
The left plot of Figure~\ref{fig:EFToI} compares this with the purity bound, showing that the corresponding ellipses have a similar tilt but the 1-loop bound~\eqref{eq:1lwb_EFToI} is a factor of a few times stronger. It was further noted in~\cite{Melville:2021lst} that this bound is already superceded by the very strong experimental constraints on primordial non-Gaussianity.

On the other hand, the authors of \cite{Grall:2020tqc} were able to generalize the amplitudes' partial wave bounds~\eqref{eq:pwb} to non-Lorentz-invariant theories, and used the results to constrain the size of the EFToI operators by studying inflaton scattering deep inside the horizon. The results were obtained numerically, but we find the allowed region to correspond approximately to the rhomboid\footnote{The bound $g_2\leq 0$ in~\eqref{eq:pwb_EFToI} does not come from perturbative unitarity, but from the fact that in the EFToI this coupling is
\begin{equation}
    g_2 = - \frac{1-c_s^2}{f_{\pi}^2c_s^2} \leq 0 \, ,
\end{equation}
with $0<c_s\leq 1$ the speed of sound and $f_{\pi}$ the symmetry-breaking scale. In deriving the purity or 1-loop wavefunction bounds, the physical origin of the action~\eqref{eq:EFToI} has been ignored, so the ellipses in Figure~\ref{fig:EFToI} extend on both sides of $g_2=0$.}
\begin{equation} \label{eq:pwb_EFToI}
    \begin{rcases}
        \left| \dfrac{g_1}{3}+g_2 \right| H^2 \lesssim \dfrac{5}{9} \\
        -\dfrac{25}{36} \lesssim g_2H^2 \leq 0
    \end{rcases} \, .
\end{equation}
This region is shown in the right plot of Figure~\ref{fig:EFToI}. We see that this constraint is much stronger than the purity or 1-loop wavefunction ones. It would be nice to understand why both cosmology-native bounds are much weaker than those derived from flat-spacetime. Either there are other cosmology-native bounds that diagnose a much earlier breakdown or the EFT in de Sitter is actually capable of consistently computing cosmological observables even when it fails to correctly predict sub-Hubble processes. 

Interestingly, all the methods give us straightforward bounds on the maximum size of the couplings, rather than an energy dependent cutoff as we saw in some of the Lorentz-invariant flat space examples. The reason for this is that in the de Sitter theory the scale invariance of the interactions fixes the scaling of $\psi_3$ to go like $k^3$, whilst the derivatives of the interaction increase the scaling with the finite $p$ at the vertex. This implies a scaling $p^2k$ as $\textbf{k}\rightarrow \infty$ and so the purity integral converges to a number. In fact, the convergence of this integral is a feature of all scale invariant interactions involving at least three derivatives. This is the characteristic behaviour in single field inflation, once the contribution to local non-Gaussianity implied by Maldacena's consistency condition \cite{Maldacena:2002vr} is correctly interpreted as an artifact of the choice of unperturbed comoving coordinates (see e.g. \cite{Pajer:2013ana}).

\subsection{Effective field theories: $ \phi\dot \phi^2$}\label{sec:phidotphi}

As we mentioned, interactions with three or more derivatives will just return results of the same form as those in Section~\ref{sec:EFToI}. However, we can find a very different behaviour when reducing the number of derivatives. If we reduce the number of derivatives too far then the interaction will have logarithmic divergences at late times. These are also called IR divergences, but we stress that they are different from the IR divergences we encountered in the calculation of the purity in that they appear already at the level of the initial wavefunction, before any trace has been performed. To avoid having to discuss these late-time divergences, we consider the interaction $\phi\dot \phi^2$. This interaction has a wavefunction coefficient that is given by 
\begin{equation}
    \psi_3 (k_1,k_2,k_3) = \dfrac{g}{ H^2} \left( \dfrac{e_3e_2}{k_T^2} + \frac{e_2^2}{k_T} -2e_3\right).
    \label{eq:dS EFT psi3}
\end{equation}
For this cubic wavefunction coefficient, the superficial degree of divergence for our purity integral goes like $k^3$, which is dictated by scale invariance for a massless field. However, unlike in the flat space case this integral does not just diverge as $\textbf{k}\rightarrow \infty$. Instead, the purity integral looks like
\begin{align}
    \gamma=1-g^2H^2\int_{k_1=0}^\infty\int_{k_2=\lvert k_1-p\rvert}^{k_1+p} \frac{dk_2dk_1}{p^2\pi^2}\frac{\left(e_3e_2+e_2^2k_T-2e_3k_T^2\right)^2}{8k_T^4e_3^2}\,.
    \label{eq:dS EFT purity integral}
\end{align}
This diverges when $k_1$ approaches $0,\,p$ and $\infty$. These three limits can be understood together as they all correspond to the ratio between two of the energies diverging: the squeezed limit. At leading order in the squeezed limit~\eqref{eq:symmetric squeezed limit}, the wavefunction coefficient~\eqref{eq:dS EFT psi3} tends to
\begin{equation}
   \lim_{k_l\to 0}\psi_3(k_1,k_2,k_3) = \frac{g}{H^2} k_s^3 \left[ \frac{1}{2} + \frac{25 - 7 \cos^2\theta}{16} \left( \frac{k_l}{k_s} \right)^2 + \mathcal{O}\left( \frac{k_l}{k_s} \right)^3 \right] \, ,
\end{equation}
which does not vanish, and certainly does not vanish quickly enough to guarantee a finite purity.
This is to be contrasted with the cubic wavefunction coefficient~\eqref{eq:EFTI psi3} from the EFT of inflation, which has squeezed limit in~\eqref{EFTsqueezed} and vanishes quickly enough to yield finite purity~\eqref{eq:EFTI purity} without a cutoff: the entanglement of hierarchically different scales is suppressed in the EFT of inflation because of the larger number of derivatives implied by the shift symmetry. 

\begin{figure}[h]
   \centering
   \includegraphics{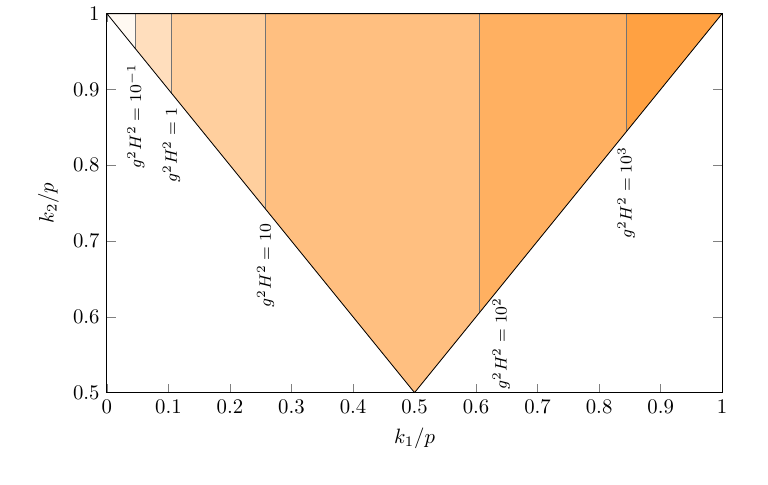}
   \caption{The allowed kinematic space of the $g \phi \dot \phi^2/2$ theory on de Sitter spacetime. The squeezed limit is approached in the top-left corner of the plot, where $k_1\ll k_2 \sim p$. For each value of the coupling $g$, a vertical line is drawn corresponding to the maximum $k_s/k_l$ allowed by perturbative unitarity, namely~\eqref{eq:dS EFT purity integral}.  For each value of the coupling $g$, the region to the left of the corresponding vertical line is excluded. Very squeezed kinematics are outside of the control of perturbation theory.}
   \label{fig:squeezed cutoff}
\end{figure}

Returning to the present case, we need to regulate the purity integral, cutting it off at a maximum value of the squeezing ratio $k_s/k_l$, where $k_s$ is the largest momentum in the problem (shortest wavelength) whilst $k_l$ is the smallest momentum (longest wavelength) and so this ratio is large. The purity~\eqref{eq:dS EFT purity integral} associated to the wavefunction coefficient~\eqref{eq:dS EFT psi3}, to leading order in the squeezing ratio~\footnote{Just as in the flat spacetime case, the integration region depends on the size of this ratio and care must be taken if we want to take this ratio close to one.}, is
\begin{align}
    \gamma = 1-\frac{g^2 H^2}{92\pi^2} \left( \frac{k_s}{k_l} \right)^3 +\mathcal{O}\left(\frac{k_s}{k_l}\right)^2.
\end{align}
We therefore see that insisting that the purity is well defined restricts the maximum value of the ratio permitted by the EFT,
\begin{align}
    \left( \frac{k_s}{k_l} \right)^3 \lesssim \left( \frac{30.8}{g H}\right)^2.
\end{align}
This behaviour, shown in Figure~\ref{fig:squeezed cutoff}, is rather different from flat space where, for fixed coupling, we could view this as a bound on the maximum energy beyond which the EFT is invalid. However, such a difference should not be surprising as, in de Sitter, the boundary observables are scale invariant. 

Therefore, it is possible to reach arbitrarily high energies withing the EFT through a rescaling and we are unable to detect such a breakdown in the perturbativity of the theory. In spite of this, we still expect perturbation theory to break down, we just need to adjust our view on the regime of validity. If we were to consider modes at vastly different scales then we would be permitting coupling between the unknown UV modes and the IR. This behaviour is something that our EFT is sensitive to as the low energy mode essentially sets a scale for us by which we can define the other energy as high. Therefore, the observed dependence on the ratio between the short and long modes is precisely what we should expect.

\subsection{Bogoliubov initial state: ${\dot \phi}^3$}\label{sec:bogo}
The $ \phi\dot \phi^2$ interaction in the Bunch-Davies vacuum discussed in the previous subsection produced a $\psi_3$ wavefunction coefficient that did not decay quickly in the squeezed limit; the purity bound then required a restriction on the momentum hierarchy that could be described by the EFT. We now consider a $g \dot \phi^3/3!$ interaction with a Bogoliubov initial state instead. These states are related to the Bunch-Davies vacuum by a Bogoliubov transformation in each Fourier mode \cite{Brandenberger:2002hs,Meerburg:2009ys,Ganc:2011dy}. In these vacua, the integrand in the purity diverges in folded configurations of momentum. Folded configurations satisfy  $k_1 + k_2 = p$ or a permutation thereof.
In the case of contact wavefunction coefficients, there is a convenient mapping between Bunch-Davies and Bogoliubov wavefunctions; for simple exchange diagrams, a similar mapping has also been explored \cite{Ghosh:2023agt,Ghosh:2024aqd}.

A Bogoliubov initial state is specified by two functions $\alpha_k$ and $\beta_k$ describing the deviation from the Bunch-Davies state for each mode $|\bfk|$. To avoid strong back-reaction problems from a large number of particles with an arbitrarily high momentum, we assume that $\beta_k$ goes to zero as $k$ becomes very large. Conversely, to keep our analysis as simple as possible, for all the modes of physical interest in our investigation we set the Bogoliubov $\alpha_k$ parameters equal to a real constant $\alpha$, and all $\beta_k$ parameters equal to a real constant $\beta$ satisfying $\abs{\alpha}^2 - \abs{\beta}^2 = 1$. This can be achieved by choosing $\alpha_k$ and $\beta_k$ constant on CMB and LSS scales, but $\beta_k \simeq 0$ on all other scales. \\
\begin{figure}[h]
   \centering
   \includegraphics{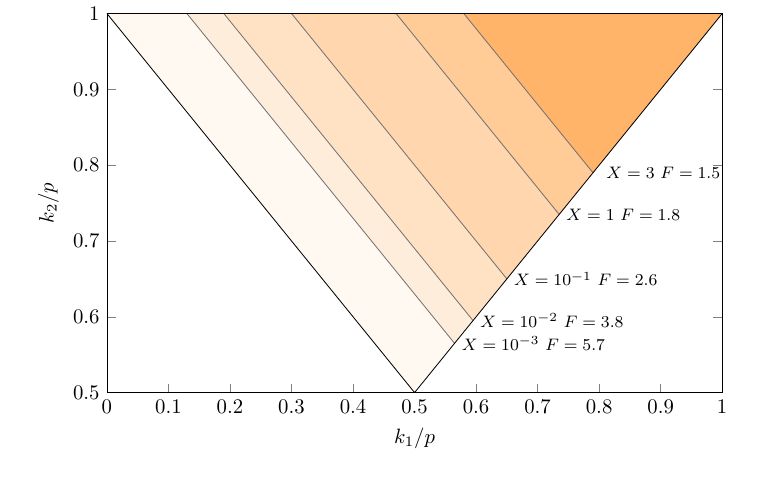}
   \caption{The kinematics allowed by the purity bound for the $g\dot\phi^3/3!$ interaction with a Bogoliubov initial state as a function of the parameter $X=g^2 H^4\alpha^2\beta^2(\alpha-\beta)^2$. The space to the left of each line is excluded by the purity bound associated to the choice of $X$, so that very folded kinematics are outside of the control of perturbation theory. The kinematics are characterised by the $F$ of the most folded triangle included in the EFT (defined in~\eqref{eq:folded regulator}), with $F=1$ corresponding to an equilateral triangle and $F\rightarrow\infty$ corresponding to the folded limit.}
   \label{fig:bogo kinematics}
\end{figure}

In this case, given the cubic wavefunction coefficient of the $g \dot \phi^3/3!$ interaction in the Bunch-Davies vacuum
\begin{equation}
   \psi_3^{\text{BD}} (k_1,k_2,k_3) = -\frac{2g}{H} \frac{k_1^2 k_2^2 k_3^2}{k_T^3} \, ,
\end{equation}
the Bogoliubov wavefunction coefficient takes the form
\begin{align}
   \psi_3 &=  \alpha^3 \psi_3^{\text{BD}} (k_1,k_2,k_3) + \alpha^2 \beta \left( \psi_3^{\text{BD}} (k_1,k_2, -k_3) + \text{ perms} \right) \nonumber \\
   & \quad + \alpha \beta^2 \left( \psi_3^{\text{BD}} (k_1,-k_2, -k_3) + \text{ perms} \right) + \beta^3 \psi_3^{\text{BD}} (-k_1,-k_2, -k_3) \, .
\end{align}
This wavefunction coefficient diverges for folded configurations of momenta unless $\beta=0$, and this divergence results in a divergence in the purity.
Much as we regulated the squeezed-limit divergence in the Bunch-Davies vacuum by restricting the squeezing ratio $k_s/k_l$, we can look for a scale-invariant regulator for the folded singularities. One option is
\begin{equation}
   \text{folded-singularity regulator: }\quad \abs{k_1 + k_2 - p} \geq F^{-1} \, \frac{k_T}{3} \quad \text{ \& perms} \label{eq:folded regulator}
\end{equation}
where $F > 1$ characterises the most folded triangles contributing to the purity.
The integral for the purity takes on a complicated form. An analytic expression exists, and when $g H^2 \ll 1$ and $F\rightarrow\infty$, so very folded kinematics are included, the purity approaches
\begin{equation}
   \gamma \approx 1 - \frac{1}{2\pi^2} \left( \frac{243 X \left(10 \log \left(3 F \right)-19\right)}{800 } F^5 + \mathcal{O}\left(F^4 \right) \right) \, ,
\end{equation}
where $X \coloneqq g^2 H^4\alpha^2 \beta^2 (\alpha-\beta)^2$.
So in the limit of weak coupling, the purity bound restricts the most folded triangles that can be described in the EFT to lie in the region of kinematic space with
\begin{align}
   F &\lesssim \frac{1}{3} \exp \left( \frac{19}{10} + \frac{1}{5} W_0\left( \frac{800 \pi^2 }{ e^{19/2} X} \right) \right) \\
   & \sim 2.3 \left[ X\left(\log \left(X^{-1}\right)-10.0\right) \right]^{-\frac{1}{5}}
\end{align}
where $W_0$ is the principal branch of the product logarithm/Lambert $W$ function and the second line follows from an asymptotic expansion of this function as $X \rightarrow 0$.
Divergences in folded kinematics are generic features of non-Bunch-Davies initial states, so the purity bound will generically require a restriction of the EFT to non-folded kinematics. The restricted kinematics enjoyed by this particular interaction are shown in Figure~\ref{fig:bogo kinematics}.

\subsection{Purity bounds on local non-Gaussianity}

In this subsection, we explore the purity bounds on local-type non-Gaussianity. Instead of working with a concrete model, we start from a phenomenological non-Gaussian bispectrum that peaks in the local shape, \eqref{loc}, we derive the corresponding cubic wavefunction coefficient and then we use it to compute the purity. This provides a simplified scenario in which the bound can be calculated very explicitly. We show that the purity bound gives a similar result to demanding that the one-loop contribution to the power spectrum is small compared to the tree-level one. 
As far as models are concerned, this setup is relevant for typical multifield models, since in single field inflation local non-Gaussianity is absent when working in physical coordinates at early times and only emerges from so-called projection effects \cite{Pajer:2013ana,Cabass:2016cgp}, which are observable dependent (e.g. they are absent to leading order for the cross correlation of spectral distortions and temperature anisotropies \cite{Cabass:2018jgj}).\\

\paragraph{Purity bounds} The bispectrum template for local-type non-Gaussianity takes the form
\begin{equation}\label{loc}
   B_3(k_1, k_2, k_3) = \frac{6}{5} \fnl \left( \frac{1}{k_1^3 \, k_2^3} \, + \text{2 perms} \right) \, ,
\end{equation}
corresponding to  wavefunction coefficients
\begin{align}\label{psi3loc}
   \psi_3(p, k_1, k_2) &= \frac{3}{5} \fnl \left( 2\psi_2(p) +  2\psi_2(k_1) + 2\psi_2(k_2)  \right)\,, \\
   2 \, \psi_2(k) &= \frac{k^3}{A} \,,
\end{align}
with $A \approx 2\pi^2 e^3 10^{-10} $ the amplitude of the primordial power spectrum \cite{Planck:2018vyg}.

One can demand that perturbation theory is valid for all scales $\kmin \leq k \leq \kmax$ visible in the CMB, which has $\kmax \approx 10^3\,\kmin$. To avoid having to discuss several special cases, it is convenient to assume $\kmin > p/2$ and $\kmax > \kmin + p$, where, as usual, $p$ is the norm of the mode whose purity we calculate.
This assumption is compatible with taking $p = \kmin$, which will yield the strongest bound.
The purity is found as usual through an integral on the hexagonal region $R$ in the $(k, k_+)$ plane of possible triangles with one side $p$ and no side shorter than $\kmin$ or longer than $\kmax$:
\begin{equation}
\begin{aligned}
   \gamma &= 1 - 2 \int_R \frac{2\pi k_1 k_2 \dd k_1 \dd k_2}{(2\pi)^3 p} \, \frac{\abs{\psi_3(p, k_1, k_2)}^2}{ 8 \psi_2 (p) \psi_2 (k_1) \psi_2 (k_2)} \\
   \Leftrightarrow \, 1 - \gamma &= \frac{4 \pi A}{(2\pi)^3 p } \left( \frac{3}{5} \fnl \right)^2 \int_R \dd k_1 \, \dd k_2 \, \left( \frac{p^3}{k_1^2 k_2^2} + \frac{k_1^4}{ p^3 k_2^2} + \frac{k_2^4}{ p^3 k_1^2} + \frac{k_1 k_2}{k_1^3} + \frac{k_1 k_2}{k_2^3} + \frac{k_1 k_2}{p^3} \right)\,.
\end{aligned}
\end{equation}
The region $R$ is bounded by the lines
\begin{align}
   k_1 &= k_2 + p\,, & k_2 &= k_1 + p\,, \\
   k_1 &= \kmin\,, & k_2 &= \kmin\,, \\
   k_1 &= \kmax\,, & k_2 &= \kmax \, .
\end{align}
The assumption $\kmin > p/2$ means that the line $k_1 + k_2 = p$ is not part of the boundary of $R$.
The strongest constraint on the parameter $\fnl$ is achieved for $p = \kmin$, in which case the result of the integral is
\begin{align}
   1 - \gamma &= \frac{4\pi A}{(2\pi)^3} \left( \frac{3}{5} \fnl \right)^2 \left( \left( 2  \frac{\kmax}{\kmin} \right)^3 - \frac{3}{2}\left( \frac{\kmax}{\kmin} \right)^2 + 6 \left( \frac{\kmax}{\kmin} \right) - \frac{267}{20}  \right. \\
   & \quad \left. + \frac{13}{5} \left( \frac{\kmax}{\kmin} \right)^{-1} + \left( \frac{\kmax}{\kmin} \right)^{-2} + 6\ln\left(  \frac{\kmax}{\kmin} - 1 \right) + 2 \ln\left( \frac{\kmax}{2 \kmin} \right) \right) \, .
\end{align}
Notably, the result scales as $(\kmax/\kmin)^3$. Taking $\kmin$ to be fixed by the size of the observable universe and $\kmax$ by the smallest mode we can measure, we recognize this factor as the number of independent modes we observe. The purity bound $\gamma \geq 0$ on $\fnl$ requires
\begin{equation}\label{fnlbound}
   \abs{\fnl} \lesssim 0.8 \, .
\end{equation}
This appears to be a surprisingly stringent constraint, especially compared to the usually quoted perturbativity bound $\fnl \Delta_\zeta \ll 1$. The difference is that the purity bound is larger by a factor of $(\kmax/\kmin)^{3/2}$. This is the square-root of the number of (independent) observed modes. So, even though each mode individually is well described by the EFT, the small mistakes in each mode add up to give a large mistake in the full purity, which ends up violating the unitarity bound.

The bound in~\eqref{fnlbound} implies that a near-future detection of local-non-Gaussianity around $\fnl\sim \mathcal{O}(1)$ would need to be explained without resorting to perturbation theory. On this point, it is worth recalling that one of the simplest models that predicts local non-Gaussianity is the curvaton model \cite{Lyth:2002my}, in which non-Gaussianity is typically computed to zeroth order in an expansion in derivatives, an approach that is known as the $\delta N$ formalism. In particular, the non-Gaussianity is related to derivatives of the number of e-foldings $N$ with respect the initial value $\phi_i$ of fields, assuming a purely homogeneous evolution. Typically $N(\phi_i)$ is computed by numerically solving the homogeneous equations of motion or by analytically solving them in some approximation, such as slow roll or that the universe is dominated by the curvaton after inflation. This approach gives non-perturbative but semi-classical results, and can therefore be safely used to compute predictions in which $\fnl$ violates our purity bound.

\paragraph{Perturbativity of the loop expansion}
Because the bound we found above is surprisingly strong, it is interesting to see if we can find other hints of the breakdown of the perturbative expansion. Indeed, as we saw previously, the one-loop contribution to the power spectrum becomes bigger than the tree-level one precisely when the purity bound is violated. To see this we recall the argument given in Section~\ref{sec3p3}. The power spectrum at one loop can be written in terms of wavefunction coefficients as in~\eqref{P1loop}. Notice that, since $\psi_3$ is real in this case, the classical loop contribution to the one-loop power spectrum in~\eqref{P1loop} is precisely the same integral we encountered in the purity, \eqref{eq:puritycubic}. Therefore, assuming that there is no precise cancellation\footnote{Precise cancellations between these two terms are known to take place in another context \cite{Lee:2023jby,Agui-Salcedo:2023wlq}.} between this last classical loop term and the quantum loops in $  \psi_{2}^{(1L)} $, we conclude that the purity bound correctly captures the breakdown of perturbation theory in the calculation of the power spectrum. This parallel with the one-loop power spectrum

One last comments concerns counterterms and renormalization. Some of the contributions to purity and to the one-loop power spectrum may be degenerate with local counterterms that need to be added to renormalize the theory and therefore one may worry about the robustness of our bound~\eqref{fnlbound}. We don't investigate this in detail because the process of renormalization in inflationary/de Sitter spacetime is subtle and not well studied. However, we don't expect that renormalization will appreciably change our purity bound because of the following argument. Renormalization is supposed to remove contributions to observables that come from modes that are \textit{outside} the validity of an EFT, such as for example arbitrarily high-energy modes. Conversely, the modes contribution to our calculation of the purity are all supposed to be well described by the EFT under investigation, which we assume is valid for all modes we observed, say, in the CMB. It would be interesting to investigate this issue in more detail.

\section{Conclusions and outlook} \label{ConclusionsSect}

In this work we have proposed a new breakdown diagnostic for perturbation theory in effective field theories (EFTs). The object of study is the purity $\gamma$ of the reduced density matrix obtained from tracing out all Fourier modes in an EFT except for one. In perturbation theory, $\gamma$ violates its unitarity bounds, $0\leq \gamma \leq 1$, either when the couplings become too strong or when one traces out modes that are hierarchically separated. This captures the fact that a typical EFT can only describe a finite range of Fourier modes and breaks down when extrapolated beyond this regime. We have shown that purity bounds on perturbative unitarity qualitatively reproduce bounds from partial wave scattering amplitudes and the general scaling of loop contributions. However, we have also noticed some important differences:
\begin{itemize}
    \item Purity, in contrast to amplitudes, does depend on the choice of fields and hence can detect problems that would be completely invisible to amplitude techniques. Since in cosmology one is interested in the correlators of certain specific fields\footnote{For example, even if inflation involved very many fields, we seem to see only ``adiabatic" fluctuations, corresponding to one specific direction in the larger field space.}, as opposed to some field-redefinition invariant quantity, this is gives purity bounds an important unique advantage compared to partial wave bounds. 
    \item Purity bounds rely on the existence of a density matrix for the system, which can be pure or mixed. This does not require any specific asymptotic behaviour at spatial or temporal infinity. This means that purity bounds can be straightforwardly employed in curved spacetime. This is in contrast to partial wave bounds, which, strictly speaking, only apply to flat spacetime. Since cosmological observables need to be computed on super-Hubble scales, it is desirable to have bounds that do not require any flat spacetime limit. 
    \item Purity bounds are sharp and give precise inequalities. This is in contrast to bounds obtained from estimates of higher order  corrections such as loop contributions. 
    \item Placing a scale invariant theory defined in de Sitter into flat space spoils this symmetry, therefore, bounds of the type seen in Section~\ref{sec:phidotphi} will be lost by considering partial waves of the equivalent flat space theory, even if the boost breaking is taken into account. As the purity is defined directly in curved spacetime, it permits a scale invariant diagnostic of the breakdown of perturbation theory.
\end{itemize}
Since our work explores a new perturbative unitarity bound, there are many avenues for future research:
\begin{itemize}
    \item It would be interesting to make more extensive comparison between purity and partial wave bounds in flat space to see if the qualitative agreement applies more generally or if there are other important differences. Gauge theories and gravity would be of particular interest in this respect.
    \item We have noticed that the large-$N$ traces $\Tr (\rho_{\rm R}^N)$, $N\gg 2$, can signal a breakdown of perturbation theory in the same regime in which the purity is still compatible with unitarity. It might be interesting to study what this implies for the perturbative computation of high-multiplicity observables, like cosmological correlators of a large number of fields.
    \item Many different quantitative measurements of entanglement have been proposed, such as von Neumann entropy, R\'enyi entropy, Tsallis entropy, etc.. It would be important to have a better understanding of what quantity is a better breakdown diagnostic in different situations, for example different cosmological models with different number of fields, IR behaviour, UV behaviour etc.
    \item Whenever the Hilbert space of a free quantum field theory factorises into eigenmodes, it should, in principle, be possible to bound the validity of perturbation theory using the purity of a single mode. This might apply, for example, to spherically symmetric spacetimes like the Schwarzschild solution. It would be interesting to see how the purity bound might apply to interacting theories on such spacetimes.
    \item Very high dimension operators in de Sitter can become large in spite of the energy suppression due to the factorial growth in the strength of the interaction coming from the infinite past. This enhancement is absent when considering the flat space limit of the theory and so purity considerations could be used to understand a more appropriate power counting scheme in the context of de Sitter. 
    \item It would be interesting\footnote{We thank the anonymous referee for suggesting this.} to see if our bounds can be made sense of in the case in which one regulates loops not with hard cutoffs, but in a different way, e.g. in dim reg. The interpretation in terms of tracing over part of the Hilbert space would then be lost, but perhaps something interesting could come out of it.
\end{itemize}

\paragraph{Acknowledgements} We would like to thank Daniel Baumann, Marine de Clerck, Thomas Colas, Jackson Fliss, Jiri Minar, Geoff Penington, David Skinner, Ronak Soni and Dong-Gang Wang for useful discussions. E.P. and C.D.P. have been supported in part by the research program VIDI with Project No. 680-47-535, which is (partly) financed by the Netherlands Organisation for Scientific Research (NWO). C.M.~is supported by Science and Technology Facilities Council (STFC) training grant ST/W507350/1. This work has been partially supported by STFC consolidated grant ST/T000694/1 and ST/X000664/1 and by the EPSRC New Horizon grant EP/V017268/1. H.G. is supported by a Postdoctoral Fellowship at National Taiwan University funded by the National Science and Technology Council (NSTC) 113-2811-M-002-073. H.G. was also supported jointly by the Science and Technology Facilities Council through a postgraduate studentship and the Cambridge Trust Vice Chancellor’s Award. H.G. was also supported by a Postdoctoral Fellowship at National Taiwan University funded by the Ministry of Education (MOE) NTU-112L4000-1. The work of C.D.P. is supported by the Scuola Normale Superiore.

\newpage
\appendix

\section{Examples of diagrammatic computations} \label{app:DiagEx}

Here we illustrate the diagrammatic formalism introduced in Section~\ref{sec:diagrams} with some examples. For simplicity, we work with theories that have $\psi_n=0$ for all except one $n\geq 3$, such that all blobs in our diagrams will have the same number of legs. Keep in mind, however, that this will not be the case for a realistic theory (even if their Lagrangian only contains one interaction), and in general diagrams can contain blobs with different numbers of legs.

\subsection{Three-point wavefunction coefficient}

We start by presenting the diagrams for a theory that only has a three-point wavefunction coefficient $\psi_3$ (and $\psi_n=0$ for $n>3$) at order $\mathcal{O}(\psi_3^2)$. The trace of $\rho_R$ was already given in~\eqref{eq:TrrhoPsi3},
\begin{equation}
    {\rm Tr}\, \rho_{\rm R} = 1 + 
    \frac{1}{12} \, \vcenter{\hbox{
}} \hspace{-25pt} \, .
\end{equation}
The reduced density matrix is
\begin{align}
    & \qquad \rho_{\rm R} = \frac{1}{\tilde{\mathcal{N}}^{\,2}} \cdot \exp{\left( - \frac{\psi_2 (\mathbf{p})}{L^3} |\phi_{\mathbf{p}}|^2 - \frac{\left(\psi_2 (\mathbf{p}) \right)^*}{L^3} |\bar{\phi}_{\mathbf{p}}|^2 \right) } \cdot \nonumber \\
    & \cdot \left( \vphantom{
    \vcenter{\hbox{%
}} \right) \, , \nonumber
\end{align}
and the trace of $\rho_R^2$ is
\begin{align}
    {\rm Tr}\, \rho_{\rm R}^2 & = 1 + 2 \left(
    \frac{1}{8} \hspace{-25pt} \vcenter{\hbox{%
}} \right) \, . \nonumber
\end{align}
We can now use
\begin{equation} \label{eq:blobsum}
\vcenter{\hbox{%
}}
\end{equation}
and the fact that inside a loop
\begin{equation} \label{eq:linesum}
\vcenter{\hbox{%
}}
\end{equation}
to rewrite the above result as\footnote{Notice that lines whose momentum is fixed by momentum conservation to be some $\bfk\in\mathcal{E}$ can be represented by either a solid or a dashed line. We use this freedom in four diagrams of~\eqref{eq:TrRho2Psi3_2}, two examples being the middle line of the first diagram, which has momentum $\bfk=\mathbf{0}$, and the middle line of the last diagram, which has $\bfk=2\bfp$.}
\begin{align}
    {\rm Tr}\, \rho_{\rm R}^2 = 1 & + 2 \left(
    \frac{1}{8} \hspace{-25pt} \vcenter{\hbox{%
}} \right) \, . \nonumber
\end{align}
Therefore, the purity at $\mathcal{O}(\psi_3^2)$ is
\begin{equation} \label{eq:psi3resb}
\begin{aligned}
    \gamma = \frac{{\rm Tr}\, \rho_{\rm R}^2}{\left( {\rm Tr}\, \rho_{\rm R} \right)^2} & = 1 - 2 \left(
    \frac{1}{2} \vcenter{\hbox{%
}} \right) \, .
\end{aligned}
\end{equation}
Taking into account the volume factors in accordance to~\eqref{eq:volumef} and going to the continuum ($L\to\infty$) we finally arrive to the result in~\eqref{eq:purityPsi3},
\begin{equation}
    \gamma = 1 
    - \vcenter{\hbox{%
}} \,\, .
\end{equation}

\subsection{Four-point wavefunction coefficient}

For a theory with only a four-point wavefunction coefficient $\psi_4$, and staying at $\mathcal{O}(\psi_4^2)$, the trace ${\rm Tr} \, \rho_R$ has contributions from both connected and disconnected diagrams. As advanced in~\eqref{eq:Trrho}, the result can be expressed as an exponential of just the connected ones (recall that we are omitting terms of $\mathcal{O}(\psi_4^3)$ and higher):
\begin{equation} \label{eq:TrrhoPsi4}
\begin{aligned}
    {\rm Tr}\, \rho_{\rm R} & = 1 + 
    \frac{1}{8} \hspace{-25pt} \vcenter{\hbox{
}}
    \hspace{3pt} \right\} \, .
\end{aligned}
\end{equation}
The reduced density matrix for this theory is
\begin{align}
    & \qquad \rho_{\rm R} = \frac{1}{\tilde{\mathcal{N}}^{\,2}} \cdot \exp{\left( - \frac{\psi_2 (\mathbf{p})}{L^3} |\phi_{\mathbf{p}}|^2 - \frac{\left(\psi_2 (\mathbf{p}) \right)^*}{L^3} |\bar{\phi}_{\mathbf{p}}|^2 \right) } \cdot \nonumber \\[5pt]
    & \qquad\quad \cdot \left\{ 1 + 
    \frac{1}{8} \hspace{-25pt} \vcenter{\hbox{%
}}} \right\} \nonumber \, .
\end{align}
The trace of $\rho_R^2$ is then
\begin{align}
    {\rm Tr}\, \rho_{\rm R}^2 = & \, 1 + 2 \left(
    \frac{1}{8} \hspace{-25pt} \vcenter{\hbox{%
}} \hspace{-25pt} \right)^2 \, , \nonumber
\end{align}
and again using~\eqref{eq:blobsum} and~\eqref{eq:linesum} we rewrite it as
\begin{align}
    {\rm Tr}\, \rho_{\rm R}^2 = 1 & + 2 \left(
    \frac{1}{8} \hspace{-25pt} \vcenter{\hbox{%
}} \hspace{-25pt} \right)^2 \, . \nonumber
\end{align}
The purity at $\mathcal{O}(\psi_3^2)$ is then
\begin{align}
    \gamma = \frac{{\rm Tr}\, \rho_{\rm R}^2}{\left( {\rm Tr}\, \rho_{\rm R} \right)^2} = 1 - 2 & \left(
    \frac{1}{6} \, \vcenter{\hbox{%
}} \right) \, , \nonumber
\end{align}
and after taking into account the volume factors and going to the continuum this becomes
\begin{equation} \label{eq:purityPsi4}
\begin{aligned}
    \gamma & = 1 
    - \frac{1}{3} \, \vcenter{\hbox{%
}} \\
    & = 1 - \frac{1}{3} \int_{\mathbf{k}_1,\mathbf{k}_2} {\frac{\left| \psi_4 (\mathbf{p},\mathbf{k}_1,\mathbf{k}_2,-\mathbf{p}-\mathbf{k}_1-\mathbf{k}_2) \right|^2+\left| \psi_4 (-\mathbf{p},-\mathbf{k}_1,-\mathbf{k}_2,\mathbf{p}+\mathbf{k}_1+\mathbf{k}_2) \right|^2}{2\text{Re}\, \psi_2(\mathbf{p}) \, 2\text{Re}\, \psi_2(\mathbf{k}_1) \, 2\text{Re}\, \psi_2(\mathbf{k}_2) \, 2\text{Re}\, \psi_2(\mathbf{p}+\mathbf{k}_1+\mathbf{k}_2)}} \, .
\end{aligned}
\end{equation}

\subsection{Six-point wavefunction coefficient}

For a theory with only a six-point wavefunction coefficient $\psi_6$ we just show the purity diagrams $D^{(c)}$. This is the lowest $\psi_n$ for which there are two diagrams at leading order $\mathcal{O}(\psi_6^2)$. They are
\begin{equation} \label{eq:purityPsi6}
    D^{(c)} = \left\{
    \frac{1}{4!} \, \vcenter{\hbox{\begin{tikzpicture}
    \coordinate (o) at (0,0);
    \coordinate (c1) at ([shift={(-30pt,0)}]o);
    \coordinate (c2) at ([shift={(30pt,0)}]o);
    \draw[thick] (c1) -- (c2);
    \draw[thick] (c1) to[out=-90,in=-90] (c2);
    \draw[thick,-><-] (c1) to[out=90,in=90] (c2);
    \draw[thick] (c1) to[out=-35,in=-145] (c2);
    \draw[thick] (c1) to[out=35,in=145] (c2);
    \draw (c1) pic {wb};
    \draw (c2) pic {bb};
    \end{tikzpicture}}} \, ,
    \frac{1}{8} \hspace{-15pt} \vcenter{\hbox{\begin{tikzpicture}
    \coordinate (o) at (0,0);
    \coordinate (c1) at ([shift={(-30pt,0)}]o);
    \coordinate (c2) at ([shift={(30pt,0)}]o);
    \draw[thick] (c1) to[out=55,in=125,distance=50pt] (c1);
    \draw[thick] (c2) to[out=55,in=125,distance=50pt] (c2);
    \draw[thick] (c1) -- (c2);
    \draw[thick] (c1) to[out=-45,in=-135] (c2);
    \draw[thick,-><-] (c1) to[out=45,in=135] (c2);
    \draw (c1) pic {wb};
    \draw (c2) pic {bb};
    \end{tikzpicture}}} \hspace{-20pt} \, \right\} \, .
\end{equation}

\newpage

\bibliographystyle{JHEP}
\bibliography{CTCRefs}

\end{document}